%% file: thesis.tex
\documentclass[12pt,normalmargins]{ut-thesis}
\providecommand{\waltonExampleOnly}[1]{}
\providecommand{\smokingAndCancerExampleOnly}[1]{}

\usepackage{microtype}
\usepackage{paralist}
\newcounter{counterone}
\newcounter{countertwo}

\usepackage{dustin_fav_packages}
\usepackage{dustin_utilities_etc}
\usepackage{dustin_fonts}
\usepackage{dustin_environments}
\usepackage{dustin_macros}
\usepackage{thesis_macros}

\usepackage{pdfpages}
\usepackage{newclude}

\usepackage{hyperref}
\hypersetup{
  colorlinks = true,
	citecolor = blue
}
\newcommand{\IMG}[1]{#1}

\newcommand{\MSFOL}{MSFOL\xspace}

\newcommand{\viproof}{interpreted formal proof\xspace}
\newcommand{\viproofs}{interpreted formal proofs\xspace}

\newcommand{\Finished}[1]{#1}

\title{Rigorous Deductive Argumentation \\ for Socially Relevant Issues}
\author{Dustin Wehr}
\degreetype{Doctor of Philosophy}
\department{Computer Science}
\gradyear{2015}

\begin{document}

\begin{preliminary}
\maketitle

%\noindent \noindent 
%{\bf \Finished{This color}} means ready to send to be read by committee.\\
%{\bf \Stable{This color}} means needs a re-reading before sending off to committee, but otherwise stable. At most (probably minor) editing left to do. \\
%{\bf \Mostly{This color}} means there are some specific things that need to be addressed, but at most a little new writing left to do. 
%{\bf \Red{This color}} means substantial work still to do.\\
%means substantial work still to do, but deferred until the Red and Green sections are completed. 

\begin{abstract}
%\TODO[abstract goes here]
The most important problems for society are describable only in vague terms, dependent on subjective positions, and missing highly relevant data. This thesis is intended to revive and further develop the view that giving non-trivial, rigorous deductive arguments concerning such problems --{\it without} eliminating the complications of vagueness, subjectivity, and uncertainty-- is, though very difficult, not problematic in principle, does not require the invention of new logics (classical first-order logic will do), and is something that more mathematically-inclined people should be pursuing. The 
%specific format and approach 
framework of \textit{interpreted formal proofs} is presented for formalizing and criticizing rigorous deductive arguments about vague, subjective, and uncertain issues, and its adequacy is supported largely by a number of major examples.
This thesis also documents progress towards a web system for collaboratively authoring and criticizing such arguments, which is the ultimate goal of this project.
%Major examples are the main offerings ; the rest is an assembly of old and elementary ideas.\footnote{For us; not so for the people who have expressed the strongest interest in this project, who mostly work in non-mathematical fields. An important goal for me, which I probably won't have touched until after the publication of my dissertation, is to design an unintimidating web-based system where intelligent people without a background in logic can contribute to formal-logic-backed arguments.} All are original arguments. One is fully formalized, verified by a first-order theorem prover, and viewable in a readable dynamic HTML format online,\footnote{http://www.cs.toronto.edu/{\textasciitilde}wehr/research\_docs/sue\_rodriguez.html} and two are mostly-formalized and presented inside. Each is or can be written as an {\it interpreted formal proof}, which is a simple presentation convention introduced in this paper --and used without name since the discovery of predicate logic-- for attaching one's vague and subjective intended semantics to the symbols of one's proof, and which is well-suited for the kinds of what-do-you-mean-by? semantics criticisms that are fundamental to arguing about vague and subjective issues. Sketches of some such criticisms for the examples are included.  
\end{abstract}

\begin{acknowledgements}
\indentoff
I have a lot of people to thank. Each group is in alphabetical order.

--My brilliant and open-minded supervisors Professor Stephen Cook and Professor Alasdair Urquhart.

--My parents, Dr. Judith Bloomer and Dr. Robert J Wehr.

--My friends, Lily Bernstein, Kate Busby, Sam Caldwell, Isabel MacKay-Clackett, Mike Markovich, Deborah Perkins-Leitman, Natalie Wiseman

\end{acknowledgements}
\newpage

\hypersetup{linkcolor=blue}
\tableofcontents

\end{preliminary}

%%%%%%%%%%%%%%%%%%%%%%%%%%%%%%%%%%%%%%%%%%%%%%%%%%%%%%%%%%%%%%%%%%%%%%%%%%%%%%%%%%%%%%%%%%%%%%%%
\chapter{\Finished{Introduction}}%  (\U{\Red{don't look at this yet!}})}
Gottfried Leibniz had a radical and idealistic dream, long before the formalization of predicate logic, that some day the rigor of mathematics would find much broader use.
\begin{aquote}{Gottfried Leibniz, 1679, {\it ``On the General Characteristic''}\cite{leibniz1976}}
{\it 
For men can be debased by all other gifts; only right reason can be nothing but wholesome. But reason will be right beyond all doubt only when it is everywhere as clear and certain as only arithmetic has been until now. Then there will be an end to that burdensome raising of objections by which one person now usually plagues another and which turns so many away from the desire to reason. When one person argues, namely, his opponent, instead of examining his argument, answers generally, thus, `How do you know that your reason is any truer than mine? What criterion of truth have you?' And if the first person persists in his argument, his hearers lack the patience to examine it. For usually many other problems have to be investigated first, and this would be the work of several weeks, following the laws of thought accepted until now. And so after much agitation, the emotions usually win out instead of reason, and we end the controversy by cutting the Gordian knot rather than untying it. 
}
\end{aquote} \index{Leibniz (quotes)}

What is understood by many mathematically-inclined people --that formal logic is {\it in principle} applicable to arguments about social, contentious, emotionally charged issues-- sounds absurd to most people, even the highly educated. The first, rather unambitious goal of this project, is to illustrate this understanding. The second goal, a very difficult and lonely one, is to investigate whether such use of rigorous deduction is worth doing, even if only in our spare time. 

%\Gray{Due to my personal experience with that difficulty, my main approach has slowly evolved to not be one of persuasion --it is too early for that-- but rather to develop a tool to minimize the difficulty of writing and criticizing rigorous deductive arguments about such issues.}

There are thousands and thousands of pages by hundreds of scholars that are tangentially related to this project; papers about vagueness in the abstract,\footnote{See \cite{sep-vagueness}, where the approach I take to reasoning in the presence of vagueness does not appear to be covered. I call my approach \U{vagueness as plurality of intended models}.} the theoretical foundations of Bayesian reasoning,\footnote{I recommend \cite{Causality}.} {\it abstract argumentation systems} \cite{Prakken2010}, etc. There is a huge amount of scholarly work on systems and tools and consideration of the theoretically-interesting corner cases, but too little serious work in which the problems take precedence over the tools used to work on them. This thesis concerns a project of the latter kind; the work is on important specific problems, attacking general theoretical problems only as-necessary. In this way, we avoid getting hung up on details that don't matter. 
%I am still hopeful that work on argumentation systems will turn out useful, but see the short Section \ref{ss:SR} for my reservations about trying to integrate it too early.

Surprisingly, it is the (normative side of the) field of Informal Logic that is probably most related to this project \cite{InformalLogic}\cite{Commitment}. For a long time now the researchers in that field have understood that dialogue-like interactions, or something similar, are essential for arguing about the problems we are concerned with here (Section \ref{s:problemdomain}). But formal logic has something to contribute here; there are too many examples where good, intelligent scientists and statisticians are given a voice on such problems, only to fail to adhere to the same standards of rigor that they follow in their professional work.\footnote{\cite{sesardic} provides a good example. There Sesardic, a philosopher, contradicts the hasty conclusions of some very reputable statisticians, essentially by applying the same Bayesian quantitative argument, but with much more care taken in constraining the values of the prior probabilities.}
\medskip

There are important commonalities between proofs in mathematics and proofs about subjective and vague concepts. For example, in both domains, we only need to axiomatize the structures we are thinking about precisely enough for the proof to go through; our proofs about numbers and sets never\footnote{Except for proofs about finite structures.} require complete characterizations, and similarly, for proofs about people, laws, moral values, etc, there is no need to fully {\it eliminate} the vagueness that is inherit in axiomatizations with multiple distinct models. That observation is materialized in this project's use of \U{top-down, minimally-reductionist formal proofs} --my name for formal proofs where one does not strive to minimize the number of fundamental (not defined) symbols,  or the number of axioms (an assertion remains an assertion until someone demands it become a proved lemma).\footnote{As a non-essential demonstration of the concept of top-down, minimally-reductionist proofs, and of the dynamic HTML output I've developed for reading arguments, here are two examples that I wrote while debugging my current system. The first is fully formally verified by a first-order theorem prover.\\
Infinitely-many primes: \href{http://www.cs.toronto.edu/~wehr/thesis/infinitely-many\_primes.html}{\Blue{http://www.cs.toronto.edu/{\textasciitilde}wehr/thesis/infinitely-many\_primes.html}} \\
5 color theorem: \href{http://www.cs.toronto.edu/~wehr/thesis/5colortheorem.html}{\Blue{http://www.cs.toronto.edu/{\textasciitilde}wehr/thesis/5colortheorem.html}}
} I believe top-down, minimally-reductionist formal proofs are the only option when reasoning faithfully about vague concepts. 

%For the remainder of this introduction, I will talk about the differences in reasoning between the two domains (mathematics and the domain of problems we are interested in here). 
There are three aspects of contentious socially-relevant questions that distinguish them from questions that are commonly considered mathematical\footnote{But note that, as this thesis will make clear, my opinion is that there is no sharp qualitative boundary between the two domains.}: vagueness\footnote{Classic examples are vague predicates expressing tallness, or baldness.}, subjectiveness\footnote{E.g. the weights of various principles of morality}, and uncertainty. None of these can be eliminated completely without changing the fundamental nature of the problems. 

With mathematics problems we can {\it usually} axiomatize structures sufficiently-precisely at the beginning of our attempt to resolve a problem (a statement to prove or disprove), whereas in reasoning about social issues one must delay the sharpening of vague definitions until necessary -- in particular, until critics of one's argument are too unclear about one's informal semantics for a symbol to be able to evaluate whether they should accept or reject an axiom that depends on that symbol (this is called a {\it semantics criticism} in Section \ref{s:criticizingVIproofs}). 
%To understand the approach to formalizing vagueness used in this project, \U{vagueness as plurality of intended models}, it is vital to think of models and the satisfaction relation as the fundamental logical concepts, as opposed to validity. Section \ref{s:VIFP} should make that clear. See the footnote at the end of Appendix \ref{s:dialog} for consideration of a Sorites problem.
%Finally, 
Of course, questions about vague concepts cannot always be answered in a particular way. What may happen is that the question has different answers depending on how it is sharpened, which is determined by the author of an argument that purports to answer that question (and sometimes, indirectly, by the critics of the argument). An illustrative example of this can be made with Newcomb's Paradox\footnote{Start at the Wikipedia page if you haven't heard of this and are curious.}; for all of the many English presentation of the problem that I have seen, it is not hard to give two reasonable formalizations that yield opposite answers, a fact that has been ignored, downplayed, or overlooked by many commentators arguing that one of the answers is the {\it right} answer (and likewise  for many puzzles or paradoxes argued about in analytic philosophy).

As with vagueness, subjectiveness demands some system of interaction between people on the two sides of an argument, and I am working on an implementation of such a system now (Section \ref{s:websystem}). 
Of course I do not mean to suggest that formal logic can help two parties with conflicting beliefs come to the same answer on, say, questions of ethics. However, where formal logic can help is to find {\it fundamental} sources of disagreement starting from disagreement on some complex question (which is progress!).

Uncertainty is the most difficult of the three complications. Sparsity of information can make it impossible to give an {\it absolutely}-strong deductive argument for or against a given proposition, and the inability to do so can easily deflate one's motivation to make a formal demonstrations. But interaction is useful here, too: In Chapter \ref{c:AIDWYC}, I give a proof that a key piece of evidence that was used to convict a man of murder has no inculpatory value. Now, I cannot say that the assumptions from which that conclusion (the proposition named $\langle$the newspaper hair evidence is neutral or exculpatory$\rangle$) follows are {\it absolutely} easy to accept, but I confidently challenge anyone to come up with a proof of the negation of that conclusion, i.e. a proof that the likelihood of the convicted man being guilty given the evidence is significantly larger than the likelihood that he is innocent given the evidence. Hence, I am claiming that my assumptions {\it are} easy to accept {\it relative} to what my opponents could come up with.

\medskip

I use a superficial extension of classical FOL in this project, and {\it for this particular project} --rigorous deductive reasoning about the kind of questions described in Section \ref{s:problemdomain}-- that seems to be the right fit. It is vital that the interface between syntax and semantics is as simple as possible, and classical FOL with Tarski semantics is the best in this respect. In Section \ref{s:choiceoflogic} I make that argument in more detail. In Chapter \ref{c:defeasible}, I take some space to explain how common forms of defeasible reasoning can be carried out in deductive logic, and how ideas from e.g. modal (or, by extension, epistemic or temporal) logic can be used as-necessary without needing to build them into the definition of the logic (complicating the semantics). 

%I may be wrong about this. That is, if it turns out that the more-concise syntax of some such logic is sufficiently-often enough helpful for reading and writing proofs, without obscuring the meaning, to compensate for the new barrier to entry that using it in the system would entail. But a great deal more experience should be decided after a great deal more experience.

\section{\Finished{What this project is and isn't}} \label{s:isandisnt}
\textbf{What it is:}

This thesis gives the foundations for, and documents progress toward, a collaborative web system intended for arguing about certain kinds of questions in the most rigorous, fair, and civil way that we know of: formal deductive proof. The main proposed uses/benefits of the system are:
\begin{PPE}
\item Making progress in arguments about questions usually considered outside of math and science.
\item Helping people find the fundamental sources of their disagreements with each other (a special case of progress).
\item Demonstrating deductive thinking to people who are not interested in mathematical problems, or are not mathematically inclined.
\item Giving mathematicians an outlet for advocacy work that utilizes their technical abilities in an essential way.\footnote{Of course statisticians have always engaged in advocacy work. By ``mathematicians'' here, I mean those in disciplines whose overwhelmingly central focus is on proving theorems.}
\item Serving as a practically-approachable ideal for rigorous, fair, and civil argumentation.
\end{PPE}
The apparent contradiction between items 3 and 4 is resolved by noting that people who are mathematically inclined and people who are not have different roles in the system. In particular, authoring a new argument requires at least one person who is familiar with basic formal logic, but contributing to new and existing arguments, or criticizing arguments, does not. Moreover, as a small number of excellent lawyers and philosophers have demonstrated over the centuries, writing natural language arguments that approximate the ideal (item 5) does not necessarily require familiarity with formal logic. 
I choose, perhaps over-generously, to interpret Leibniz's (latest, and most pessimistic) writing describing his imagined \textit{universal characteristic}, with those notes in mind:
\begin{aquote}{Gottfried Leibniz, 1706\footnotemark}
{\it It is true that in the past I planned a new way of calculating
suitable for matters which have nothing in common with mathematics,
and if this kind of logic were put into practice, every reasoning,
even probabilistic ones, would be like that of the mathematician: if
need be, the lesser minds which had application and good will could,
if not accompany the greatest minds, then at least follow them. For
one could always say: let us calculate, and judge correctly through
this, as much as the data and reason can provide us with the means for
it. But I do not know if I will ever be in a position to carry out
such a project, which requires more than one hand; and it even seems
that mankind is still not mature enough to lay claim to the advantages
which this method could provide.}
\end{aquote} \index{Leibniz (quotes)}
\footnotetext{From translation of a letter to Sophia of Hanover \cite{leibniz}}

\bigskip
\noindent \textbf{What it isn't:}

This project is not about using logic for discovery, as in the \textit{axiomatic method}. And it is not concerned with developing general or elegant mathematical theories (including theories presented as ``logics''). An important premise of this project's approach is that an informal yes/no question must be set before formalization begins (though it may later be modified), and only making progress on the question matters (where a more-precise formulation of the question is progress). Of course, abstracting out common axiomatizations for reuse is still a good idea, but as with writing software libraries, it should not be done preemptively. This preoccupation with constructing relatively-elegant, widely-applicable theories, is one of two factors to which I attribute the lack of success of the Logical Positivists' project, the other being their focus on questions outside of the Problem Domain (Section \ref{s:problemdomain}).

%The logical positivists worked to varying degrees to expand the applications of formal logic to areas outside traditional mathematics, especially science. According to the most vocal of modern commentators, they were largely unsuccessful. I attribute that failure to a preoccupation with constructing relatively-elegant, widely-applicable theories, and to their focus on questions outside of the Problem Domain (Section \ref{s:problemdomain}).

This project is fundamentally \textit{normative}. It is not concerned with descriptive modelling of argumentation, as in \textit{abstract argumentation systems}\cite{Prakken2010}. There is no interest here in modelling real legal reasoning, for example, nor in assisting it. But there is great interest here in depicting what legal reasoning \textit{should} be like in an ideal environment with lots of time available (e.g. for severe criminal cases, in which there is the rest of the convicted person's life to argue that they are innocent with sufficiently-large probability\footnote{Noting that a convict is adversely effected for the remainder of their life, even if they are released.}).

\section{\Finished{Role of formal logic}}
\indent\indent Remove formal logic from this project and there is no benefit over our current system of arguing with each other through papers and blog posts and shouting. Those mediums are easier to work in, and superior if one is interested in persuasion. Indeed, interpreted formal proofs (the format for arguments and criticisms advocated in this thesis -- see Section \ref{s:VIFP}) can be made more persuasive to most people if converted to an informal argument that mixes natural language and mathematics in the normal way we use in conference and journal papers. The problem with that, from the point of view of this project, is that unsound, invalid, misleading, unfair, and otherwise bad arguments benefit from the lax regulations as much {\it \U{or more than}} good arguments do. This project uses deductive formal logic because it is our best tool for forcing the weaknesses of arguments to be exposed. Thus {\bf the role of formal logic in this project is regulatory, and nothing more than that.} The success of this project rides on the regulatory benefit outweighing the overhead of formalization.

I have caught omissions in my own reasoning thanks to the constraints of formal deductive logic, things that never occurred to me in thinking and talking about an issue for years, resulting in my having to temper my opinion, sometimes temporarily and sometimes long term. I have gained respect for my opponents on {\it every} issue that I have attempted arguments about, having been forced to consider all the subtle details (e.g. Canada's lifetime ban against blood donations from men who have had sex with men,\footnote{The efficacy of our system for preventing contaminated blood from ending up in the stock of blood donations relies not just on tests for HIV, Hepatitis, etc, but also on self-disclosure of known infections and risk-factors. If the ban is lifted, is there non-negligible probability of a significant increase in the rate of people lying in the self-disclosure part of the system?  Reasoning deductively, one {\it must} consider this non-obvious question, and I have found no way to derive my target conclusion (that, with additional safeguards, the ban should be lifted) without making an assumption that is not far from explicitly answering the question `no'. We could imagine, for example, that after lifting the ban, some homophobic HIV-positive person intentionally donates blood in retaliation. That may seem far-fetched, but it \textit{must} nonetheless be ruled out (with high probability), one way or another. That strictness imposed by formal deductive logic should be reassuring.} assisted suicide in Canada, the evidence for anthropogenic global warming\footnote{The closest thing we have to a strong deductive argument, that I have found, comes from the Berkeley Earth Surface Temperature project, which has mostly been ridiculed by climate science researchers, who simply view it as making no significant advancement in climate science, ignoring or not valuing the fact that it seeks to minimize the use of argument from expert opinion.}). It is hard to write a person off as ignorant or stupid, relative to oneself, after a great struggle to find acceptable formal assumptions from which it follows logically that they are wrong. 
%\textbf{Incidentally, I \textit{have} experienced that the process of finding a rigorous deductive argument has often resulted in my having to temper my opinion, at least temporarily.} 
It is hard to overstate this advantage of rigorous deduction. Among other things it provides a force for compromise -- a force in the same democratic spirit as ``I could be wrong'', but with much greater discriminatory power.

\section{\Finished{Preface to examples}}

On one hand, the better the quality of the example arguments in Chapters \ref{c:SR}-\ref{c:assistedsuicide}, the more seriously this project and its theoretical ideas will be taken. There is some basis for that; the expected value of the project and theoretical ideas are much harder to assess, so instead one might choose the heuristic of assessing the author, via the remaining material.

I am urging you to resist that temptation. These are not exemplars of interpreted formal proofs. I am not a genius, nor a real statistician, nor an expert in criminal law or human biology. I am not even an especially good mathematician. But once the web system (Chapter \ref{s:websystem}) is ready, so that the project has as good a chance at gaining traction as possible, I will persuade some such people to collaborate with me on new arguments or to author their own. Much, much better examples are still to come.

\medskip

This, like almost any thesis, is not intended to be read from start to finish. You are encouraged to skip ahead after reading at least Sections \ref{s:problemdomain} and \ref{s:VIFP}. The most complete and accessible example is the Sue Rodriguez argument, which should be read in HTML, though a static version is given in Chapter \ref{c:SR}. The most complete and accessible example written in \LaTeX\ is Chapter \ref{c:berkeley}.

\section{Related Work}
Unfortunately this project does not fit well within any current area of research. On the other hand, it would be impossible without certain firmly established work, especially the fundamentals of classical predicate logic and Bayesian reasoning/statistics. I have chosen to cite related work predominately when it is contextually relevant, throughout this thesis. E.g. in Section \ref{s:isandisnt} I briefly talk about the field of Abstract Argumentation Theory, which was not helpful in this project, and in Section \ref{s:carneades} I briefly cover one of the implemented projects that has come out of the Informal Logic community. 

%If this were a dissertation in Philosophy, I would start with Plato. My coverage of related work begins with Leibniz, whose aspirations were {\it very} close to the goals of this project, indeed the closest I have found. Unfortunately his vision absolutely required predicate logic, which was not discovered and understood until two centuries after his death. I won't go further back because his re

%In summary, Leibniz's aspirations were {\it very} close to the goals of this project, and indeed the closest I have found. Unfortunately his vision absolutely required predicate logic, which was not discovered and understood until two centuries after his death. 

\subsection{Logical Positivism, Logical Empiricism, and Analytic Philosophy Since the Early 1900s}
In general in analytic philosophy, there has been an immense amount of writing \textit{about} the hypothetical application of formal logic to vague and subjective issues, but very little of the application itself. 

The first scholars to have the requisite technical framework of predicate logic, and to attempt to expand its scope to matters outside of traditional mathematics, were analytic philosophers of the early-to-mid 1900s, especially those associated with the movements of Logical Positivism and Logical Empiricism. Unusually in philosophy, there seems to be a general agreement that they failed. See, for example, \textit{The Heritage of Logical Positivism}\cite{HeritageOfLogicalPositivism}. Before explaining why they failed, it is important to note that they didn't fail at {\it this} project. Geoffrey Sayre-McCord writes in \textit{Logical Positivism and the Demise of ``Moral Science''}\footnote{From the previously-cited compendium.}:
\begin{quote}
{\ldots}most of the Logical Positivists were convinced that moral theory is nonsense. They thought their arguments showed that there really is no such thing as ``moral science.'' Moral language, they maintained, is not used to report facts, rather it is simply a tool used to manipulate the behavior both of ourselves and of others.
\end{quote}
That is, most of the positivists shied away from working on questions affected by moral relativism, instead spending their formalization efforts on questions from science. 
% We engage in moral discourse, Reichenback emphasized, because ``our fellow men are conditioned to respond to words as instruments of our will.''

Considering the few analytic philosophers who made serious attempts to reason in formal logic about specific problems outside of mathematics, I attribute their lack of progress to two main factors:
\begin{PPE}
\item Preoccupation with constructing elegant, widely-applicable theories. 
A premise of my approach is that an informal yes/no question must be fixed before formalization begins. Hence, the development of general theories about subjective and vague matters is explicitly not a goal; only making progress on the question matters. Of course, abstracting out common axiomatizations for reuse is still a good idea, but as with writing software libraries, it should not be done preemptively.
\item Working with examples outside of the problem domain I outlined in Section \ref{s:problemdomain}. Because there is less promise of discovering mathematically interesting material in the formal investigation of a question about a vague and subjective issue, the motivation for the very difficult work involved in rigorous reasoning must come entirely from elsewhere, namely from the question itself; we must be convinced that there is no easier way to make progress on the question, and that making progress on the question is indeed worth the work. %Leibniz knew the danger of being insufficiently conscious of this point; speculating about why the project he envisioned had not been taken up by others earlier, like Descartes, he wrote:
%\begin{aquote}{Gottfried Leibniz, 1679, \textit{On the General Characteristic} \cite{leibniz1976}}
%{\it 
%The true reason for this straying from the portal of knowledge is, I believe, that principles usually seem dry and not very attractive and are therefore dismissed with a mere taste.}
%\end{aquote}
 %It is related to the fear of being labeled a philosopher, which on one is u
\end{PPE}

\subsection{Mathematics and Theoretical Computer Science}
The stigma in mathematics against working toward progress on value-laden issues has grown over time. Doing so gets one's work labeled as philosophy. The stigma is not surprising, given what has passed as good work in contemporary philosophy, but on the other hand it is a clear fallacy of association to condemn a subject of study on account of the people who have managed, so far, to get paid to work on it. Nonetheless, disrepute is the current state of things, and has been for many decades.

As a consequence, the most related work in mathematics differs from the work required for this project in one very important respect: It is intensionally reductionist. We find this in Game Theory and Decision Theory, for example. In these fields, real social problems are used to inspire and motivate interesting mathematical problems, but little more than that, aside from extremely rare situations when the simplifying assumptions of models are met, and countless disastrous situations in which the mathematics is used without meeting the simplifying assumptions. 

I should mention, however, that these fields are progressing, with the simplifying assumptions for some problems becoming more and more palatable. Perhaps we will see strategy-proof mechanisms  employed for kidney exchange programs in the future\cite{KidneyExchange}, for example. Though, I doubt it. I expect that researchers will continue to ignore factors that would make their model too messy and unwieldy, and ignore possible changes to their models that would result in mathematically-uninteresting solutions to the real problem (since then there would be nothing to publish).\footnote{Minimizing the number of people who die waiting for kidney transplants, for example. The sophisticated algorithms found in the Algorithmic Game Theory literature implicitly use a model in which, for example, there is no possibility of legislative solutions that mandate participation in a kidney exchange (say, for the hospital to be eligible for federal funding) while criminalizing or penalizing lying (see previous source for details about why hospitals might lie).}

\subsection{Informal Logic, Defeasible Reasoning, and Intentional Logics}
The idea to adopt an asymmetric dialog system (author vs critics) came from work on argumentation systems in Informal Logic; see \cite{Commitment} for one of many sources. 

Chapter \ref{c:defeasible} is devoted to explaining the inappropriateness of defeasible logic for this project, and Section \ref{s:choiceoflogic} does the same for intentional logics. All work that I have found on argumentation about subjective and vague questions uses defeasible or intentional logics. The vast majority of it is concerned with the construction of formal systems for hypothetical use (i.e. never seriously applied) in reasoning about subjective and vague questions. Unfortunately, these systems are not useful for the task of writing an isolated deductive argument, as they come with the overhead of their own syntax and semantics, and because, as I hope the examples I provide will convey, one cannot expect to find a formal system of axioms that is completely adequate for one's argument; each argument requires at least a slightly new system in order to be formulated in the most natural way, and formulation in the most natural way is vital for approaching the goal of \textit{locally-refutable proofs}(page \pageref{p:locally-refutable}).

\chapter{\Finished{Proofs and critiques}}
%\minitoc
\section{\Finished{Problem domain}} \label{s:problemdomain}
\include*{problem_domain}

%\section{Related and apparently-related work \TODO[(Steve says to criticize as I go)]}
%Other discussion of related work appears in Sections \ref{s:Bayesian} and \ref{s:Vagueness}.
%\external{ProofsAndCritiques/RelatedWork}

\section{\Finished{Interpreted proofs and critiques}} \label{s:VIFP}
\include*{vaguely_interpreted_formal_proofs}

\section{\Finished{Choice of logic}} \label{s:choiceoflogic}
\include*{choice_of_logic}

\section{\Finished{Implementation for reading and verifying interpreted formal proofs}} \label{s:implementation}

A good approximation of the format in which I intend interpreted formal proofs to be read --in an effort to make reading them less effortful and tedious-- can be seen in any of these examples:
\begin{PPE}
\item Sue Rodriguez's case at the Supreme Court of Canada: \url{http://www.cs.toronto.edu/~wehr/thesis/sue_rodriguez.html}
\item Assisted suicide should be legalized in Canada: \url{http://www.cs.toronto.edu/~wehr/thesis/assisted_suicide_msfol.html}
\item Walton's intentionally-fallacious argument that no one should get married: \url{http://www.cs.toronto.edu/~wehr/thesis/walton_marriage.html}
\item Infinitely-many primes: \url{http://www.cs.toronto.edu/~wehr/thesis/infinitely-many_primes.html} 
\item 5 color theorem: \url{http://www.cs.toronto.edu/~wehr/thesis/5colortheorem.html}
\item High-level proof of G{\"o}del's Second Incompleteness Theorem: \url{http://www.cs.toronto.edu/~wehr/thesis/G2b.html}
%\item \TODO[Update and include link to G{\"o}del's Ontological Proof and criticism?]
%\item \TODO[Update and include link to Knowability Paradox proof and criticism?]
\end{PPE}

\index{HTML proof documents}
\noindent \noindent An interpreted formal proof is implemented as an HTML document with the following structure:
\begin{PPI}
\item A sequence of declarations, each of which is a new symbol introduction (with definitions being a special case), axiom, or lemma. An axiom is either an Assumption, Simplifying Assumption, Assertion (intended to be uncontroversial), Claim (author is a prepared to prove, once challenged), or Quasi-definition (a symbol introduction together with an axiom that is not technically a syntactic definition, but plays a definition-like role).
\item The statement of a lemma or theorem $A$, the \textit{goal} of the interpreted formal proof, which uses only the previously-introduced symbols.
\item A collapsible proof of $A$, which is another interpreted formal proof whose immediate-child declarations, combined with those that preceded the statement of $A$, entail $A$, where the entailment is verified by a first-order theorem prover\footnote{``preceded'' means the declaration is in its scope, where scope is defined as in many programming languages. For example, if a lemma $A_2$ immediately follows a lemma $A_1$, then the proof of $A_2$ can use $A_1$ but nothing introduced in the proof of $A_1$, and the proof of $A_1$ cannot use $A_2$.} (see below for more detail). Note that this is slightly atypical in that one may delay introducing axioms and primitive symbols until just before they are used in a proof. The purpose of this is to lessen the effect of an interpreted formal proof starting with an overwhelming number of symbol introductions and axioms before it even gets to the statement of the goal sentence. Instead, the declarations that must precede the statement of a lemma are just the symbol introductions for the symbols that are explicitly used in the lemma. 
\item Each lemma is the goal of an interpreted formal proof or else it has an informal natural language proof.
\end{PPI}

Such HTML documents will be a central part of the web system described in Section \ref{s:websystem}. 
The main advantages over reading in \LaTeX/PDF are these:
\begin{PPI}
\item Collapsible sections of text (implemented). This is helpful as an author if you want to hide nasty parts of the proof by default, and for readers it is helpful for decluttering the screen once they are satisfied with a proof/justification of some lemma/claim, or once they have sufficiently-internalized the syntactic definition or informal semantic description of a symbol.
\item Pop-up references on cursor hover (implemented). Hovering the mouse cursor over occurrences of symbols will reveal information from their initial declaration. This is more useful for proofs about socially-relevant issues than it usually would be for proofs in mathematics, because of the much higher ratio of \[\frac{\text{number of fundamental symbols with no standard meaning}}{\text{length of proof}}\]
%This feature will be revamped; as of now, it is of little use 
\item Reader comments (implemented): As a temporary stand-in for the plans of Section \ref{s:websystem}, readers can attach annotations (e.g. for criticisms) to any part of an argument (via AnnotateIt.org), or post (nested) comments at the bottom of the page (via DISQUS.com).
\item Renameable symbols (in the works). If the name chosen by the author is not conducive to your reading, then change it!
\item Multiline display of formulas (in the works). As in some programming languages, there will be a standard format, in terms of where white space is placed, for displaying the structure of formulas across multiple lines, so that an author need only indicate with a checkbox whether the children of a subterm should be displayed on different lines.    
%\item In the works: In-place editing and in-place criticizing.
\end{PPI}

The displayed syntax need not be the same as the input syntax. In particular, two distinct symbols can display the same way. Hovering the cursor over an occurrence will reveal which version it is. This takes care of most of the use cases for overloading (where e.g. a function symbol can have multiple function types) while remaining in standard many-sorted first-order logic.

Most of the examples listed above are written in a formal language\footnote{The exceptions are examples 5 and 6, which I wrote before implementing the formal language.}, which gets translated to HTML and to instances of theorem proving problems in many-sorted first order logic, specifically the TFF (\U{t}yped \U{f}irst-order \U{f}ormula) language of TPTP\cite{tptp}. Those problems can be solved automatically by first-order theorem provers, although when there are a very large number of axioms and definitions, as in the Assisted Suicide argument (Chapter \ref{c:assistedsuicide}), it is sometimes necessary to tell the prover which axioms to use for each lemma.\footnote{This is due to the non-goal-directed nature of the saturation-based first-order theorem provers that I have used; it is possible that a backwards theorem prover, perhaps even a cut-free proof search, would work better in such cases, but I have not yet found a good, easy to set up implementation.} I used SNARK\cite{snark} for type checking and sometimes short proofs, CVC4\cite{cvc4} for model finding and sometimes proof search, and 
%Z3\cite{Z3} and 
Vampire\cite{vampire} for fast proof search and sometimes countermodels, all via the System on TPTP web interface\footnote{\url{http://www.cs.miami.edu/~tptp/cgi-bin/SystemOnTPTP}} (except Vampire was also easy to setup locally). \index{First-order theorem provers}

\section{\Finished{Toy example: Walton's fallacious argument demonstrating equivocation via ``variability of strictness of standards''}} \label{s:walton}
\include*{walton_marriage}

\subsection{\Finished{Formal criticism of Walton's marriage argument}}
\include*{walton_formal_critique}

%%%%%%%%%%%%%%%%%%%%%%%%%%%%%%%%%%%%%%%%%%%%%%%%%%%%%%%%%%%%%%%%%%%%%%%%%%%%%%%%%%%%%%%%%%%%%%%%
\chapter{\Finished{Classical deductive formalization of defeasible reasoning}} \label{c:defeasible}
\include*{defeasible_reasoning_chapter}
\chapter{\Finished{Example: Sue Rodriguez's supreme court case}} \label{c:SR}
This argument is meant to be read in a browser, and can be found at:
\[ \text{\url{http://www.cs.toronto.edu/~wehr/thesis/sue_rodriguez.html}} \]
I include an inferior static version here just 
%to keep track of the approximate effective length of this thesis, or 
in case you have a printed copy and you strongly prefer to read on paper.
\IMG{\includepdf[pages={-},pagecommand={}]{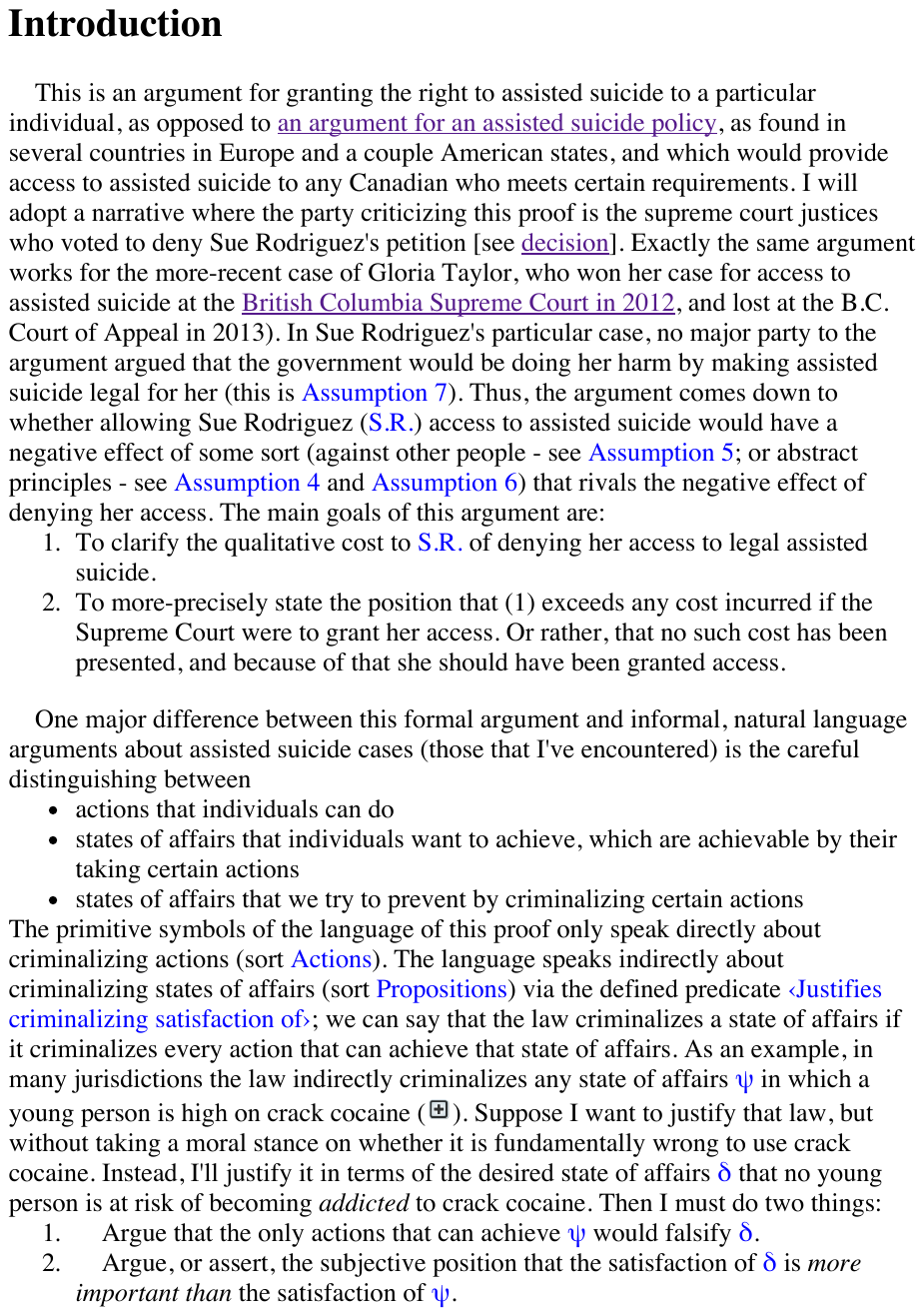}}
%\IMG{\includegraphics{Examples/SueRodriguez/SueRodriguezSupremeCourt_small_font.pdf}}
%\section{\Blue{Formal criticism}} \label{s:SRFormalCrit}

%%%%%%%%%%%%%%%%%%%%%%%%%%%%%%%%%%%%%%%%%%%%%%%%%%%%%%%%%%%%%%%%%%%%%%%%%%%%%%%%%%%%%%%%%%%%%%%%
% I didn't actually reread this...
\chapter{\Finished{Example: Berkeley gender bias lawsuit}} \label{c:berkeley}
\include*{berkeley_gender_bias_example_MSFOL}

\chapter{\Finished{Example: Leighton Hay's wrongful conviction}} \label{c:AIDWYC}

\include*{LeightonHayExampleFeb2014}

%%%%%%%%%%%%%%%%%%%%%%%%%%%%%%%%%%%%%%%%%%%%%%%%%%%%%%%%%%%%%%%%%%%%%%%%%%%%%%%%%%%%%%%%%%%%%%%%
\chapter{\Finished{Example: Arguing that smoking causes cancer in 1950}} \label{c:smoking}

\include*{SmokingAndCancerExample_feb2014}
\indenton
\chapter{\Finished{Example: Assisted suicide should be legalized in Canada}} \label{c:assistedsuicide}
%\section{Intro and guide to understanding argument}
%\external{Examples/AssistedSuicide/AssistedSuicideIntro}
%
%\section{Partially-expanded HTML argument}
This is the most complex and ambitious of all the examples in this thesis. Not all of the many essentially-boolean-algebra lemmas have been formally verified using a first-order theorem prover, although any one of them can be checked easily by hand. Those remaining lemmas are stated as Claims.

This argument is meant to be read in a browser, and can be found at:
\[ \text{\url{http://www.cs.toronto.edu/~wehr/thesis/assisted_suicide_msfol.html}} \]
I include an inferior static version here just 
%to keep track of the approximate effective length of this thesis, or 
in case you have a printed copy and you strongly prefer to read on paper.
%\includepdf[landscape=true,pages={-}]{Examples/AssistedSuicide/CanadianAssistedSuicidePolicy.pdf}
\IMG{\includepdf[landscape=false,pages={-},pagecommand={}]{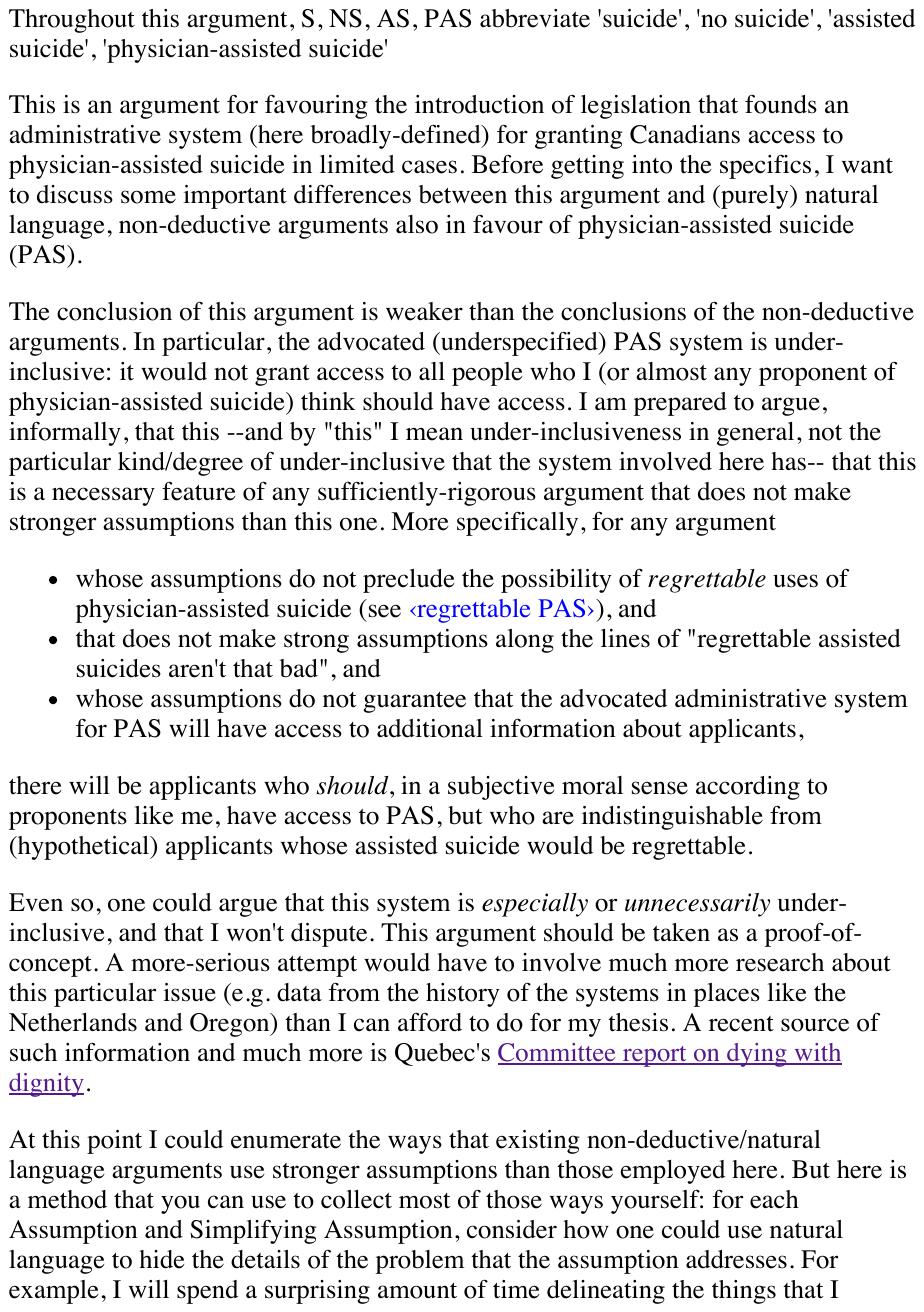}}
% ,pagecommand={} inserts page numbers

%%%%%%%%%%%%%%%%%%%%%%%%%%%%%%%%%%%%%%%%%%%%%%%%%%%%%%%%%%%%%%%%%%%%%%%%%%%%%%%%%%%%%%%%%%%%%%%%
\chapter{\Finished{Ongoing work}}
\section{\Finished{Web system for collaborative authoring and criticizing of interpreted formal proofs, and a minimal dialogue system}} \label{s:websystem}

Initially I thought that a sophisticated dialogue system, with rules designed to ensure progress under certain assumptions, would be essential to move forward with this project. With more experience writing interpreted formal proofs, however, it became clear that reasoning faithfully about complicated inelegant structures was already so onerous a task that it would be asking too much of authors to require the extra work of demonstrating that their argumentative moves make progress. This has led me to shift to a relatively simple and lax model of interaction. The end of this chapter contains some notes about the issue of progress.

%\TODO[maybe take more from workshop paper and put it here]\\

%Vagueness and subjectiveness demand some system of interaction between people on the two sides of an argument. In reasoning about social issues one must delay the sharpening of vague definitions until necessary -- in particular, until critics of one's argument are too unclear about one's informal semantics for a symbol to be able to evaluate whether they should accept or reject an axiom that depends on that symbol (this is called a \textit{semantics criticism} in Section \ref{s:VIFP}). 

\subsection{\Finished{Related work from Informal Logic}} \label{s:carneades}
{\bf Carneades:}

\indenton

Carneades is a web application in active development ``which provides software tools based on a common computational model of argument graphs useful for policy deliberations and claims processing.''\cite{CarneadesWebapp} It is the application, of those I am aware of, that is most related to the one I am working on, although its focus on propositional defeasible reasoning makes it still only weakly related. In more detail, the principle developer Thomas F. Gordon describes Carneades as a collaborative, online system for (quoting from \cite{CarneadesWebapp}):
\begin{PPI}
\item modeling legal norms and argumentation schemes
\item (re)constructing arguments in an argument graph
\item visualizing, browsing and navigating argument graphs
\item critically evaluating arguments
\item forming opinions, participating in polls and ranking stakeholders by degrees of agreement
\item obtaining clear explanations, using argument graphs, of the differential effects of alternative policies or legal theories in particular cases
\end{PPI}
So Carneades is a software system with very broad intended applications, but it is not misleading to say, more concisely, that it is a tool suite for supporting the practice of deliberate 
%\footnote{The developers do not use the word ``deliberate'', but a few minutes using the system make it clear that it would not be useful for } 
defeasible argumentation. 
%in the same sense that Adobe Photoshop is a tool to support the creation and editing of 2D images.

I tried out Carneades. As of 15 Aug 2014, the program comes with only one example, called ``Copyright in the Knowledge Economy''. The instructions for the ``guided tour'' of the example (described as an ``opinion formation and polling tool'') certainly have a lot in common with the goals of this thesis:
\begin{quote}
\textit{It guides you step by step through the arguments on all sides of a complex policy debate, providing you with an overview of the issues, positions and arguments in a systematic way. The tool can help you to form your own opinion, if you don't yet have one, or to critically evaluate and reconsider your preexisting opinion, if you do. The tool also enables you to compare your responses with published positions of some stakeholders, such as the official positions of political parties. This can help you to find persons and organizations which best represent or share your views and interests. }
\end{quote}

However, examining the argument itself one finds that, aside from the tree structure, it is not a great departure from typical natural language arguments. Here is a prototypical example, where ``exceptions'' means copyright exceptions:\\

\indentoff

\texttt{Q4. Should certain categories of exceptions be made mandatory to ensure more legal certainty and better protection of beneficiaries of exceptions?}\\

\texttt{pro Argument \#1: The permitted exceptions should be harmonised so that they are available in all Member States.}
\begin{PPI}
\item \texttt{pro Argument \#1:}
	\begin{PPI}
		\item \texttt{Performing the action of harmonizing the exceptions and giving }\\ 
		\texttt{precedence to community law over contracts would achieve a state in which it easier for 
		researchers and students to work in more than one Member State.}
		\item \texttt{Harmonizing the copyright exceptions would make it easier for researchers and students to work in more than one Member State.}
		\item \texttt{Achieving the goal of making it easier for researchers and students to work in more than one Member State would promote the values of efficiency, legal certainty, scientific research and education.}
		\item \texttt{In the circumstances: Researchers and students increasingly work in more than one Member State. The patchy availability of exceptions makes their work difficult, because what is lawful in one country is probably unlawful in another. The situation is made worse by the provision of most Member States that contracts, governing the use of digital material, automatically overrides statute law.}
	\end{PPI}
\item	\texttt{con Argument \#2:}
	\begin{PPI}
		\item \texttt{It is essential that the basic principle of freedom of contract be recognized and preserved by any copyright legislation.}
		\item \texttt{Harmonizing copyright exceptions would impair the freedom of contract.}
		\item \texttt{Impairing the freedom of contract would demote the values of innovation and the dissemination of knowledge and information.}
		\item \texttt{Currently, the lack of harmonization of copyright exceptions facilitates the freedom of contract.}
	\end{PPI}
\end{PPI}

In the argument graphs approach they take, formal logic is not imposed on arguments. Instead, the dialogue features, together with the fundamental notions of an argument attacking or defending a proposition, must be used by one arguer to try to make the other's reasoning seem less sound.

%\IMG{\includepdf[pages={-}]{ProofsAndCritiques/RelatedWork/Carneades_-_Copyright_in_the_Knowledge_Economy_full.pdf}}

%\medskip
%
%Carneades uses a defeasible reasoning framework, and as a result is weakly opinionated; 

%The developers of Carneades would love to see their system used by lawyers or policy makes for \textit{any} aspect of their work. Most software systems have a similar philosophy: they are built to nonjudgementally improve the efficiency of people doing certain kinds of work. They are unopinionated.

%\bigskip
%{\bf ArgueWeb}

\subsection{\Finished{Design of a web system}}
\indenton
This section describes work in progress.

Interpreted formal proofs are written using a web-based IDE (integrated development environment), where the document is tree-structured except for some leaves that contain natural language text, which is scanned for symbol ids to insert references. There is an auto-complete feature for already-declared symbols. \U{Declarations} (axiom, lemma, or new symbol introduction) can be tagged, to make groups of declarations, in order to more-concisely specify a subset of declarations that should be used to prove a lemma\footnote{As I mentioned earlier in Section \ref{s:implementation}, this has so-far been necessary when there are a large number of declarations, due to the non-goal-directed nature of the saturation-based first order theorem provers that I have used.}.

An author of an interpreted formal proof may grant edit privileges to other users, and simultaneous multi-collaborator editing, versioning, and unbounded undo will be implemented with the help of a \textit{realtime framework}, such as the Google Drive Realtime API. %\footnote{The Google Drive API is also used for almost all persistent data storage, to eliminate the need for a database. Among its other benefits are versioning and unbounded undo operations.}.
The vision is that interested people who don't know formal logic will be able to contribute to interpreted formal proofs by writing and improving the many required sections of natural language text, both for the language interpretation guide entries and for introductions to arguments.

When a user begins to criticize an author's interpreted formal proof, a dialogue data structure is created. It stores, first of all, the critic's current critique of each axiom, which includes one of the following stances:
\begin{PPI}
\item accept - the critic commits to having only personal interpretations of the current language that satisfy the axiom. 
\item weakly reject - the critic commits to having some personal interpretations of the current language that satisfy the axiom, and some that falsify it. 
\item strongly reject - the critic commits to having only personal interpretations of the current language that falsify the axiom.
\item semantics criticism - the critic submits at least one symbol used in the axiom whose language interpretation guide entry is too vague for them to evaluate the axiom - that is, to take one of the previous three stances. 
\end{PPI}
A critique of any of the latter three categories may have attached to it declarations that are owned by the critic (see Section \ref{s:criticizingVIproofs} for details). The dialogue data structure also stores responses to those critiques by the author of the interpreted formal proof, which may also include new declarations. The author may make changes to the original declarations of their interpreted formal proof that are local only to one dialogue. 

The greatest foreseen challenge is when the author wishes to make a change to their interpreted formal proof that affects all ongoing dialogues\footnote{Note that dialogues will happen slowly, since the work is being done in users' free time.} (hereafter: a change to their interpreted formal proof's \textit{root document}), especially for changes that are not a direct response to a criticism. That includes improvements initiated by the author to the wordings of language interpretation guide entries (which will happen very often), as well as major structural changes to the proof to fix an earlier-made poor formalization decision (i.e. refactoring, as it's called in software engineering). Such changes can subtly affect the meaning of critiques, or in the worst case render current dialogues unintelligible. My current best idea to address this is as follows. When an author wishes to make a change to a root document that is currently involved in at least one dialogue, they are asked to make a claim about the change's effect on each ongoing dialogue, in particular whether they expect the change to require certain specific kinds of adaptations by the critics of the current state of their criticism. Fortunately, it should not ever be necessary to make adaptations to the entire dialogue, since the author's change to the root document can be recorded in each dialogue and viewed in the same way as a change that was initiated as a response to a critique.

\section{\Finished{Obstacles for this project}}

%\begin{aquote}{Gottfried Leibniz, 1706\footnotemark}
%{\it ``It is true that in the past I planned a new way of calculating
%suitable for matters which have nothing in common with mathematics,
%and if this kind of logic were put into practice, every reasoning,
%even probabilistic ones, would be like that of the mathematician: if
%need be, the lesser minds which had application and good will could,
%if not accompany the greatest minds, then at least follow them. For
%one could always say: let us calculate, and judge correctly through
%this, as much as the data and reason can provide us with the means for
%it. But I do not know if I will ever be in a position to carry out
%such a project, which requires more than one hand; and it even seems
%that mankind is still not mature enough to lay claim to the advantages
%which this method could provide.''}
%\end{aquote}
%\footnotetext{From translation of a letter to Sophia of Hanover \cite{leibniz}}

%\textbf{Difficulty of writing and reading interpreted formal proofs}\\

\begin{aquote}{Gottfried Leibniz, 1679, \textit{On the General Characteristic} \cite{leibniz1976}}
\textit{ 
The true reason for this straying from the portal of knowledge\footnote{Leibniz is referring to his imagined \textit{characteristica universalis}, or any similar project.} is, I believe, that principles usually seem dry and not very attractive and are therefore dismissed with a mere taste.}
\end{aquote} \index{Leibniz (quotes)}

I have come to believe that there is only one significant obstacle against this project gaining interest, which is the difficulty of writing and reading interpreted formal proofs. The proofs in this paper are tedious to read, even in HTML, and they were tedious to write as well. 
% This is already in the Logical Positivists section
% Leibniz was aware of this problem:
%\begin{aquote}{Gottfried Leibniz, 1679, {\it ``On the General Characteristic''}\cite{leibniz1976}}
%{\it The true reason for this straying from the portal of knowledge is, I believe, that principles usually seem dry and not very attractive and are therefore dismissed with a mere taste.}
%\end{aquote}
I do not have a perfect solution in mind for this problem. My current approach is to eliminate barriers to entry for use of the software system (e.g. programming experience shouldn't be necessary to author or criticize an argument, and one should be able to get started immediately, on any operating system, without installation), and to minimize editing friction as much as possible (e.g. type-aware autosuggest, as found in modern IDEs for strongly-typed programming languages). 

For a long time I have been concerned about the practical effect of authors and critics who do not argue ``in good faith'', according to the ideals set out in Section \ref{s:VIFP}. I no longer believe that uncooperative \textit{authors} will be a significant concern. Due to the fundamental difficulty of writing an argument as a computer-readable formal deductive proof (even with an excellent user interface), I expect to have interest only from authors who will want to write their arguments so that they are as strong as possible according to the critics whose reasoning they respect the most. In contrast to authoring a new interpreted formal proof, writing a simple criticism of an existing  proof (Section \ref{s:criticizingVIproofs}) is by design not difficult, so there is greater potential for frivolous criticisms and other time-wasting uncooperative behavior. However, formal logic provides a natural notion of progress by which most earnest criticisms will ``make progress'', namely the proof-theoretic strengthening of a system of axioms, and the growing proof-theoretic independence relations that result from a critic weakly-rejecting an axiom, or an author weakly-rejecting an axiom proposed by a critic (recall this means a commitment to the independence of the axiom, which has consequences of independence of other sentences). Of course, when an author's intended interpretations of their language are infinite structures, this notion of progress is not necessarily terminating, even when the language is never non-conservatively extended, since for non-propositional first-order languages $\cL$, there are infinite families of $\cL$-theories $\Gam_1 \subset \Gam_2 \subset \cdots$ such that $\Gam_{i+1}$ is strictly stronger than $\Gam_i$ for all $i$. I do not expect that technical possibility to arise in practice \textit{except} when it is accompanied by non-conservative extensions of the language. There is a correspondence between the non-conservative extensions case and the familiar experience in informal argumentation where a dialogue is never ending due to the \textit{scope} of the argument being repeatedly expanded. The theoretical issue is that it may be difficult to distinguish by any practical technical test between those non-conservative language extensions that are necessary to state a criticism, and those that are merely stalling. Whether this occurs often in practice, and if so how it effects the goals of the project, remains to be seen.

\bibliographystyle{alpha}
\bibliography{thesis_refs}

%\printindex

\end{document}

%% file: problem_domain.tex
%!TEX root = ../thesis.tex
\indenton
\indent \indent Provided the uncertainty involved in a problem is not too great, or that it {\it is} too great but one side of the argument has the burden of proof, it is my view, from several years of working on this project and a decade of thoughts leading up to it, that the main impediments to rigorous deductive reasoning about socially relevant issues are
\begin{inparaenum}[\itshape a\upshape)]
\item conventional mathematical modeling difficulty;
\item conventional mathematical problem solving difficulty\footnote{Two of my proofs, the Leighton Hay argument and the smoking/cancer argument, are currently contingent on the truth of mathematical statements that I cannot easily prove. This is my attitude about such statements: there are mathematicians out there who can easily prove or disprove them, but I think it would be premature to call upon them until proofs of the statements have actually been demanded by critics (called a {\it mathematics detail criticism} in the paper). In the meantime, I give some empirical evidence of their truth (in this case, numerical evaluation of a complicated integral, without error bounds). Most importantly, there are other, more-subjective axioms of the proof that are much easier targets for criticism. It may even be wise to build in some precedence in the rules for criticizing an interpreted formal proof, whereby under certain conditions (which aren't obvious to me at present) one must accept the axioms that involve vague and subjective concepts before demanding a proof of a purely-mathematical claim (of course, one should always be able to present disproofs). \label{f:unprovedclaims}}; and
\item tedium\footnote{This has been the hardest of the three for me to cope with. My hope is that this impediment will be reduced by making the construction of such arguments a collaborative, social process on the web, with an editor having auto-suggestion and other features of modern IDEs (Section \ref{s:websystem}).}.
\end{inparaenum}
These are strong impediments. For that reason, I think it is worthwhile to describe the questions that I think are best-suited for rigorous deductive reasoning. These are {\it contentious questions with ample time available}. Typical sources of such problems are public policy and law.  
Without ample time, it may be detrimental to insist on deductive reasoning; as pointed out in many places, when complete heuristic reasoning and incomplete deductive reasoning are the only options, it is probably best to go with the former.
Without contentiousness, there is little motive for employing fallacious reasoning and rhetoric to advance one's position, and this, I think, defeats much of the benefit of using formal logic (or some approximation of it, as appears in mathematics journals). At the same time, lack of contentiousness does not proportionally reduce the work required for rigor, so we are left with less expected benefit relative to cost. Leibniz was conscious of this point:
\begin{aquote}{Gottfried Leibniz, 1678, Letter to Walter von Tschirnhaus\cite{leibniz1976}}
\textit{I certainly believe that it is useful to depart from rigorous demonstration in geometry because errors are easily avoided there, but in metaphysical and ethical matters I think we should follow the greatest rigor, since error is very easy here. Yet if we had an established characteristic\footnote{Leibniz is referring to the practical system/method that he envisioned, but was unable to devise.} we might reason as safely in metaphysics as in mathematics.}
\end{aquote}
In contrast, some prominent Logical Positivists seem to have thought that this is not a crucial constraint (e.g. Hans Reichenbach's work on axiomatizing the theory of relativity).

%% file: vaguely_interpreted_formal_proofs.tex
%!TEX root = ../thesis.tex
\newcommand{\Typ}{\mathcal{T}}

\indenton

%\Gray{I've gone back to ``vaguely'' because that is the real difference between this use of formal logic and the use of formal logic in mathematics. All formal proofs can reasonably be considered ``informally-interpreted'', so that is not an ideal name.}

This section defines an elaboration of a kind of document that most teachers of first-order logic have used at least implicitly. The point is just to make concrete and explicit a bridge between the formal and informal, providing a particular way, which is amenable to criticism, for an author of a proof to describe their intended semantics in the metalanguage. 

The definition of {\it interpreted formal proof} is tailored for classical many-sorted first order logic, but it will be clear that a similar definition can be given for any logic that has a somewhat Tarski-like semantics, including the usual untyped classical first order logic, or fancier versions of many-sorted first-order logic.\footnote{Earlier versions of this thesis included the syntax and semantics of such a fancier logic. That logic is a little more convenient for formalization, but I discarded it because it introduces another barrier to entry for users of the system, and because its reduction to many-sorted first order logic --the language of resolution theorem provers-- introduced too many usually-unuseful axioms that drastically slowed down proof search.} A very minor extension of the usual definition of many-sorted first order logic (where sorts must be interpreted as disjoint sets that partition the universe) with easily-eliminable sort operators is used here and in the examples.
% it includes easily-eliminable sort operators and partial functions/undefinedness, the latter based on \cite{farmer-simple}. 
A \U{language} is just a set of symbols, each of which is designated a constant, predicate, function, sort, or sort-operator symbol. A \U{signature} is a language together with an assignment of types to the symbols (or, in the case of sort operators, just an assignment of arities).

There are four kinds of formal axioms that appear in an interpreted formal proof:
\begin{LPPI}
\item An \U{assumption} imposes a significant constraint on the semantics of vague symbols (most symbols other than common mathematical ones), even when the semantics of the mathematical symbols are completely fixed.
\item A \U{claim} does not impose a significant constraint on the semantics of vague symbols. It is a proposition that the author of the proof is claiming would be formally provable upon adding sufficiently-many uncontroversial axioms to the theory.
\item A \U{simplifying assumption} is a special kind of an assumption, although what counts as a simplifying assumption is vague. The author of the proof uses it in the same way as in the sciences; it is an assumption that implies an acknowledged inaccuracy, or other technically-unjustifiable constraint, that is useful for the sake of saving time in the argument, and that the author believes does not bias the results. 
\item A \U{definition} is, as usual, an axiom that completely determines the interpretation of a new symbol in terms of the interpretations of previously-introduced symbols.
\end{LPPI}

A \U{language interpretation guide} $g$ for (the language of) a given signature is simply a function that maps each symbol in the language to some natural language text that describes, often vaguely, what the author intends to count as {\it an} intended interpretation of the symbol. Due to the vagueness in the problems we are interested in, a set of axioms will have many intended models. Typically $g(s)$ will be between the length of a sentence and a long paragraph.

A signature's language has {\it sort symbols}, which structures must interpret as disjoint subsets of the universe. A language can also have {\it sort operator symbols}, which are second order function symbols that can only be applied to sorts. In this project sort operators have a nonvital role, used for uniformly assigning names and meanings to sorts that are definable as a function of simpler sorts, when that function is used multiple times and/or is applied to vague sorts (i.e. sorts in $\lav$, introduced below).\footnote{For example, if our proof only needs the power set of one mathematical sort $S$ (in $\lama$), then using a sort operator would have little benefit over just introducing another mathematical sort symbol named $2^S$. Arguably one cannot say the same if $S$ is a vague sort (in $\lav$), since then we would have to introduce $2^S$ as a vague sort as well, and there is some value for readers in minimizing the number of vague symbols when possible.} 
A signature assigns sorts to its constants, and types to its function and predicate symbols. In this project, types are mostly used as a lightweight way of formally restricting the domain on which the informal semantics of a symbol must be given (by the language interpretation guide). To see why they are beneficial, suppose that we didn't have them, e.g. that we were using normal FOL. For the sake of clarity, we would nonetheless usually need to specify types either informally in the language interpretation guide, or formally as axioms. In the first case, we inflate the entries of the language interpretation guide with text that rarely needs to be changed as an argument progresses, and that often can be remembered sufficiently after reading it only once. In the second case, we clutter the set of interesting axioms (e.g. the non-obvious and controversial axioms) with uninteresting typing axioms.

A \U{sentence label} is one of $\{\fassum, \fsimp, \fclaim, \fdefn, \fgoal\}$, where $\fassum$ is short for {\it assumption} and $\fsimp$ is short for {\it simplifying assumption}. A \U{symbol label} is one of $\{\svague, \smath, \sdef\}$. 

An \U{interpreted formal proof} is given by the following components. Intuitively, $\lama$ is for symbols that should have the same interpretation in all models.
\begin{LPPI} 
\item A signature $\Sig$.
\item A set of well-typed $\Sig$-sentences \G called the {\it axioms}. 
\item An assignment of symbol labels to the symbols of $\Sig$. If $\la$ is the language of $\Sig$, then for each symbol label $l$ we write $\la_{l}$ for the symbols assigned label $l$.
\item An assignment of sentence labels to the elements of \G, with one sentence labeled $\fgoal$. For each sentence label $l$ we write $\G_{l}$ for the sentences in $\G$ labeled $l$.
\item An assignment of one of the sentence labels $\fassum$ or $\fsimp$ to each type assignment of $\Sig$. These typing declarations can be viewed as sentences too, and though they will usually be regular assumptions (labeled $\fassum$), occasionally it's useful to make one a simplifying assumption (labeled $\fsimp$).
\item The sentences in $\Gdefn$ define the constant, function, and predicate symbols in $\lad$. Function and predicate symbol definitions have a form like $\forall x_1{:}S_1.\ldots.\forall x_k{:}S_k.\ f(x_1,\ldots,x_k) = t$ where $t$ can be a term or formula (in the latter case, replace $=$ with $\iff$) and the $S_i$ are sorts. 
\item $\lav, \lama, \lad$ are disjoint languages, $\lav$ does not contain any sort-operator symbols,\footnote{I suppose that restriction could be lifted, but I haven't had any desire for vague sort operators in all the time I've worked on this project.} and $\lad$ contains neither sort nor sort-operator symbols\footnote{Another inessential constraint, which I've added simply so that I don't have to include something in the grammar for defining sorts or sort-operators in terms of other sorts and sort operators}.
\item $g$ is a language interpretation guide for a subset of the language of $\Sig$ that includes $\lav$ and $\lama$. So, giving explicit informal semantics for a defined symbol is optional\footnote{\label{f:definitionsemantics}This may change after experience with criticizing proofs on the web (Section \ref{s:websystem}), as the intended semantics of a defined symbol can be very obscure relative to the intended semantics of the symbols used in the definition, despite the fact that the former semantics is completely determined by the latter.}.
\item Optionally for each axiom, a natural language translation of the axiom.\footnote{This was added late to the definition of \viproofs, as it appears to introduce another source of semantics problems (see ``translation criticism'' below). The dilemma is that natural language translations will be demanded by readers of \viproofs as they are usually easier to read, and simply positing that such translations, when given, should not be trusted for veracity, will not make it so. If this becomes a problem, adding features of English to the logic may be useful. On the other hand, see Footnote \ref{fn:translation} on page \pageref{fn:translation} about how this is a special case of language interpretation guide entries for defined symbols.}
\item $\Ggoal$ is provable from $\Gassum \cup \Gsimp \cup \Gclaim \cup \Gdefn$.
\item For each $\psi \in \Gclaim$, any reader in the intended audience of the proof can come up with a set of $\lama$-sentences $\Delta$, which are true with respect to the (informal) semantics given by $g$, such that $\Gassum \cup \Gdefn \cup \Gsimp \cup \Delta$ proves $\psi$. The first paragraph of the next section gives a more-precise condition.
\end{LPPI}

\subsection{\Finished{Criticizing \viproofs}} \label{s:criticizingVIproofs}

$\lama$ is intended to be used mostly for established mathematical structures, but in general for structures that both sides of an argument agree upon sufficiently well that they are {\it effectively objective with respect to} $\Gclaim$. For each person $p$ in the intended audience of the proof, let $\Delta_p$ be the set of $\lama$-sentences that $p$ can eventually and permanently recognize as true with respect to the informal semantics given by $g$. Then we should have that $\bigcap\limits_{p \in \text{audience}} \!\!\!\!\!\!\Delta_p$ is consistent and when combined with $\Gassum \cup \Gdefn \cup \Gsimp$ proves every claim in $\Gclaim$. If that is not the case, then there is some symbol in $\lama$ that should be in $\lav$, or else the intended audience is too broad.
\medskip

The purpose of the language interpretation guide is for the author to convey to readers what they consider to be an acceptable interpretation of the language. 
\textbf{Subjectiveness results in different readers interpreting the language differently, and vagueness results in each reader having multiple interpretations that are acceptable to them.} Nonetheless, an ideal language interpretation guide is detailed enough that readers will be able to conceive of a vague set of {\it personal $\Sig$-structures} that is precise enough for them to be able to accept or reject each assumption (element of $\Gassum \cup \Gsimp$) independent of the other axioms. 
When that is not the case, the reader should raise a \U{semantics criticism} (defined below), which is analogous to asking ``What do you mean by X?''.

In more detail, to review an interpreted proof $\pi$ with signature $\Sig$ and language $\la$, you read the language interpretation guide $g$, and the axioms $\Gam$, and either accept $\pi$ or criticize it in one of the following ways:
\begin{PPE} \label{d:reviewproof}
\item[(1)]  \U{Semantic criticism}: Give $\phi \in \Gam$ and at least one symbol $s$ of $\lav$ that occurs in $\phi$, and report that $g(s)$ is not clear enough for you to {\it evaluate} $\phi$, which means to conclude that all, some, or none of your personal $\Sig$-structures satisfy $\phi$. If you cannot resolve this criticism using communication with the author in the metalanguage, then you should submit a $\Sig$-sentence $\psi$ to the author, which is interpreted by the author as the question: Is $\psi$'s truth compatible with the intended semantics given by $g$? 
\item[(2)] \U{Classification criticism}: Criticize the inclusion of a symbol in $\lama$, if necessary by doing the same as in (1) but for $\lama$. This is the mechanism by which one can insist that vague terms be recognized as such. The same can be done when $\phi$ is a type assignment or sort constraint, in which case $\psi$ is a $\Sig$-sentence that uses sort symbols as unary predicate symbols.
\item[(3)] \U{Mathematics detail criticism}: Ask for some claim in $\Gclaim$ to be proved from simpler claims (about $\lama$ interpretations). 
\item[(4)] \U{Subjective criticism}: Reject some sentence $\phi \in \Gassum \cup \Gsimp$, which means to conclude that at least one of your personal $\la$-structures falsifies $\phi$. If you wish to communicate this to the author, you should additionally communicate one of the following:
  \begin{PPE}
  \item \U{Strongly reject $\phi$} : Tentative commitment to $\neg \phi$, i.e. that all of your personal $\Sig$-structures falsify $\phi$.
  \item \U{Weakly reject $\phi$} : Tentative commitment to the independence of $\phi$, i.e.  that $\phi$ is also satisfied by at least one of your personal $\Sig$-structures. Intuitively, this means that $\phi$ corresponds to a simplifying assumption that you are not willing to adopt.
  \end{PPE}
\item[(5)] \U{Translation criticism}: Criticize as \textit{misleading} the informal natural language text attached to an axiom.\footnote{\label{fn:translation}This is actually a special case of criticizing the optional semantic description attached to a definition, since one can always replace an axiom $A$ with a new defined 0-ary predicate symbol $P_A \iff A$ and the new axiom $P_A$.}
\end{PPE} 

In the context of its intended audience, we say that an interpreted formal proof is \U{locally-refutable} if no member of the intended audience raises semantic or classification criticisms when reviewing it. \label{p:locally-refutable} A locally-refutable proof has the desirable property that by using the language interpretation guide $g$, any member of the audience can evaluate each of the axioms of the proof independently of the other axioms. Local-refutability is the ideal for interpreted formal proofs. It is a property that is strongly lacking in most mathematical arguments in economics or game theory, for example, and in every sophisticated abuse of statistics. When an argument is far from locally-refutable, it is hard to criticize in a standard, methodical manner, and that ease of criticism is a central goal of this project.

%% file: choice_of_logic.tex
%!TEX root = ../thesis.tex

For this project I use a superficial extension of classical many-sorted first-order logic (MSFOL), itself a superficial extension of classical first-order logic (FOL), and {\it for this particular project} --rigorous deductive reasoning about the kind of questions described in Section \ref{s:problemdomain}-- I believe that is the right choice:
\begin{quote}
\textbf{Claim:} Nothing simpler than classical FOL will suffice, and nothing significantly more complex is necessary.
\end{quote}

\textbf{Note 1:} \label{p:whynodeductive} Though there are many vocal advocates of nonclassical logics, the scope of the previous claim --this project only-- is narrow enough that there may be no serious dispute of it, putting it outside of the problem domain described in Section \ref{s:problemdomain} due to lack of contentiousness. I will therefore not put the effort into giving a rigorous deductive argument for the claim, and indeed my argument will be neither rigorous nor deductive. That said, I believe that I could find such a formalization without significantly weakening my position in the process. %\textbf{Incidentally, I \textit{have} experienced that the process of finding a rigorous deductive argument has often resulted in my having to temper my opinion, at least temporarily.} I think it hard to overstate that advantage of rigorous deduction. Among other things it provides a force for compromise -- a force in the same democratic spirit as ``I could be wrong'', but with much greater discriminatory power.

\textbf{Note 2:} I do believe that the claim holds more broadly --that nonclassical logics should generally be framed and thought of as mathematical theories, as opposed to the grandiose framing as \textit{alternatives to classical logic}\footnote{Which has social implications that I won't go into, e.g. giving outsiders the impression that formal logic is controversial or in flux.}-- but it is not relevant to this project to argue that here. Maria Manzano in \cite{Manzano} gave arguments in support of this, demonstrating in one book that second-order logic, type theory, modal and dynamic logics can be naturally and usefully simulated in MSFOL. Her book appears to have been mostly ignored by researchers in applied nonclassical logic, and according to \cite{ManzanoReview} she made too strong a claim about the usefulness of translations to MSFOL. I do not advocate that researchers should use the syntax of MSFOL in all situations, since often a custom syntax is more concise, and allows for more-natural statements of metatheorems and simpler programming of automated theorem proving tools (e.g. if one is programming a decision procedure for some version of propositional temporal logic, it would be silly and inefficient to use a problem encoding with bound variables).

\medskip

Back to the topic of this section: the choice of logic for \textit{this project}. I will focus on the second part of the above Claim, as arguments about the need for the expressiveness of FOL, in particular its predicate symbols and variables, are commonplace. 

The main factor compelling the use of FOL is this: It is vital for this project that the interface between syntax and semantics is as simple and transparent as possible, and FOL with Tarski semantics is the best in this respect. I know of no \textit{substantial} dispute of the relative simplicity and transparency of Tarski semantics. There are criticisms of material implication, of course, but such criticisms stem from misuse of classical logic, in particular use motivated by a desire to extract or impose some sort of meaning from validity that is different from its characterization in terms of truth-with-respect-to-structures. Unfortunately, there are no official directions for the use of FOL, but if there were, they would clearly imply that to reason formally using other sorts of implication, e.g. relevant implication, causality, or conditional probability, one should develop a mathematical theory of relevant implication, causality, or conditional probability, and formalize it as a first-order theory. Since such mathematical theories, without fail, are more complex and nuanced than Tarski semantics, we have no substantial dispute of the relative simplicity and transparency of FOL with Tarski semantics.

%That said, in this paper, which is long enough already, I do not give the argument the space it deserves. In my dissertation, I take some space to explain how common forms of defeasible reasoning can be carried out in deductive logic, and how ideas from e.g. modal or epistemic or temporal logic can be used as-necessary without needing to build them into the definition of the logic (complicating the semantics). On the other hand, it may turn out that the more-concise syntax of some such such logic is often enough helpful for reading and writing proofs, without obscuring the meaning, that it would be wise to adopt it for the system. That, I think, should be decided after a great deal more experience.

\medskip

The second factor compelling the use of classical FOL, part of which is alluded to in the previous paragraph's counter-criticism of criticisms of material implication, is this: extensions or similarly-expressive alternatives to FOL are either (1) unuseful or conflicting with the goals of this project, or (2) useful and compatible with the goals of this project, but simulating their benefits for specific proofs is not hard, and so the added complexity of complicating the logic is not justified. This claim is clarified by the following 
  
%\TODO[Modifying this so the simulations are proof-dependent. See talk about this from slides.]\\

%I will take it as agreed upon that we should be using a logic that is at least as expressive as first order logic. I can argue for that, but no one yet, in conversation, has disputed it, and I do not expect that any readers will, making the writing of the argument not worthwhile (a standard similar to the contentiousness requirement given in \ref{s:problemdomain}).

\noindent \textbf{Hypothesis:}\\
Let $L$ be any nonclassical alternative to \MSFOL with its own semantics. I hypothesize that the difficulty of converting a proof in $L$ to an equally-readable proof in \MSFOL, or of extending the definition of \MSFOL to accommodate the useful features of $L$\footnote{Which I have done, and undone, several times, ending with MSFOL plus type operators.}, is proportional to the difficulty, compared to \MSFOL, of interpreting sets of $L$-sentences using $L$'s semantics. Furthermore, whenever there is no added difficulty of interpreting sets of sentences, or when the features of $L$ allow us to write {\it easier} to interpret (sets of) sentences,\footnote{As I believe is the case for type operators (AKA parametric polymorphism), subtyping, and \cite{farmer-simple}-style partial functions, although the latter two features have drawbacks when it comes to proof search using currently existing theorem provers.} 
%(and without making it too much easier to write deceptively-simple-looking theories whose meaning is complex)
then I hypothesize we can already simulate those features with low overhead in \MSFOL, and/or we can easily extend our definition of \MSFOL to accommodate those features, without straying much from Tarski semantics (a ``superficial extension'').

I should clarify that the Hypothesis is not an argument against the study of defeasible and nonclassical logics in general, because its force depends on some uncommon aspects of this project: 
\begin{PPI}
\item Sentences about vague, subjective concepts and uncertain knowledge are by nature already more difficult to interpret than sentences about traditional mathematical concepts and certain knowledge.
\item With the top-down, minimally-reductionist approach to proofs that I advocate, there are fewer syntactic definitions (more non-defined symbols) and more axioms, making ease of interpretation more important than it usually is in applications of formal logic.
\item Theory-construction is not a goal; we only care about individual proofs. Because of this, simulation of the used features of another logic $L$ can be \textit{proof specific}, often making the task much easier than a general translation of all $L$-sentences. 
\end{PPI}

\medskip

In Chapter \ref{c:defeasible} I will explain why the arguments used to motivate defeasible logic do not apply to this project, and give some minimal examples of formalizing defeasible reasoning in deductive logic.\footnote{The major interpreted formal proofs that take up the bulk of this thesis contain more complex examples, although they are not labeled as such.} In contrast to there, where I argue that there is nothing to be gained from using defeasible logic for this project, here I will briefly argue that something would be lost. 

\begin{quote}
\textbf{Claim 1:} The requirements of formal deduction make interpreted formal proof versions of \textit{all} defeasible arguments less psychologically persuasive, though bad defeasible arguments suffer worse than good ones. If defeasible logic were permitted, there would be much less incentive to ever do the extra work required for formal deduction, as bad defeasible arguments can be more psychologically persuasive than good deductive ones.
\end{quote}

\begin{quote}
\textbf{Claim 2:} Argumentation with defeasible logic is not profoundly different from natural language argumentation (e.g. see Section \ref{s:carneades} about the online defeasible argumentation system Carneades), whereas argumentation with deductive logic is. The novelty of interpreted formal proofs about socially-relevant issues is an essential motivation for this thesis, so asking whether defeasible logic should be used reduces to asking whether this thesis should be written at all. Thus, if you are convinced that this thesis was adequately motivated, then you should be convinced that declining defeasible logic was adequately motivated as well.
\end{quote}

%I will continue this argument for the case of defeasible logics in Chapter \ref{s:defeasible}. In short, defeasible logics are 

\bigskip

The remainder of this section, far from considering all alternatives to FOL other than defeasible logics, is devoted to a consideration of modal logics. This is for concreteness, and because, as I said earlier (page \ref{p:whynodeductive}), it is not clear that a deductive argument for why I've chosen FOL is called for. Moreover, {\it many} of the popular extensions of FOL are modal logics, and modal logics are especially popular in philosophy, which is the field that has historically taken the greatest interest in formal reasoning about contentious issues. 

The case against modal logic for this project comes down to two points:
\begin{PPI}
\item[(i)] Modal logic introduces another syntactic device by which semantically-complex sentences can be written in very simple terms. Note that this property is the main \textit{advantage} of modal logic for other applications (other than this project). Note also that FOL (and modal first-order logic, by extension) is not without the capacity to disguise semantically-complex sentences, namely by using deeply nested definitions (but see Footnote \ref{f:definitionsemantics} on page \pageref{f:definitionsemantics} for discussion of how to address that). 
\item[(ii)] Modal logics are easy to simulate in MSFOL, in a natural way, by formalizing possible-world semantics. Hence, in the worst case we would be writing slightly more-verbose sentences (and even that can be mitigated using syntactic definitions). Also, more can be expressed in the translated MSFOL language than can in the language of modal logic.
\end{PPI}
First let's look at the general simulation, and after that I will use an example to illustrate (i). This is a standard simulation that can be found, in more detail, in \cite{blackburn}.

Let $\cL$ be a signature for a many-sorted first-order modal logic with one or more $\Box$-like modal connectives $\Box_1,\ldots,\Box_k$ and with corresponding $\loz$-like connectives $\loz_j$. The corresponding MSFOL signature $\cL'$ for the simulation is the same as $\cL$ except:
\begin{PPI}
\item It has an extra sort $\cW$ for worlds. The variables $w, w_1, w_2, \ldots$ are reserved for $\cW$.
\item For each $\Box$-like modal connective $\Box_j$, it has an extra predicate symbol $R_j : \cW \times \cW \to \BB$ for the corresponding reachability relation.
\item Each function or predicate symbol $f$ in $\cL$ whose domain type is $S_1 \times \ldots \times S_n$ has domain type $\cW \times S_1 \times \ldots \times S_n$ in $\cL'$. That is, an extra argument is added for the world with respect to which the predicate or function is being evaluated.
\end{PPI}
Then $\cL$-sentences are easily translated to $\cL'$-sentences by the following function $\trans{\cdot}$, which given an $\cL$-formula $A$ produces an $\cL'$ formula with the same number and sorts of free variables as $A$, plus exactly one free variable $w$ of sort $\cW$. The syntax $t[w \mapsto w']$ means to substitute variable $w'$ for free occurrences of variable $w$ in $t$. 
\begin{PPI}
\item $\trans{P(t_1,\ldots,t_n} = P(w,\trans{t_1},\ldots,\trans{t_n})$ for $P$ a predicate symbol, including $=$.
\item $\trans{f(t_1,\ldots,t_n} = f(w,\trans{t_1},\ldots,\trans{t_n})$ for $f$ a function symbol.
\item $\trans{A \implies B} = \trans{A} \implies \trans{B}$ and similarly for the other boolean connectives. 
\item $\trans{\Box_j A} = \forall w_i.\, R_j(w_i,w) \implies \trans{A}[w \mapsto w_i]$, where $w_i$ does not occur in $A$. Thus, $\Box_j A$ holds in the current world $w$ iff $A$ holds in every world that is $R_j$-reachable from $w$. Note that in most presentations of frame semantics, the order of the arguments to the reachability relation $R$ is swapped. I use this order so that we can display the translated formula more neatly as $\forall w_i R_j w.\, \trans{A}[w \mapsto w_i]$. 
\item $\trans{\loz_j A} = \exists w_i.\ R_j(w_i,w) \wedge \trans{A}[w \mapsto w_i]$, where $w_i$ does not occur in $A$. We can display such a formula more neatly as $\exists w_i R_j w.\, \trans{A}[w \mapsto w_i]$.
\end{PPI}

For any modal logic $L$ that can be characterized by frame semantics, I claim there is a recursive (and usually finite) set of $\cL'$-sentences $\Psi_{L'}$ that capture validity for $L$. That is, such that for any $\cL$-sentence $B$ and set of $\cL$-sentences $A_1, A_2,\ldots$, have 
\[A_1, A_2,\ldots \models_{L} B \quad \text{ iff }\quad \Psi_{L'},\ \forall w.\,\trans{A_1}, \forall w.\trans{A_2}, \ldots \models \forall w.\trans{B} \]
where $\models_L$ is entailment for $L$ and $\models$ is entailment for MSFOL.

Note that, according to the definition of interpreted formal proof, language interpretation guide entries must be given for the worlds sort $\cW$ and the reachability relations $R_i$. Those components of the modal logic semantics are usually \textit{not} given explicit intended interpretations in applications of modal logic in philosophy, and I claim this is the \textit{only} reason why modal logic ``paradoxes'' in philosophy do not get resolved; the intended semantics remains too vague. 
%\TODO[-----I'm not quite finished with the rest, which uses Fitch's Paradox of Knowability to illustrate point (i) above.-----]

As an example, I will use Fitch's Paradox of Knowability. It is a proof in the language of a propositional modal logic with either one $\Box$-like modal connective $K$ or two $\Box$-like modal connectives $K$ and $\Box$. I will use the latter formulation, since it is easy to convert to the former by identifying $K$ and $\Box$. In either case, only one $\loz$-like connective is used, and it is connected to $\Box$. The axioms are substitution instances of the following axiom schema, where $\phi,\phi_1,\phi_2$ are metavariables that range over formulas, except for schema ($\Box$-Valid) where they range over provable formulas. In every presentation of the proof that I have seen, an (overly-simple) English translation of each sentence is provided. I am giving the ones from \cite{knowability}, except for ($K\loz$-connection) for which I give two versions.

\[
\begin{array}{lll}
\text{($K$-Factivity)} & K \phi \implies \phi & \text{``If a proposition is known, then} \\
& & \text{ it is true.''} \\
\text{($K\wedge$-Distributivity)} & K(\phi_1 \wedge \phi_2) \implies K \phi_1 \wedge K \phi_2 & \text{``If a conjunction is known, then } \\
& & \text{its conjuncts are also known.''} \\
\text{($\Box\loz$-Connection)} & \Box(\neg \phi) \implies (\neg\loz \phi)  & \text{One direction of ``$\Box \neg \phi$ is logically } \\
& & \text{equivalent to $\neg \loz \phi$.'' The SEP} \\
& & \text{article\cite{knowabilitysep} translates it as ``necessarily}\\
& & \text{$\neg \phi$ entails that $\phi$ is impossible.''} \\
\text{(Knowability Principle)} & \phi \implies \loz K \phi & \text{``Every truth is knowable.''}
\end{array}
\]
\ \ ($\Box$-Valid) \ \qquad \qquad \qquad $\Box \phi$, if $\phi$ is provable from $K$-Factivity and $K\wedge$-Distributivity \\

From those axiom schema and the rules of classical propositional logic, it follows that for any formula $\phi$, 
\[(G) \qquad\qquad\qquad \phi \implies K \phi \qquad\qquad\qquad \text{``Every truth is known'' (the unwanted conclusion)} \]

I will argue that the proof is circular when $K$ is \textbf{S5}-like\footnote{More accurately, when the corresponding reachability relation $R_K$ is trivial in that every world is reachable from every other.}, whereas in weaker modal logics $G$ simply does not mean ``Every truth is known,'' and the meanings of the (Knowability Principle) and ($K$-Factivity) are far from clear (see second-to-last paragraph of this section for why they are unclear). My position is that the ongoing philosophical discussion about this proof is completely reliant on underspecified semantics.

Since the translation I gave is for first-order modal logic, I'll briefly describe how to convert to that form. Any nonlogical language can be used, since the proof is really a meta-level proof, but it is convenient to use a minimal language with a single 0-ary predicate symbol $P$, in which case we can take the goal sentence $G$ to be $P \implies K P$. Then, the substitution instances of the above axiom schema needed to derive $G$ and justify the instance of ($\Box$-Valid)\footnote{That is, the third and fourth axioms prove $\neg K \neg G$, which is the prerequisite needed to use $\Box \neg K \neg G$ as an axiom.} are:
\[
\begin{array}{ll}
\text{($K$-Factivity)} & K \neg K P \implies \neg K P \\
\text{($K\wedge$-Distributivity)} & K(P \wedge \neg K P) \implies K P \wedge K \neg K P \\
\text{($\Box\loz$-Connection)} & \Box(\neg K \neg G) \implies (\neg\loz K \neg G) \\
\text{(Knowability Principle)} & \neg G \implies \loz K \neg G  \\
\text{($\Box$-Valid)} & \Box \neg K \neg G 
\end{array}
\]
And the universal closures of the translations of those axioms into MSFOL prove $\forall w. \trans{G}$. In fact, it is easy to see that the universal closures of the translations of ($K\wedge$-Distributivity) and ($\Box\loz$-Connection) are theorems of MSFOL, and ($\Box$-Valid) can be made a lemma provable from the other axioms, since it is a logical consequence of the (Knowability Principle) and ($\Box\loz$-Connection)\footnote{Using the fact that  if $\forall w. \phi(w)$ is valid for any MSFOL-formula $\phi$, then $\forall w. \forall w'R_{\Box}w. \phi(w)$ is valid also.}. Therefore, I will give the translations only for the remaining two axioms, since they are the only ones that can be criticized. Note that $\trans{G} = P(w) \implies \forall w'R_{K}w. P(w')$.

\begin{PPE}
\item $\forall w.\ \left[ \forall w_1R_{K}w.\, \neg \forall w_2R_{K}w_1.\, P(w_2) \right] \implies \neg \forall w'R_{K}w.\, P(w')$, i.e.\\ $\forall w.\ \left[ \forall w_1R_{K}w.\, \exists w_2R_{K}w_1.\, \neg P(w_2) \right] \implies \exists w'R_{K}w.\, \neg P(w')$. Suppose that for every world $w_1$ that is $K$-reachable from the current world, there exists a world $w_2$ that is $K$-reachable from $w_1$ in which $P$ is false. Then there exists a world $K$-reachable from the current world in which $P$ is false.
\item $\forall w.\ \neg \trans{G} \implies \exists w_1R_{\Box}w.\, \forall w_2R_{K}w_1.\, \neg \trans{G}[w \mapsto w_2]$
\end{PPE}

\newcommand{\NI}{\textsf{NI}_{\Box}}

For the class of models that satisfy $\forall w_1,w_2.\, R_K(w_1,w_2)$, those axioms are equivalent to
\begin{PPE}
\item $\neg \forall w.\, P(w) \implies \neg \forall w.\, P(w)$, a tautology.
\item $\forall w.\ \neg \trans{G} \implies \left(\NI(w) \wedge \forall w_2. \neg \trans{G}[w \mapsto w_2] \right)$ where $\NI(w)$ (`\textbf{n}ot \textbf{i}solated') abbreviates $\exists w_1.\, R_{\Box}(w_1,w)$. For the models under consideration, $\forall w_2. \neg \trans{G}[w \mapsto w_2]$ is false, so the consequent in the implication is false, and the sentence further reduces to $\forall w.\ \trans{G}$.
%That sentence
%$(\forall w. \trans{G} ) \vee (\forall w. \neg \trans{G})$. 
%Expanding $\trans{G}$ in the right disjunct yields $\forall w. P(w) \wedge \neg \forall w'R_{K}w.\, P(w')$, which reduces to a contradiction. Thus the Knowability Principle instance for this class of models is equivalent to $\forall w. \trans{G}$.
\end{PPE}
So, for that class of models the (Knowability Principle) instance is equivalent to $G$, i.e. the argument is circular.

Thus, we can assume that at least some of the author's intended models of the axioms falsify $\forall w_1,w_2.\, R_K(w_1,w_2)$. But then we can no longer collapse quantifiers; the two axioms
\begin{PPE}
\item $\forall w.\ \left[ \forall w_1R_{K}w.\, \exists w_2R_{K}w_1.\, \neg P(w_2) \right] \implies \exists w'R_{K}w.\, \neg P(w')$.
\item $\forall w.\ \neg \trans{G} \implies \exists w_1R_{\Box}w.\, \forall w_2R_{K}w_1.\, \neg \trans{G}[w \mapsto w_2]$
\end{PPE} 
cannot obviously be simplified. And without having more-precise semantics for $R_K, R_\Box, $ and $\cW$, I claim one cannot evaluate whether those axioms are satisfied for the author's intended interpretations. Now let's see why it would be hard to give any interesting interpretation of those axioms. The ($K$-Factivity) axiom schema $K \phi \implies \phi$ seems acceptable according to the naive interpretation of $K$ when written in the language of modal logic. However, for instances when $\phi$ itself contains modal connectives, the naive interpretation of the sentence ``If $\phi$ is known, then it is true'' is plainly wrong. The reason is that $\phi$ talks about different objects of the universe (namely, different worlds) when evaluated at different worlds; the sentence should be read as ``If $\phi$ is known, then a certain formula related to $\phi$ is true''. The same goes for the (Knowability Principle) $\phi \implies \loz K \phi$; it is fine to interpret it as ``Every truth is knowable'' only if $\phi$ contains no modal connectives. Otherwise, one must say the more verbose ``If $\phi$ is true, then a certain formula related to $\phi$ is knowable.'' This point is similar to the criticism given by Kvanvig in a number of papers beginning with \cite{kvanvig1995} (and later an entire book!).

Of course we did not {\it need} to translate to MSFOL to make the criticism of the previous two paragraphs. We could have just used the language of frame semantics. But that is missing the point. The criticism explained why the axioms seemed reasonable, which is a requirement for a strong rebuttal in philosophy, and I have demonstrated that it is not hard to do this in MSFOL. But my position --the position of the system advocated in this thesis-- is that this is asking too much of a critic. My work in the criticism should have ended much earlier, at the point just before considering the two cases of whether or not there are intended interpretations that falsify $\forall w_1,w_2.\, R_K(w_1,w_2)$. At that point, I should simply make a semantics criticism with either the ($K$-Factivity) or (Knowability Principle) instance and one of the symbols $R_K, R_\Box, $ or $\cW$. The virtue of insisting on writing the axioms in MSFOL is just this: it forces the \textit{author} of the proof to reveal the complexity of their axioms, rather than putting that burden on the critic.

%% file: walton_marriage.tex
%\tableofcontents
}

\newcommand{\holds}[1]{\prefnone{Holds}{#1}}

% These use more-concise syntax by putting some of the args in subscript or supercript.
\newcommand{\does}[2]{\optdefn{#1}{\text{Does}}{\text{Does}_{#1}(#2)}}
\newcommand{\getmarried}[2]{\optdefn{#1}{\text{getMarriedTo}}{\text{getMarriedTo}_{#1}^{#2}}}
\newcommand{\promise}[2]{\optdefn{#1}{\text{makePromise}}{\text{makePromise}^{#2}(#1)}}
\newcommand{\tlivewith}{\text{LiveWith}}
\newcommand{\livewith}[3]{\optdefn{#1}{\tlivewith}{\tlivewith_{#1,#2}^{#3}}}
\newcommand{\tcompatible}{\text{Compatible}}
\newcommand{\compatible}[3]{\optdefn{#1}{\tcompatible}{\tcompatible_{#1,#2}^{#3}}}
\newcommand{\tsafelypredict}{\text{CanSafelyPredict}}
\newcommand{\safelypredict}[3]{\optdefn{#1}{\tsafelypredict}{\tsafelypredict_{#1}^{#3}(#2)}}
\newcommand{\datedeath}[1]{\optdefn{#1}{\text{dateOfDeath}}{\text{dateOfDeath}_{#1}}}
\newcommand{\yearof}[1]{\optdefn{#1}{\text{year}}{\text{year}(#1)}}
\newcommand{\shouldnot}[2]{\optdefn{#1}{\text{ShouldNotDo}}{\text{ShouldNotDo}_{#1}(#2)}}

\newcommand{\tlivetilldeath}{\text{liveWithTillDeath}}
\newcommand{\livetilldeath}[3]{\tlivetilldeath_{#1,#2}^{#3}}

\newcommand{\firstdeath}[2]{\text{firstDeath}_{#1,#2}}

\newcommand{\tcompatdeath}{\text{compatibleTillDeath}}
\newcommand{\compatdeath}[3]{\optdefn{#1}{\tcompatdeath}{\tcompatdeath_{#1,#2}^{#3}}}

\newcommand{\ptype}{:}
\newcommand{\tto}{\rightarrow}

%\subsection*{Walton's fallacious marriage argument demonstrating equivocation via ``variability of strictness of standards''}

\indentoff

\newcommand{\sA}{A}
\newcommand{\sP}{P}
\newcommand{\sY}{Y}
\newcommand{\sD}{D}
\newcommand{\sS}{*}
\newcommand{\sF}{\Psi}

%This section provides an example of an \textit{interpreted formal proof} and criticism against it. 
This example is also available for reading in HTML:\\
\url{http://www.cs.toronto.edu/~wehr/thesis/walton_marriage.html}
\medskip

Here is the informal argument, verbatim from \cite{InformalLogic}:
\begin{PPE}
\item Getting married involves promising to live with a person for the rest of your life.
\item Nobody can safely predict compatibility with another person for life.
\item One should not make a promise unless one can safely predict that one will keep it.
\item If two people aren't compatible, they can't live together.
\item One should not promise to do something one cannot do.
\item Therefore, nobody should ever get married. 
\end{PPE}
Lines 1-4 of the informal argument correspond to Assumptions \ref{a:marriageinvolves}-\ref{a:compatlive} below. Line 5 is redundant, and line 6 corresponds to the proved conclusion, Proposition \ref{e:marriageconclusion} below.

\medskip

Sorts: 
\begin{PPI}
\item $\sP$ is for people.
\item $\sD$ for dates (e.g. 1 Sept 1998). 
\item $\sA$ for potential actions that are associated with a particular date, but not a particular person (like verbs).  
\item $\sF$ contains a subset of the $\pair{\text{formula},\text{object assignment}}$ pairs (see Definitions \ref{viqd:livetilldeath} and \ref{viqd:compatdeath}). The intended interpretation of this symbol would be simpler if we introduced constant symbols for each element of $\sP$ and $\sD$, in which case I could just identify $\sF$ with a particular finite set of sentences (of the form of the formulas on the right side of $\iff$ in Definitions \ref{viqd:livetilldeath} and \ref{viqd:compatdeath}). 
%\item $\sF$ contains a set of ``propositions'', each element of which is tied to the truth value of a particular $\pair{\text{formula},\text{object assignment}}$ pair, via the predicate symbol $\holds{}$ (see Assumptions \ref{viqd:livetilldeath} and \ref{viqd:compatdeath}). The intended interpretation of this symbol would be simpler if we introduce constant symbols for each element of $\sP$ and $\sD$.
\end{PPI}
%I use $\sS$ for the ``type'' of the range of predicate symbols, i.e. $\{\text{true}, \text{false}  \}$, but that is just for syntactic it is not actually a sort symbol

%C : things that a person can do
%
Predicate symbols:
\[
\begin{array}{rcl}
\does{}{} &\ptype& \sP \times \sA \tto \sB \\
\livewith{}{}{} &\ptype& \sP \times \sP \times \sD \times \sD \tto \sB \\
\holds{} &\ptype& \sF \tto \sB \\
\compatible{}{}{} &\ptype& \sP \times \sP \times \sD \times \sD \tto \sB \\
\safelypredict{}{}{} &\ptype& \sP \times \sF \times \sD \tto \sB  \\
\shouldnot{}{} &\ptype& \sP \times \sA \tto \sB  \\
\leq &\ptype& \sD \times \sD \tto \sB  \\
\end{array}
\]
Function symbols:
\[
\begin{array}{rcl}
\getmarried{}{} &:& \sP \times \sD \tto \sA \\
\promise{}{} &:& \sF \times \sD \tto \sA \\
\livetilldeath{}{}{} &:& \sP \times \sP \times \sD \tto \sF \\
\compatdeath{}{}{} &:& \sP \times \sP \times \sD \tto \sF \\
\datedeath{} &:& \sP \tto \sD \\
\textsf{min} &:& \sD \times \sD \tto \sD 
\end{array}
\]

\indentoff

Style notes:
\begin{PPI}
\item The following variables are reserved for the following types: $d$ for $\sD$, $a$ for $\sA$, $p$ and $q$ for $\sP$, and $\psi$ for $\sF$. Similarly for the primed and subscripted versions of those variables.
\item I leave out leading universal quantifiers.
\item To improve readability, when a function symbol takes arguments of type $\sP$, I put the arguments in the subscript, as in $\does{p}{a}$, and when a function symbol takes one or more arguments of type $\sD$, I put them in the superscript, as in $\livewith{p}{q}{d}{d'}$.
\end{PPI}

Formalization notes;
\begin{PPI}
\item It is not hard to correct the argument for the objection that it clearly doesn't work when $p$ and $q$ are near death, in which case it's especially reasonable to reject Assumption \ref{a:nopredict}. I haven't done so since the argument has other, more-serious flaws. 
\item If after reading the next bullet list about the informal semantics,  you would, like me, still reject Assumption \ref{a:nopromise} for being too broad, then move the sentence to the position of a premise of the goal sentence (Equation \ref{e:marriageconclusion}). 
\item In retrospect, it would have been more-economical to make $\livetilldeath{p}{q}{d}$ and $\compatdeath{p}{q}{d}$ primitive instead of $\livewith{p}{q}{d,d'}$ and $\compatible{p}{q}{d,d'}$, but the way I've done it is more faithful to Walton's presentation.
%\item I have been loose with some aspects of the formalization of time but again that 
\end{PPI}

Here is a sketch of part of the informal semantics (i.e. a language interpretation guide):
\begin{PPI}
\item  $\livewith{p}{q}{d}{d'}$ iff $p$ and $q$ are both alive and share the same main residence during the period from $d$ to $d'$.
\item $\promise{\psi}{d}$ means to make an utterance, on date $d$, like ``I promise that $A$'', which is directed at someone with the intention of their interpreting it as a sincere and literal statement. 
\item The semantics of $\compatdeath{p}{q}{d}$ and $\livetilldeath{p}{q}{d}$ are essentially determined by the semantics of the other symbols by Definitions \ref{viqd:livetilldeath} and \ref{viqd:compatdeath}. 
\item The informal semantics for the other symbols (except for $\compatible{}{}{}$ and $\safelypredict{}{}{}$, which correspond to the terms in the informal argument that are used with varying ``strictness of standards'') are not surprising and not hard to flesh out.
\end{PPI}

%$\firstdeath{p_1}{p_2} $
\medskip

\begin{videfn} The year when the first of $p$ or $q$ dies.
\[ 
\firstdeath{p}{q} := \textsf{min}(\ \datedeath{p},\ \datedeath{q} \ ) 
\]
\end{videfn}
%\begin{defn}
%For all $p,q,d$ there is a proposition $\psi:\Psi$ such that $\psi$ holds iff $p$ and $q$ live together from the date $d$ until one of them dies.
%\[ 
%\forall p,q,d.\, \exists \psi.\, \holds{\psi} \, \iff \, \forall y{:}\sY.\ \yearof{d} \leq y \leq \firstdeath{p}{q} \implies \livewith{p}{q}{y}
%\]
%%There is a ternary relation $\psi:\Psi$ such that for all $p,q,d$, $\pair{p,q,d} \in \psi$ iff  $p$ and $q$ live together from the date $d$ until one of them dies.
%%\[ 
%%\exists \psi.\, \forall p,q,d.\, \holds{\psi, p, q, d} \iff \forall y{:}\sY.\ \yearof{d} \leq y \leq \firstdeath{p}{q} \implies \livewith{p}{q}{y}
%%\]
%\end{defn}
%\begin{defn} 
%There is a ternary relation $\tcompatdeath:\Psi$ such that for all $p,q,d$, have $\pair{p,q,d} \in \tcompatdeath$ iff  $p$ and $q$ are compatible from the date $d$ until one of them dies.
%\[ 
%\forall p,q,d.\, \holds{\tcompatdeath, p, q, d} \, \iff \, \forall y{:}\sY.\ \yearof{d} \leq y \leq \firstdeath{p}{q} \implies \compatible{p}{q}{y}
%\]
%%$p$ and $q$ are compatible from date $d$ until one of them dies.
%%\[
%%\compatdeath{p}{q}{d} \iff \forall y{:}\sY.\ \yearof{d} \leq y \leq \firstdeath{p}{q} \implies \compatible{p}{q}{y}
%%\]
%\end{defn}

\bigskip

\begin{viqdefn} \label{viqd:livetilldeath}
For all $p,q,d$ there is a proposition $\livetilldeath{p}{q}{d}:\Psi$ that holds iff $p$ and $q$ live together from the date $d$ until one of them dies.
\[ 
\forall p,q,d.\, \holds{\livetilldeath{p}{q}{d}} \, \iff \, \livewith{p}{q}{d, \firstdeath{p}{q}}
\]
\end{viqdefn}
\begin{viqdefn} \label{viqd:compatdeath}
%\begin{defn} 
For all $p,q,d$ there is a proposition $\compatdeath{p}{q}{d}:\Psi$ that holds iff $p$ and $q$ are compatible from the date $d$ until one of them dies.
\[ 
\forall p,q,d.\, \holds{\compatdeath{p}{q}{d}} \, \iff \,  \compatible{p}{q}{d,\firstdeath{p}{q}}
\]
\end{viqdefn}

\begin{viassert} \label{a:marriageinvolves}
If $p$ marries $q$ (on date $d$), then $p$ makes the promise (on date $d$) that they will live together until one of them dies.
\[\does{p}{\getmarried{q}{d}} \implies \does{p}{\promise{\livetilldeath{p}{q}{d}}{d}}\]
\end{viassert}
\begin{assumption} \label{a:nopredict}
Roughly: You can't safely predict that two people will be compatible till death. More precisely: No person $p'$, on any date, can safely predict that two people $p$ and $q$ will be compatible from that date until one of their deaths.
\[ \neg \safelypredict{p'}{\compatdeath{p}{q}{d}}{d} \]
\end{assumption}
\begin{assumption} \label{a:nopromise}
If $p$ cannot (on date $d$) safely predict that $\phi$ will be true, then $p$ should not (on date $d$) promise that $\phi$ will be true.\footnote{There is a sense in which it would be more technically correct to write ``is true'' at both places where I wrote ``will be true'', since the truth value of a sentence does not depend on time, but on the other hand ``will be true'' is consistent with common usage of English, where one can say ``I think $A$ will be true'' in order to convey the meaning ``I think $A$ is true, but we won't know for sure until some point in the future.'' }
\[ \neg \safelypredict{p}{\psi}{d} \implies \shouldnot{p}{\promise{\psi}{d}} \]
\end{assumption}
\begin{assumption} \label{a:compatlive}
Two people who are incompatible during a period cannot live together during that period. %If $p$ and $q$ are not compatible during the period $[d,d']$, then $p$ cannot live with $q$ during $[d,d']$.
\[ \neg \compatible{p}{q}{d,d'} \implies \neg \livewith{p}{q}{d,d'} \]
\end{assumption}
\begin{viassert} \label{a:predictimplies}
If $p$ (on date $d$) can safely predict that $\psi$ will hold, and $\psi$ implies $\psi'$, then $p$ (on date $d$) can safely predict that $\psi'$ will hold.
\[\safelypredict{p}{\psi}{d} \wedge (\holds{\psi} \implies \holds{\psi'}) \implies \safelypredict{p}{\psi'}{d}\]
\end{viassert}
\begin{viassert} \label{a:shouldNotDoDoes}
If $p$ should not do action $a$, and doing $a'$ requires doing $a$, then $p$ should not do $a'$.
\[\shouldnot{p}{a} \wedge \left(\does{p}{a'} \implies \does{p}{a}\right) \implies \shouldnot{p}{a'}\]
\end{viassert}

\bigskip

\newcommand{\persprob}[4]{\optdefn{#1}{\text{Prob}}{\text{Prob}_{#1}^{#3}(#2,#4)}}

\medskip

The axioms prove the following, which is the goal sentence:
\begin{prop} \label{e:marriageconclusion}
 $\shouldnot{p}{\getmarried{q}{d}} $
\end{prop}
\begin{proof}
Let $p,q,d$ be arbitrary. Assertions \ref{a:marriageinvolves} and \ref{a:shouldNotDoDoes} imply
\[ \shouldnot{p}{\promise{\livetilldeath{p}{q}{d}}{d}}  \implies \shouldnot{p}{\getmarried{q}{d}} \]
Hence it suffices to prove 
\begin{equation*} 
\shouldnot{p}{\promise{\livetilldeath{p}{q}{d}}{d}} 
\end{equation*}
An instance of Assumption \ref{a:nopromise} is: 
\[ \neg \safelypredict{p}{\livetilldeath{p}{q}{d}}{d} \implies \shouldnot{p}{\promise{\livetilldeath{p}{q}{d}}{d}} \]
And so it suffices to prove
\begin{equation} \label{e:GOAL}
\neg \safelypredict{p}{\livetilldeath{p}{q}{d}}{d}
\end{equation}
An instance of Assumption \ref{a:nopredict} gives:
\begin{equation} \label{e:notpredictcompat}
\neg \safelypredict{p}{\compatdeath{p}{q}{d}}{d} 
\end{equation}
Using Definitions \ref{viqd:livetilldeath}, \ref{viqd:compatdeath} and Axiom \ref{a:compatlive} we can derive:
\begin{equation} \label{e:holdslemma}
\holds{\livetilldeath{p}{q}{d}} \implies \holds{\compatdeath{p}{q}{d}}
\end{equation}
%From the previous conclusion and Assumption \ref{a:predictimplies}, it suffices to show:
%\begin{equation} 
%\neg \safelypredict{p}{\compatdeath{p}{q}{d}}{d}
%\end{equation}
Finally, from (\ref{e:notpredictcompat}), (\ref{e:holdslemma}) and Assertion \ref{a:predictimplies}, (\ref{e:GOAL}) follows.
\end{proof}

Assume that the issues I mentioned above under ``Formalization notes'' have been dealt with. Provided the audience takes the marriage vows seriously (e.g. imagine they are all devout Catholics), you should be able to fill in the language interpretation guide, starting from the sketch I gave above, in such a way that it would be hard for any audience member to reject any of the axioms except for exactly one of Assumptions \ref{a:nopredict} or \ref{a:compatlive}, and possibly Assumption \ref{a:nopromise}. 

Let's suppose that we accept Assumption \ref{a:nopromise}. So we focus on Assumptions \ref{a:nopredict} and \ref{a:compatlive}:
\[ \text{Assumption \ref{a:nopredict}:\quad} \neg \safelypredict{p}{\compatdeath{p}{q}{d}}{d} \]
\[ \text{Assumption \ref{a:compatlive}:\quad} \neg \compatible{p}{q}{d,d'} \implies \neg \livewith{p}{q}{d,d'} \]
Furthermore, suppose that I raise a semantics criticism against the symbol $\safelypredict{}{}{}$, and that our dialogue rules allow me, as a critic, to suggest an extension of the language and axioms. I suggest the introduction of a predicate symbol for personal probability assessment on a particular date,\footnote{Similar to Bayesian probability, with the intended semantics given in terms of betting games, but without the convention that a probability is assigned to every proposition.} together with the new sort symbol $[0,1]$ for the real interval $[0,1]$\footnote{Or the rationals between $[0,1]$, or even a finite set such as $\{0,.01,.02,\ldots, .98, .99, 1\}$ would suffice.}: 
\[ \persprob{}{}{}{} \ptype \sP \times \sF \times \sD \times [0,1]   \]
The intended semantics is for $\persprob{p}{\phi}{d}{x}$ to mean that on date $d$, person $p$ thinks $\phi$ is or will be true with probability at least $x$.
And I suggest the axiom:\footnote{If you change .8 to a value much closer to 1, then Assumption \ref{a:nopromise} becomes easy to dispute.} 
%\[ \safelypredict{p}{\phi}{d} \implies \exists \alpha{:}[0,1].\ \alpha \geq .8 \wedge \persprob{p}{\phi}{d}{\alpha} \]
\[ \safelypredict{p}{\phi}{d} \implies \persprob{p}{\phi}{d}{.8} \]
With that in place, it would be harder for the author of the proof to equivocate about the meaning of $\safelypredict{}{}{}$. Now, the only symbol whose intended semantics is too vague (whose definition is too incomplete) for us to evaluate Assumptions \ref{a:nopredict} and \ref{a:compatlive} is $\compatible{p}{q}{d,d'}$. And that brings us to the serious flaw in the informal argument argument. For Assumption \ref{a:nopredict} to be true under a given interpretation, the semantic definition of $\compatible{p}{q}{d,d'}$ needs to be fairly strong, but for Assumption \ref{a:compatlive} to be true under a given interpretation, the semantic definition of $\compatible{p}{q}{d,d'}$ needs to be quite weak (meaning its extension is large). For example, I would not be abusing the dialog rules if I rejected Assumption \ref{a:compatlive} for any definition of $\compatible{p}{q}{d,d'}$ that is much stronger than this:
\begin{quote}
$p$ and $q$ are compatible during $[d,d']$ unless one of them poses a physical danger to the other, or one of them makes an effective legal action to remove the other from the household.
\end{quote}
And, for such a weak definition of $\compatible{p}{q}{d,d'}$, I would have no difficulty justifying my rejection of Assumption \ref{a:nopredict}.

\waltonExampleOnly{

%% file: walton_formal_critique.tex
%!TEX root = ../../thesis.tex
\newcommand{\tmurderous}{\text{Murderous}}
\newcommand{\murderous}[3]{\tmurderous_{#1,#2}^{#3}}

\newcommand{\tgetalong}{\text{GetAlongOk}}
\newcommand{\getalong}[3]{\tgetalong_{#1,#2}^{#3}}

%\TODO[Idea: show that the persuassive informal criticism from earlier section is over-simple.]

\indenton

Recall from Section \ref{s:walton} that $\tcompatdeath$ is defined in terms of $\compatible{}{}{}$	and some symbols whose descriptions are clear. The three disputable assumptions were:

\begin{assumption} \label{a:nopredict2}
No person $p'$, on any date, can safely predict that two people $p$ and $q$ will be compatible from that date until one of their deaths.
\[ \neg \safelypredict{p'}{\compatdeath{p}{q}{d}}{d} \]
\end{assumption}
\begin{assumption} \label{a:nopromise2}
If $p$ cannot (on date $d$) safely predict that $\phi$ will be true, then $p$ should not (on date $d$) make a promise that $\phi$ will be true.
% this footnote was printed in previous section
%\footnote{There is a sense in which it would be more technically correct to write ``is true'' at both places where I wrote ``will be true'', since the truth value of a sentence does not depend on time, but on the other hand ``will be true'' is consistent with common usage of English, where one can say ``I think $A$ will be true'' in order to convey the meaning ``I think $A$ is true, but we won't know for sure until some point in the future.'' }
\[ \neg \safelypredict{p}{\psi}{d} \implies \shouldnot{p}{\promise{\psi}{d}} \]
\end{assumption}
\begin{assumption} \label{a:compatlive2}
Two people who are incompatible during a period cannot live together during that period.
\[ \neg \compatible{p}{q}{d,d'} \implies \neg \livewith{p}{q}{d,d'} \]
\end{assumption}

The initial semantic description (language interpretation guide entry) entry for $\tsafelypredict$ (which is empty, so the only hint we had for interpreting the symbol was the name of the symbol) is too vague for me to evaluate Assumption \ref{a:nopromise2}. Specifically, how much confidence must $p$ have in $\psi$'s truth in order to ``safely predict'' $\psi$? We can say the same for Assumption \ref{a:nopredict2} with either symbol $\tcompatible$ or $\tsafelypredict$, or for Assumption \ref{a:compatlive2} with symbol $\tcompatible$, although we should prefer Axioms \ref{a:nopromise2} or \ref{a:compatlive2} since each depends on only one too-vague symbol.\footnote{This can be useful if informal communication fails, since one can more-easily use the author's acceptance of the axiom to deduce constraints on the meaning of the too-vague symbol\index{Deducing constraints on the meaning of a too-vague symbol}. In particular, one can sometimes, for the sake of making a criticism, simplify an axiom by partially evaluating it using the parts of the author's language interpretation guide that \textit{are} sufficiently precise.} But supposing I choose Assumption \ref{a:nopromise2}, then by the definition of criticizing an \viproof (Section \ref{s:VIFP}) there seems to be only one productive thing to do, which is to make the semantics criticism $\pair{\text{Assumption \ref{a:nopromise2}}, \tsafelypredict}$. I will then communicate with the author directly, suggesting they change the semantic description of $\tsafelypredict$ to something like ``If $p$ can safely predict $\psi$ on a given date $d$ (i.e. $\safelypredict{p}{d}{\psi}$), then on that date $p$ has credence at least $X$ that $\psi$ is or will be true,'' where $X$ is some fixed constant.

Of course the author may reject that suggestion, and instead, for example, add prose to the semantic description of $\tsafelypredict$ that, being still too vague, does not actually help me interpret $\tsafelypredict$ well enough to evaluate Assumption \ref{a:nopromise2}. In that case, I will introduce some new symbols, which are under my control, along with an axiom $A$, also under my control, that uses the new and old symbols and expresses the above Bayesian interpretation of $\tsafelypredict$. The author can then accept (unlikely, given the previous failure using informal communication), weakly reject, or strongly reject $A$. This formalizes our disagreement about the meaning of $\tsafelypredict$, and documents it for later readers of the argument. Suppose the author accepts my suggestion, say for $X = .9$. Then I can accept Assumption \ref{a:nopromise2}. 

The author's semantic description of $\tcompatible$ is still too vague for me to evaluate the other two axioms. Once again, according to the definition of criticizing an \viproof, it seems the only productive thing for me to do is make the semantics criticism $\pair{\text{Assumption \ref{a:compatlive2}}, \tcompatible}$ or $\pair{\text{Assumption \ref{a:nopredict2}}, \tcompatible}$. 

%
%\TODO[Note about literal definition of $\tlivewith$, and how you're supposed to be ungenerous in evaluating axioms.]
%
%\TODO[Note about unsoundness of Assumption \ref{a:compatlive2}. Suggest changing it by using till-death versions]

As concluded in Section \ref{s:walton}, if the author clarifies the semantics of $\tcompatible$ and it is very weak (its extension is large), then I can make a \textit{subjective criticism} of Assumption \ref{a:nopredict2} and \textit{strongly reject} it\footnote{Meaning all of my personal interpretations of the language falsify the axiom. Note that only requires that all of my intended interpretations have at least one tuple $\pair{p',p,q,d}$ for which the formula is false.}, and if the author clarifies the semantics of $\tcompatible$ and it is (at least) moderately-strong, then I can make a subjective criticism of Assumption \ref{a:compatlive2} and strongly reject it. Finally, if the author makes his intended semantics for $\tcompatible$ somewhere between ``very weak'' and ``moderately strong'', then I can reject {\it both} of Assumptions \ref{a:nopredict2} and \ref{a:compatlive2}.
Any of those three scenarios would be good places to end the dialogue. 

%\[
%\begin{array}{rr}
%\text{Very weak $\tcompatible
%
%\end{array}
%\]
%

\newcommand{\veryweakcompat}{\text{CompatVeryWeak}}
\newcommand{\modstrongcompat}{\text{CompatModeratelyStrong}}

More likely (in this scenario with such an uncooperative author), the author would see the vulnerability, and avoid clarifying the semantics of $\tcompatible$ enough that I can make a semantics criticism. In that case, I would formalize the idea of the previous paragraph in the following way. I introduce predicates
\[
\begin{array}{rcl}
\tgetalong &\ptype& \sP \times \sP \times \sD \times \sD \tto \sB \\
\tmurderous &\ptype& \sP \times \sP \times \sD \times \sD \tto \sB 
\end{array}
\]
My language interpretation guide entry for $\getalong{p}{q}{d_1,d_2}$ says that $p$ and $q$ get along OK during the period $[d_1,d_2]$, and the entry for $\murderous{p}{q}{d_1,d_2}$ says that $p$ and $q$ will try to kill each other if they come into contact during $[d_1,d_2]$. Additionally, I introduce two defined 0-ary predicate symbols in order to give names to two sentences:
\[
\begin{array}{rcl}
\veryweakcompat &\ceq& \forall p,q,d_1,d_2.\ \neg \murderous{p}{q}{d_1,d_2} \implies \compatible{p}{q}{d_1,d_2} \\
\modstrongcompat &\ceq& \forall p,q,d_1,d_2.\ \compatible{p}{q}{d_1,d_2} \implies \getalong{p}{q}{d_1,d_2} 
\end{array}
\]

Finally, I introduce the following axioms (which I accept), which formally describe my above stated positions on the author's two remaining controversial assumptions for a range of possible sharpenings of $\tcompatible$ (since the author has not made his intended semantics precise). These axioms imply that I reject at least one of those two assumptions:
%\[ \forall p,q,d_1,d_2.\ \murderous{p}{q}{d_1}{d_2} \implies \neg \getalong{p}{q}{d_1}{d_2} \]
\[ \neg \veryweakcompat \vee \neg \modstrongcompat \] 
\[ \veryweakcompat \implies \left( \text{Assump \ref{a:compatlive2}} \wedge \neg(\text{Assump \ref{a:nopredict2}}) \right)\]
\[ \modstrongcompat \implies \left( \text{Assump \ref{a:nopredict2}} \wedge \neg(\text{Assump \ref{a:compatlive2}}) \right)\]
\[ \left(\neg \modstrongcompat \wedge \neg \veryweakcompat \right)   \implies \left( \neg(\text{Assump \ref{a:nopredict2}}) \wedge \neg(\text{Assump \ref{a:compatlive2}}) \right)\]
%
%\Blue{TODO?: Picture indicating vulnerability of argument}

\subsection{\Finished{Literal, ungenerous interpretation of (non-simplifying) assumptions}}
There is another problematic axiom in argument, Assumption \ref{a:compatlive2}, that is easily fixable and which according to the directions for criticizing an \viproof, {\it should} be criticized even if the critic knows it is fixable. One simple acceptable way for the author to respond is by changing the label of the axiom from Assumption to Simplifying Assumption.

Instances of Assumption \ref{a:compatlive2} for which $d$ and $d'$ are close should be weakly rejected. Even for a weak definition of compatible (but not quite as weak as $\veryweakcompat$), there would exist two people who are to that extent strongly incompatible and yet manage to live together for a few days. The author should address the criticism, and here are two quick ways of doing so according to the rules:
\begin{PPE}
\item Make Assumption \ref{a:compatlive2} a Simplifying Assumption, and amend the natural language text associated with it to describe the sense in which it is a simplifying assumption.
\item Add a hypothesis such as $d + 365 \leq d'$ to Assumption \ref{a:compatlive2} (introducing sort $\NN$ and symbols $+ : \sD \times \NN \to \sD$ and $365 : \NN$ \footnote{365 is larger than necessary, but it is not important.}), yielding:
\[ \left[ d + 365 \leq d' \ \wedge \ \neg \compatible{p}{q}{d,d'} \right] \ \implies \ \neg \livewith{p}{q}{d,d'} \] 
\end{PPE}
With option 2, we can also use the new symbols to formalize an assumption that has the same purpose as the informal constraint on the sort $\sP$ for people that says $\sP$ includes only people who are not near death (given in its semantic description in Section \ref{s:walton}), as follows:
\begin{simplifyingassumption} \label{a:noearlydeathaftermarriage}
When two people get married, they both live for at least a year after.
\[\does{p}{\getmarried{q}{d}} \implies d + 365 \leq \firstdeath{p}{q} \]
\end{simplifyingassumption}

\hlinesep

Now, what was the point of this pedantry? Essentially, it is the application of a safety principle. The method of formal deduction and criticism with \viproofs that I advocate may not be robust unless a critic can insist on having technical problems fixed without having to justify why it is important to do so. Otherwise, disputes about meaningful matters will sometimes devolve into unending arguments about argumentation itself. In Leibniz's words:

%\TODO[This quote might work better in the ``Which Logic'' section]
%\begin{aquote}{Gottfried Leibniz, 1679, {\it ``On the General Characteristic''}\cite{leibniz1976}}
%{\it 
%For men can be debased by all other gifts; only right reason can be nothing but wholesome. But reason will be right beyond all doubt only when it is everywhere as clear and certain as only arithmetic has been until now. Then there will be an end to that burdensome raising of objections by which one person now usually plagues another and which turns so many away from the desire to reason. When one person argues, namely, his opponent, instead of examining his argument, answers generally, thus, ``How do you know that your reason is any truer than mine? What criterion of truth have you?'' And if the first person persists in his argument, his hearers lack the patience to examine it. For usually many other problems have to be investigated first, and this would be the work of several weeks, following the laws of thought accepted until now. And so after much agitation, the emotions usually win out instead of reason, and we end the controversy by cutting the Gordian knot rather than untying it. 
%}
%\end{aquote}
\begin{aquote}{Gottfried Leibniz, 1679, {\it ``On the General Characteristic''}\cite{leibniz1976}}
{\it 
...Then there will be an end to that burdensome raising of objections by which one person now usually plagues another and which turns so many away from the desire to reason. When one person argues, namely, his opponent, instead of examining his argument, answers generally, thus, ``How do you know that your reason is any truer than mine? What criterion of truth have you?''
}
\end{aquote}

%\ldots about how formal deduction is not robust unless we insist 
%Essentially, it is a guard against a certain against the lack of robustness in 

%% file: defeasible_reasoning_chapter.tex
%!TEX root = ../thesis.tex
The purpose of this chapter is twofold. First, to share the high-level ideas of some reusable formalization patterns that I have used in the course of writing examples. Second, as an extension of Section \ref{s:choiceoflogic} that addresses, by example, objections against the foundations of this project along the lines of deduction being inappropriate for real-world reasoning. The following quote from \cite{ImportanceOfNonmonotonicity} is an example of such an objection. I have inserted numbers (n) for the purpose of commenting following the quote.
\begin{quote}
\indent {\it Standard propositional and predicate logics are monotonic. That is, if a proposition is
logically implied by some set of propositions, then it is also implied by every superset
of the initial set. (0) Another way of describing monotonicity is to say that once
something is determined to be true, it remains true. (1) No additional information can
cause conclusions to be modified or withdrawn. (2) There is no way to presume
something to be the case until there is information to the contrary. (3) There are no rules
of thumb, or general rules, which allow conclusions to be drawn which may be faulty,
but are nonetheless better than indecision. (4) Classical logic offers no theory about when
to prefer one belief to another in general, and provides no language for stating which
beliefs to prefer given that certain things are known in a particular case.}

\indent {\it (5) The subject matter of classical logic is truth, not decision making. The central concern of logic is logical consequence: which propositions are necessarily true given that
other propositions are true. (6) Monotonic logic is very useful when we want to know
what must be the case if something else is known to be true. It is less useful when we
know very little about some domain with certainty, or can discover the facts only by
extending resources, if at all. (7) Monotonic logic alone provides us with an infinite
number of conditional statements of the form "this would be true if that is true",
which is of little help in making decisions when we are unable to establish with
certainty the truth or falsity of the alternative premises.}\cite{ImportanceOfNonmonotonicity}
\end{quote}

Already at (0) we have a classic indicator of problems to come: reference to unqualified, non-relative truth is often not meaningful in formal logic, and definitely is not meaningful for classical FOL, which is defined in terms of truth-with-respect-to-structures. This ambiguous use of ``truth'' leads to all sorts of confusion and equivocation, and should be banished whenever one is debating the merits of one logic over another.

Points (1)-(4) are a straw man argument. The author implicitly conflates the {\it use of classical logic via first-order theories} with the {\it definition of classical FOL}; the former is the author's desired target, and the latter is the straw man. 
%The author implicitly equates the {\it use} of classical logic with the {\it definition} of classical logic; the former is the author's desired target, and the latter is the straw man. 
The fact is that none of these complaints about the literal, technical definition of classical predicate logic apply to {\it first-order theories}, which are what the author should really be attacking.

Points (5)-(7) are an innocent, understandable, and common oversimplification that I believe leads to an unfortunate  misconception about the scope of classical predicate logic, or even formal logic in general, especially among people unfamiliar with formal logic. (5): It would be more accurate to say that the subject matter of classical predicate logic is \textit{semantics}, with relative-truth being an important special case. Perhaps this is easier seen in the formulations of MSFOL that treat the set of truth values as just another sort, so that the boolean connectives are just very common function symbols. (6): It is true that when there is a great deal of uncertainty in a domain, those boolean function symbols are used a little less, with function symbols for Bayesian reasoning having a larger role, but that is hardly a criticism of classical logic. (7): In fact conditionals remain just as essential in domains with a lot of uncertainty. They are our main tool for excluding from consideration the structures that we are not interested in reasoning about, and they are just as useful when those structures contain e.g. Bayesian distributions, interpretations of defeasible legal statutes, etc.\label{p:conditionalunderattack}
%A typical example is a lemma $\Gam \vdash A \implies B$ where $A$ expresses that a Bayesian distribution satisfies certain constraints, and $B$ is a logical consequence of the constraints (given  background I hope the examples in this thesis will illustrate that.

\section{\Finished{Argument from expert opinion}} \label{s:ExpertOpinion}
\include*{expert_opinion}

\section{\Finished{Bayesian reasoning}} \label{s:Bayesian}
\include*{bayesian_reasoning}

\section{\Finished{Theory comparison}}
The comparison of theories (or models, explanations of evidence) is a broad category of defeasible reasoning in the physical and social sciences, as well as in criminal law. The Leighton Hay argument (Chapter \ref{c:AIDWYC}) and the smoking-causes-cancer argument (Chapter \ref{c:smoking}) are in this category. Both have the following form:
\begin{PPE}
\item[(a)] Specification of one's desired consequence of a comparison being sufficiently-favourable to one's preferred model. For example, that some scientific theory should be abandoned, that some public policy should be put into place, or that some person accused of a crime should be declared guilty.
\item[(b)] Formal definition of when one theory is better than another (or ``much better'' as in the smoking-causes-cancer argument; the strength needed depends on how much force part (a) requires of the comparison), together with a proposition that part (d) suffices to justify part (a).
\item[(c)] Formalization of the two or more competing theories.
\item[(d)]	Deductive proof that one theory is better (or ``much better'') than the others, according to the definition.
%\item Proposition that part 4 suffices to justifies part 1.
\end{PPE}
%Depending on how exactly In this form, part 3 and part 2 are main defeasible part. since any problem  
(c) is defeasible in that the formalization of a theory might be unfaithful to the intent of the proponents of the theory, a kind of straw man argument. (b) may be unfair (bias), untrustworthy (variance/bias), or otherwise inappropriate. One type of potential flaw, which is warned about especially often in discussion of the limitations of Bayesian reasoning, is that (i) the definition is a reasonable one based on how well the theories explain or predict the available evidence, but not all evidence relevant to the decision (a) is included. In contrast, the concern that has received the most technical attention in statistics (both the Bayesian and frequentist schools) is whether (ii) the definition is reasonable \textit{for the available evidence}. Finally, even when all the significantly-relevant evidence is considered and the comparison relation is reasonable, it may be that (iii) the comparison is too weak to justify (a). 
A critic could try to show any of those three types of flaws in the smoking-causes-cancer or Leighton Hay examples, though I believe types (i) and (iii) would be the most fruitful for them. See Sections \ref{s:refinements} and \ref{s:criticismOfLHArg} for some specific criticisms of the smoking-causes-cancer and Leighton Hay arguments, respectively.

%My final remarks concern the flexibility of FOL/interpreted formal proofs in comparison to the tools of statistics. This is an observaton that is technically no more substantial than saying that analysis is more flexible than calculus, but it has had some unexpected consequences for me while working on the major examples in this thesis. 

% could be better but meh
\section{\Finished{Costs/benefits analysis}}
There is a common belief in the social sciences that mathematical/logical methods necessarily oversimplify social issues, and because of this they are inappropriate for reasoning about such issues. The first part is true, although it is equally true of natural language argumentation. My main concern here is to argue that the second part is not supported by the first, provided results are reported in a disciplined way.

The Sue Rodriguez argument (Chapter \ref{c:SR}) and the physician-assisted suicide argument (Chapter \ref{c:assistedsuicide}) can both be construed as costs/benefits analyses. In both examples, only a subset of the apriori-relevant factors/concerns are considered.

In the Sue Rodriguez argument this is very explicit, since all the goal sentence says is that a certain set of 4 concerns do not justify ruling against her. There remains the possibility that I excluded some concern that is highly relevant, in fact so relevant that with its inclusion there is a strong argument to be made for the negation of the goal sentence \textit{modified to include the 5th concern}. It is tempting to consider my original 4-concern argument invalidated by the hypothetical new 5-concern argument, however that is not technically correct if one treats the arguments in the same way as mathematics arguments; it is not necessarily inconsistent/unreasonable to accept both arguments. Here is an analogy from math: We are interested in whether a majority of elements of a finite set $S$ have a property $P$. On the way, we prove the answer is no if we replace $S$ with a certain subset $S_1 \subset S$, then later we prove the answer is yes if we replace $S$ with a certain $S_2$ such that $S_1 \subset S_2 \subset S$, and then no again for a certain $S_3$ such that $S_1 \subset S_2 \subset S_3 \subset S$. There is no inconsistency, and moreover we are slowly getting closer to the truth about $S$. 

The assisted suicide argument is structured a little differently at its top level, in that its goal sentence is not relative to the simplification (the simplification being that only a subset of the apriori-relevant factors are considered). The argument's Assumption 1 that 
\quad Main Lemma\footnote{You do not need to lookup what this is for the purpose of this discussion} $\implies \lb\text{should pass}\rb$\footnote{Which is a 0-ary predicate symbol that means legislation should be passed that introduces an assisted suicide system which is consistent with the constraints given in the argument.}\quad
implicitly includes the simplifying assumption that it is only necessary to consider the direct effects of the assisted suicide system on individual people, and not, for example, on the culture of Canadian society or on groups of people. If a critic goes on to argue the negation of the goal sentence, by including a strictly larger set of factors, but without assuming anything in conflict with the assumptions of my argument about individual people, then they would indeed need to reject Assumption 1 in order to be consistent/reasonable.

It would be very easy to modify the assisted suicide argument so that its goal sentence has the same relative form as the Sue Rodriguez goal sentence\footnote{Basically just delete Assumption 1 and make the Main Lemma be the goal sentence}, and vice-versa. Moreover, in any case, the goal sentence does not even give the meaning of the proof, which is properly given by $\bigwedge_{A \in \text{Axioms}} A \implies \text{goal sentence}$. Nonetheless, I think the difference in the two forms of goal sentences is important, if only because of the tendency for proofs to be reported in terms of their main conclusion, with the axioms left tacit. That practice needs to be actively discouraged to answer the oversimplification concern. Fortunately, all that requires is familiarity with FOL and reading the proof document itself. The same cannot be said for defeasible logics, or logics that offer very simple syntax to express complex semantics (see Section \ref{s:choiceoflogic}).

%With both of those examples, I included as many factors/concerns as my time would allow. There remains the possibility that I excluded some factor that is highly relevant, in fact so relevant that with its inclusion, there is a strong argument to be made for the negation of the goal sentence (modified to include the 5th concern).

\section{\Finished{Counterfactual reasoning, hypothetical scenarios}}
There is a great deal of literature giving general mathematical theories/systems (sometimes called ``logics'') for modeling counterfactual reasoning, i.e. reasoning about what would have happened had something been different. Despite the abstract and sometimes-unreal nature of such hypothetical scenarios, we can still reason together deductively, as we are often able to describe such scenarios in such a way that our individual understandings are similar enough that the differences have no significant effect on the argument. 

The two arguments about the Berkeley gender bias case (Chapter \ref{c:berkeley}) are examples of deductively-formalized counterfactual reasoning. They have (or can easily be put into) the form:
\begin{PPI}
\item[(i)] Formalize constraints that are (supposedly) sound with respect to an informal theory/explanation/model that one wishes to attack.
\item[(ii)] Formalize constraints that one believes should be required to hold regardless of the theory.  
\item[(iii)] Deductively show that the constraints together are inconsistent.
\end{PPI}
It is defeasible in that one's type (i) constraints may misrepresent the informal theory, and in that one's type (ii) constraints may be rejected by opponents.
For example, both of the Berkeley arguments make the simplifying (and technically probably wrong) assumption that in the particular pools of applicants to each department, the males and females are equally qualified. That is a type (ii) constraint. The remainder of this section will examine those two arguments in more detail in relation to the itemized form above.

In the first argument, I require a definition of gender prejudice that is weaker than financial forces, in the sense that if all means of discrimination had been removed, then there would not have been a large change in the \textit{total} number of applicants accepted to each department. That is a type (ii) constraint. There is a type (i) constraint that the observed gender biases were caused by prejudice, so that when all means of discrimination are removed, each department should accept men and women at approximately the same rate. Finally, there is the similar type (i) constraint that if all means of discrimination had been removed, it should be possible (under the other constraints) that the overall\footnote{Meaning across all departments.} acceptance rate for women relative to men improves. Part (iii) consists of proving the negation of that constraint, i.e. that the overall acceptance rate for women gets worse for any admissions round that satisfies the other constraints.

The second argument can alternatively be construed as a theory comparison argument, but I'll describe it in the above form.
%The main type (ii) assumption is that it should be possible to have a pool of applicants (satisfying the equally-qualified assumption) of the same size as the real one, in which men and women apply to each department at close to the same rate, and to have an admissions round in which the gender-specific acceptance rates are close to the same as what they actually were. (so if there was gender discrimination, then there should still be gender discrimination). However, we can derive  
The main type (i) constraint is effectively that the test used to infer gender discrimination --whether there is a significant bias in favor of men in the overall acceptance rate-- should not be sensitive to whether or not the genders apply to different departments at different rates. The main type (ii) assumption is that we can assess that sensitivity by considering an arbitrary pool of applicants of the same size in which men and women apply to each department at close to the same rate, and an arbitrary admissions round in which the gender-specific acceptance rates are close to the same as what they actually were (so if there was gender discrimination in reality, then there should still be gender discrimination in the hypothetical case). In more detail, the main type (i) constraint says that the numeric constraints on the hypothetical admission round should not be enough to guarantee that the test's answer changes from ``gender discrimination'' to ``no gender discrimination''; part (iii) consists of proving the negation of that constraint.

%can be a change in the ratio of men to women accepted in each department, but there should not be a .    If there is gender discrimination, then bias should change in favour of women when the means of discrimination are removed.   

\section{\Finished{Multiplicity of reasons}}
The most defeasible of defeasible argument types has this form: some $n$ \textit{reasons} $R_1, \ldots, R_n$ are given for a proposition $G$. It plays a central role in the Carneades system (see Section \ref{s:carneades}), for example. Except for the qualitatively-different $n=1$ case\footnote{i.e. the general use of conditionals, discussed on page \ref{p:conditionalunderattack}.} it does not appear in any of the examples in this thesis, and nor should it; the opinion I advocate in this thesis is that, for the problems in the intended problem domain (Section \ref{s:problemdomain}), the argument form is inappropriate for anything but fast and speculative reasoning.
\medskip

\noindent It is tempting to try to deductively formalize such an argument as 
\begin{PPI}
\item[] Simplifying Assumption: $R_1 \wedge \ldots \wedge R_n \implies G$ \footnote{Recall from the beginning of this chapter that implication is interpreted classically, so that to accept this simplifying assumption (Section \ref{s:VIFP}) means nothing more or less than that you are willing to exclude from your set of personal $\cL$-interpretations any interpretations that satisfy all the $R_i$ but falsify $A$}
\item[] Axiom 1: $R_1$ 
\item[] \ldots
\item[] Axiom $n$: $R_n$
\end{PPI}
However, that formalization is not faithful to the intended meaning of the defeasible argument. It is too fragile. In the framework of deduction, an effective criticism against any one of those $n+1$ axioms is as good as an effective criticism against them all. In the framework of defeasible reasoning, in contrast, a criticism that effectively argues against only one of the $R_i$ is regarded as only \textit{weakening} the criticized argument. For fast and speculative reasoning, that is a good thing. But when there is adequate time available, the problem is serious, and it is hard to make progress, that principle of defeasible argumentation puts too great a burden on the critic. The burden should be on the argument's author to formalize the sense in which the acceptability of each $R_i$ contributes to the acceptability of $G$.\footnote{Of course, a defeasible logic may provide some sophisticated schemas for formalizing that kind of relationship, but those can just as well be made into reusable first-order theories.}

Nonetheless, the pattern from the previous paragraph \textit{can be useful} when the axioms are properly interpreted deductively. For example, suppose that the Crown prosecutors of the Canadian government publish a high-level interpreted formal proof that some person is guilty of a murder. A good such argument will employ Bayesian reasoning in some places, but at the top level it could have the above propositional structure. Let's say the goal sentence is a 0-ary predicate symbol $G$ whose language interpretation guide says something along the lines of ``the suspect is guilty''. The prosecution has evidence linking the murder weapon to the suspect, eye-witness evidence identifying the suspect at the scene of the crime, and DNA evidence of the suspect's blood collected at the scene of the crime. We'll make $R_1,R_2,R_3$ be 0-ary predicate symbols. The prosecution's language interpretation guide entries for them are:
\begin{PPI}
\item	$R_1$: The murder weapon belonged to the suspect.
\item $R_2$: The eye-witness correctly saw the suspect fleeing the scene of the crime.
\item $R_3$: Blood belonging to the suspect was found 10 feet from the victim.
\end{PPI}
Each $R_i$ is a lemma proved from other assumptions, and they are connected to the goal sentence by the simplifying assumption $R_1 \wedge R_2 \wedge R_3 \implies G$. This is interesting, surprisingly, when we consider what it means for the prosecution to put forward such a simplifying assumption, and for the defense to accept it. From the prosecution's perspective, it is useful since it simplifies their task to giving arguments for each of the $R_i$ independently, but it is also risky since it introduces fragility to their argument -- the defense only needs to argue against the weakest of the $R_i$. It is worthwhile for the prosecution if they strongly believe in $R_1,R_2,R_3$, and have no other strong inculpatory evidence . From the defense's perspective, the simplifying assumption is useful since it allows them to focus on refuting the weakest of the $R_i$, but it is also a concession, since it \textit{is} possible that $R_1,R_2,R_3$ are true and $G$ is false\footnote{For example, it may be that the actual murderer was an associate of the suspect who had access to the suspect's gun, both were at the scene of the crime (but the murderer was no seen by the witness), and the suspect was injured by the murderer while trying to defend the victim (hence the suspect's blood).}. It is worthwhile for the defense to accept the simplifying assumption if they think they can give a strong argument that at least one of the $R_i$ is false, and they have no strong exculpatory evidence that isn't related to the $R_i$.

%% file: expert_opinion.tex
%!TEX root = ../thesis.tex
When I explain my work to intelligent people outside of mathematical disciplines --students of law, politics, philosophy; incidentally the people who take the greatest interest in it-- the most difficult task is explaining the (lack of) practical effects of the limitation to deductive reasoning. 

Consider the following excerpt from a recent paper \cite{WaltonFindingTheLogic} by Walton, which advocates the use of defeasible logic:
\begin{quote}
{\it The most widely useful argumentation schemes that fit arguments in everyday conversational 
argumentation are defeasible ones }[citation omitted]{\it. A good example is 
argument from expert opinion. This scheme is not well modelled by a deductive interpretation. 
Basing it on an absolutely universal generalization, to the effect that what an expert says is 
always true, does not yield a useful logical model. Indeed such a deductive model would make 
the scheme into a fallacious form of argument by making it unalterably rigid. In practice, 
evaluating an argument from expert opinion is best carried out by seeing how well it survives the 
testing procedure of critical questioning }[citation omitted].
\end{quote}
That quote is an unusually respectable one of its kind, due to Walton's explicitly saying ``everyday conversational argumentation'' and ``Basing it on an absolutely universal generalization''. But it still resembles a kind of straw man argument that is often used in the motivation for defeasible logic:
%\TODO[Edit first sentence, since he is respectable enough to say ``Basing it on an absolutely universal generalization'', making the straw man somewhat explicit, which many other authors don't.] \\
%Ironically, Walton has employed a straw man argument here (but on the other hand, he does say that he is talking about ``everyday conversational argumentation'', which is certainly not my interest). 
implicitly the proponent suggests that in a deductive logic framework, to formalize an argument from expert opinion, or an instance of inductive or abductive reasoning, or an instance of default reasoning, the only option is to assert a general rule that has obvious fallacious instances. A slightly more dignified criticism of deductive logic makes the implicit suggestion that in a deductive framework, the only option is to formulate a concise and elegant schema for some type of defeasible argument, which suffices to justify all and only the ``good'' instances of that argument type (the problem being that no such schema exists). I say that it is slightly more dignified because it is sometimes an innocent instance of academics' often-very-productive instinct to generalize, and to obsess about elegance and wide applicability.  

In contrast, for arguing about contentious and important issues when plenty of time is available, I think it is a good idea to insist on the use of deductive logic, particularly for the sake of obtaining {\it locally-refutable proofs} (Section \ref{s:criticizingVIproofs} page \pageref{p:locally-refutable}). Of course, that ``restriction'' does not preclude the use of defeasible reasoning --which is easy to do in deductive logic, provided you don't insist on elegant, widely-applicable schema-- but rather just makes it stand out, often as the weakest part of an argument. 
I claim, moreover, that any really solid use of a defeasible reasoning pattern, such as appeal to expertise, can be formulated, with enough effort and perhaps a little creativity, as a really solid deductive argument. Two examples of special classes of deductively-strong appeals to expertise follow.

\begin{example}
A man, John Doe, is on trial for vehicular homicide. A forensics expert testifies that from his examination of the skid marks on the road and the tires of Doe's car after the accident, he is ``certain'' (subtext: as certain as he ever is on a judgement like this), that Doe's car was traveling at least 75mph just before the point where the skid marks begin. The proposition that the prosecuting attorney wants to use is a formalization of ``John Doe's car was traveling at least 75mph just before it began to skid''. In the initial version of the attorney's informally-interpreted proof, she uses a 0-ary predicate symbol $X$ to represent that statement (i.e. the assertion that the statement is true), and another 0-ary predicate symbol $Y$ for a statement that quotes the full record of the testimony of the expert witness and asserts that it is in fact what the expert said. She then includes the axioms $Y$ and $Y \implies X$.% in the axiom set $\Gsu$. 

Why am I calling this a deductively strong appeal to expertise? It is because the prosecuting attorney, with the help of her expert witness, is quite prepared to replace those two axioms with a longer proof from a larger set of axioms, each of which is much more trustworthy than $Y \implies X$. Those axioms include axioms about measurements taken by the investigators, which can be checked against crime scene photos and evidence collected at the scene, as well as the physics-based assumptions that the expert uses to derive a lower bound on the speed of the car, e.g. from the length of the skid marks, upper bounds on the force of friction of the tires against the pavement (as a function of the distance along the skid), weight of the car, wind resistance, etc. 
\end{example}
%\TODO[Todo: example of the nature of the best kind of appeal to expertise: expert opinion about inferring bounds on the speed of a car deduced form skid marks on the road. how this is easily formalized deductively as a pair of axioms $A$ and $A \implies B$, where the author is prepared to replace $A \implies B$ with a complex argument giving the expert's reasoning.]
%A forensics expert testifies at a trial, where a man, John Doe, is accused of vehicular homicide, that from the skid marks on the road, and the state of the tires, he is as certain as he ever is on a judgement like this, that the man's car was traveling between 75 and 85mph just before the point where the skid marks begin.  The proposition that the prosecuting attorney wants to use is a formalization of ``John Doe's car was traveling between 75 and 85 mph just before he applied his breaks''. Initially, we use a propositional variable $X$ to represent that statement (i.e. the assertion that the statement is true), and another propositional variable $Y$ for a statement that quotes the full record of the testimony of the expert witness, and asserts that it is in fact what the expert said.

\begin{example}
Frustrated with accusations of bias and lack of rigor, the climate change experts involved in the Fifth Assessment Report of the UN Intergovernmental Panel on Climate Change take steps to clarify the meaning of their highest-certainty 10-year extrapolations. From those extrapolations\footnote{By ``extrapolations'', I have in mind conditional statements, possibly with many premises, e.g. ``IF at least $N$ tons of oil are burned in the next 10 years AND no major geoengineering project is initiated, then $\ldots$''}, they formalize a family of increasingly-weak assertions $A_1,\ldots,A_k$, in enough detail that independent third parties, after 10 years, can check whether the assertions hold. Together with their government's politicians, they then sign a contract, which stipulates a family of increasingly-lucrative sets of legal entitlements $S_1,\ldots,S_k$ for oil-producing nations (e.g. which permit high levels of pollution), such that $S_i$ is awarded if $A_i$ turns out false. In the aftermath of the IPCC's move, for those highest-certainty extrapolations, the accusations of bias and lack of rigor fall off.  

This example demonstrates a distinction between two kinds of appeal to expertise. As in the previous example, to formalize their argument, I could start by using $X$ for the extrapolation of the expert, i.e. one of the assertions $A_i$. And again we will have axioms $Y$ and $Y \implies X$, for some additional 0-ary predicate symbol $Y$ that says something along the lines that the IPCC experts said $A_i$ will almost certainly turn out true. The legal entitlements $S_i$ justify our strengthening of $g(Y)$ from some elaboration of ``The IPCC experts say that $A_i$ will almost certainly turn out true'', to some elaboration of ``The IPCC experts say, {\it and clearly believe}, that $A_i$ will almost certainly turn out true''. 

At first reading, the difference between the two versions of $g(Y)$ may seem too informal to be meaningful. And indeed, with or without the entitlements, it would be acceptable to make a semantics criticism (page \pageref{d:reviewproof}) about the second version of $g(Y)$. What makes the difference between the two versions of $g(Y)$ meaningful is that, without the entitlements, I do not see the author of the argument being able to adequately formalize ``clearly believe'' (they will get stuck after a sequence of rigor criticisms and other dialog moves, made with the intention of forcing them to clarify what they mean), whereas with the entitlements it is a simple matter, since the IPCC scientists will also be arguing elsewhere that, {\it assuming} $A_i$, a policy should be put into place that would conflict with the legal entitlements $S_i$. In other words, essentially all that needs to be claimed is that the governments who employ the experts have a strong desire to avoid the granting of the entitlements $S_i$. 
\end{example}

%% file: bayesian_reasoning.tex
%!TEX root = ../thesis.tex
%\subsection{The difficulty with Bayesian reasoning for this project}
This section is about how to criticize arguments that use Bayesian reasoning. I use the phrase ``subjective probability assumptions'' to refer to the informal class of assumptions that includes priors, bounds on conditional probabilities, and choices of parametric models. 

Bayesian reasoning and statistics feature prominently in several of my major examples. The literature on the problem of interpreting Bayesian subjective probability assumptions, and the (sometimes insubstantial) Bayesian vs Frequentist debate, is vast (see \cite{BayesianPublicPolicy} and \cite{BayesiansFrequentistsScientists} for refreshingly concrete and pragmatic perspectives).
%What I plan to do with this section is explain how I formalize subjective probabilities, and compare my approach to others mentioned in the literature. 
Without surveying all the motivations and interest in the interpretation problem, let us make more precise why it is a problem for this project that cannot be easily dismissed. 

There are subjective assumptions that seem normal and obviously necessary, such as some of the assumptions I make in my arguments about assisted suicide, and generally the kind of assumptions one must make in order to derive anything with nontrivial ethical ramifications. And then there are subjective probabilities, which make most of us at least a little uneasy. If you say that your prior for a murder suspect's guilt is $x$, what recourse do I have if I think that prior is unreasonable? And how do I make precise what it is that I am disagreeing with? Interpretations involving betting ratios/dutch books\cite{sep-epistemology-bayesian} help with the latter problem, but not obviously with the former. 

I claim that the origin of this uneasiness is the same as that caused by using explicit real-valued utility functions to reason about people's subjective values. I handle those cases in the same way. The issue in both cases is with the mixture of qualitative and quantitative subjective assumptions.

To be more concrete, let's look at an example of a typical Bayesian probability assumption, from a famous case in England in which a criminal defense team was allowed to have a statistics expert present a Bayesian analysis to a jury - one of only a few times this has ever been allowed in a jury case\cite{DonnellyRvAdams}\cite{StatisticsInTheLaw}. Most of the details of the case are not important for our purposes; a woman was assaulted by a man, the assailant left DNA evidence, and years later a man was prosecuted, and ultimately convicted, after a ``cold hit''\footnote{Meaning the man's DNA was run through the DNA database before he was a suspect.} on London's DNA database made him the suspect. Please note that we won't be focusing on the DNA aspect of the case, which is the most interesting and contentious part (see \cite{ColdHit}, \cite{DonnellyDNA}; later on I will release an interpreted formal proof about this case). The only reason I mention the DNA aspect is for point 1 on page \pageref{ThoughtExperimentObservations}. The probability assumption we are focusing on is the instantiation (or bounding, as I would prefer) of two of the defense team's fundamental parameters involved in estimating the likelihood ratio:
\[
\frac{\text{Pr(\textsf{evidence} \ | \ suspect guilty)}}{\text{Pr(\textsf{evidence} \ | \ suspect innocent\footnotemark)}}
\] \footnotetext{Here ``innocent'' means {\it factually innocent}, as opposed to {\it legally not guilty}.}
Namely these two:
\[
c_1 = \text{Pr(victim failed to identify suspect in police lineup \ | \ suspect guilty)}
\]
\[
c_2 = \text{Pr(victim failed to identify suspect in police lineup \ | \ suspect innocent)}
\]
Now suppose I make assumptions that constrain the parameters of the argument, including those two, enough to imply that the likelihood ratio is large, suggesting --contrary to my opponent's intuition, let's assume-- that the suspect is guilty. One contributor to that numeric result is that I make an assumption that implies $c_2/c_1$ is upperbounded by a particular number close to 1, so that the failure to identify the suspect cannot contribute much to making the likelihood ratio large. My opponent will want to reject that assumption, believing that the ratio should be much larger than 1.

Moreover, suppose my opponent believes that my assumption about $c_2/c_1$ is {\it unreasonable}. Since these are probabilities representing my credence about whether certain unrepeatable events happened in the past, it is hard for my opponent to argue that I am being unreasonable in an objective sense (this is the crux of the Bayesian interpretation problem). Some Bayesians will answer that my assumption is {\it not} unreasonable provided it is consistent with my other probability assumptions and standard probability axioms. 
%\cite{subjectivebayesian}. 
Ultimately I agree with that position, and I offer no magic solution that would enable us to {\it resolve} situations of fundamental uncertainty. However, in the context of this project, there is something more I can do to at least {\it state} my position in a clear way.%; that is, to formalize the sense in which you think I'm being unreasonable.  
%but not with the often-implicit conclusion that there is little more to be said or done about the problem\cite{bayesiandefeatist}. 

Before demonstrating my recommendation, I'll briefly remind the reader how the clarification of a position is usually done in Bayesian reasoning. One introduces additional random variables representing evidence or environmental factors, with their own (usually brief) informal interpretations. Then one introduces new independence and conditional probability assumptions to derive a bound on the new likelihood ratio (new because now the meaning of \textsf{evidence} has changed). This is indeed the right thing to do when one can precisely describe the domains of the new random variables, and when one has some reason to be confident in the new probability assumptions. But those conditions are often hard to meet, and when they are not met, the additions may introduce more noise and obfuscation into the argument than they do clarification. Consider, for example, trying to formalize a model of the victim forgetting the look of the assailant over time (over a year passed between the crime and the lineup), which is complicated by the dependence on both the look of the (unknown!) assailant and the victim's general ability to recall faces.

My recommendation involves adding much more detail to the informal semantics, only adding detail to the formal mathematical model once agreement has been established for the informal semantics. Suppose I formulate my assumption about the ratio $c_2 / c_1$ as follows. I make two assumptions:
\begin{PPE}
\item[A1:] We can model all our knowledge that is significantly relevant to $c_2 / c_1$ with the following thought experiment, which consists of two scenarios. First, I specify large, precise sets of white men $M$ and women $W$. In scenario 1, random members $m$ of $M$ and $w$ of $W$ appear in the exact locations of the assailant and victim in the real crime, under similar lighting and weather conditions. Time resumes and a crime may or may not happen, depending on $m$ and $w$. Any instances of the thought experiment in which events do not transpire in a way sufficiently similar to how they did with the actual crime are ignored. $w$ is shown a police lineup containing $m$ (after a time delay equal to the one in reality), where the other men in the lineup are selected from $M - m$ in a fixed manner (which I would explain in detail) that is similar to how they were selected in reality (roughly based on looking similar to $m$). Then $c_1$ is the probability that $w$ fails to identify $m$. Scenario 2 is the same except that two distinct random members $m,m'$ of $M$ are sampled, with $m'$ representing an innocent suspect, and $w$ is shown a lineup containing $m'$ instead of $m$. Then $c_2$ is the probability that $w$ fails to identify $m'$.
\item[A2:] The second assumption says that $c_2 / c_1$ is upperbounded by a particular number close to 1.
%
%\TODO[Use language of a \U{thought experiment} without language of having to set it up. Use only the runs of the experiment where the outcome turns out the same as in the present case. Initial condition is that a random man and random woman from some specific sets are in the same locations as the true assailant and victim were in, during the same time of night. There is a second random man from the same set as well, who represents an innocent suspect. We imagine the victim being shown two lineups independently, for the two men, where the lineups were constructed in a way similiar to how they were in the present case.]
%\item \ldots
\end{PPE}
We cannot perform such an experiment, since doing so would be unethical and impractical. Although that verbose pair of assumptions does not get us significantly closer to an objective {\it test}, it does get us significantly closer to an objective {\it description}. The motivation is to be more precise, and a few observations about the thought experiment will convey that:
\begin{PPE} 
\item \label{ThoughtExperimentObservations}The innocent suspect and the assailant are sampled almost-independently of each other (except for the constraint that they are distinct). This is actually a significant modeling assumption! Because, even if we assume the suspect Adams is innocent, we know his DNA test profile is the same as the assailant's -- and it is {\it plausible} that such genetic similarity makes two men significantly more likely to look significantly similar to each other.\footnote{We already know the genetic similarity makes them at least a little more likely to look similar to each other because. For example, siblings are much more likely to have identical DNA test profiles, and two men with identical DNA profiles have the same race\cite{truthmachine}.} 
\item Assuming $W$ and $M$ are large and varied, I do not use any of the (mostly noisy) knowledge that we have of the victim and suspect. This amounts to an implicit assumption that such knowledge is irrelevant.
\item I have fixed a method of constructing police lineups. In my opponent's favour, I have made no assumption that the lineup was flawed in a way that would make the victim less likely to select a guilty suspect and/or more likely to select an innocent suspect.
\item We cannot test my second assumption A2, but A2 {\it does} have the virtue of being a relatively-objective assumption (relative to how precisely I specify the thought experiment). I claim this is valuable.
\end{PPE}

If the thought experiment is actually carried out, then this reasoning strategy is essentially {\it Empirical Bayes}\cite{BayesiansFrequentistsScientists}\footnote{See reference. In short, this is the approach of using loosely-related data to set or constrain priors and other parameters, under the assumption that the data is not chosen in a biased way.}. 
Note the structural/qualitative character of assumption A1. It amounts to my modeling assumptions. In contrast A2 has a numeric character. This fits my general recommendation: split a subjective probability assumption into two parts: (A1) a subjective, qualitative part, and (A2) an objective, quantitative part. 
Doing that separation in a way that makes A1 acceptable to both sides of an argument often requires the description of elaborate (but sufficiently precise) thought experiments that will never be carried out, so I should reiterate that this is not a strategy for directly resolving disputes. 

So what is gained? We can now move forward with our disagreement on $c_2 / c_1$, either by coming to agreement on an A1-type assumption, or by failing to do so. 
\textit{If we fail to agree on A1, then the quantitative part of the probability assumption was likely obscuring the fundamental source of disagreement more than anything it was contributing. If we agree on A1, then we are left with two (inconsistent) versions of A2-type assumptions, which are as uncertain as the original probability assumption, but are more objective and precise, and thus easier for other people to judge for themselves.}

%\TODO[Get different example? This might be complicated by the fact that Adams is likely to look like the assailant even if he's innocent. NO!! THAT MAKES IT MORE INTERESTING]

%What recourse do you have?   suggesting that you aren't satisfied with the 

%\blockquote{
%Bayesian arguments often present situations where you are confident that you must reject some assumption, but without 
%\end{blockquote}

%% file: berkeley_gender_bias_example_MSFOL.tex
%!TEX root = ../../thesis.tex

% NOTE: this version has some extra changes that berkeley_gender_bias_example.tex doesn't have, besides to move to \QQu. For example, I removed quotes for the LIG entries.

\newcommand{\QQu}{\QQ^{\bot}}

\newcommandx{\App}[2][1={},2={}]{
\ifthenelse{\isempty{#1} \or \isempty{#2}}
                    {\textsf{App}^{#1#2}}
                    {\textsf{App}^{#1,#2}}  }
\newcommand{\Apps}{\textsf{App}}
\newcommandx{\Acc}[2][1={},2={}]{
\ifthenelse{\isempty{#1} \or \isempty{#2}}
                    {\textsf{Acc}^{#1#2}}
                    {\textsf{Acc}^{#1,#2}}  }
\newcommandx{\AccH}[2][1={},2={}]{
\ifthenelse{\isempty{#1} \or \isempty{#2}}
                    {\textsf{Acc}_{\textsf{hyp}}^{#1#2}}
                    {\textsf{Acc}_{\textsf{hyp}}^{#1,#2}}  }
\newcommand{\AppliedTo}{\textsf{AppliedTo}}
\newcommand{\AppM}{\App[m]}
\newcommand{\AppF}{\App[f]}
\newcommand{\AS}{\mathcal{A}}
\newcommand{\gendirrel}{\pair{\textsf{gender uncor with ability in each dept}}}
\newcommand{\dismiss}{\pair{\textsf{lawsuit should be dismissed}}}
\newcommand{\biasonly}{\pair{\textsf{bias only evidence}}}
\newcommand{\discrim}{\pair{\textsf{gender discrimination likely occured}}}

\newcommand{\accrate}[1]{\textsf{accRate}^{#1}}
\newcommand{\accrateh}[1]{\textsf{accRate}^{#1}_{\text{hyp}}}

The following table summarizes UC Berkeley's Fall 1973 admissions data for its six largest departments. Across all six departments, the acceptance rates for men and women are about 44.5\% and 30.4\% respectively. The large observed bias prompted a lawsuit against the university, alleging gender discrimination.\footnote{The data given is apparently the only data that has been made public. The lawsuit was based on the data from all 101 graduate departments, which showed a pattern similar to what the data from the 6 largest shows.} In \cite{SexBiasBerkeley} it was argued that the observed bias was actually due to a tendency of women to disproportionately apply to departments that have high rejection rates for both sexes.

\[ 
\begin{array}{cccc}
 &  \text{Male}  & \text{Female} & \text{Total} \\
   \begin{array}{c}
    \text{Department} \\
     D_1  \\
     D_2  \\
     D_3  \\
     D_4  \\
     D_5  \\
     D_6  
   \end{array}
  & 
  % 1198 men accepted, 2691 applied, 44.5% acceptance rate
   \begin{array}{cc}
     \text{Applied} & \text{Accepted} \\
	 825 & 512 \ (62\%) \\
	560 & 353 \ (63\%)  \\
	325 & 120 \ (37\%)  \\
	417 & 138 \ (33\%)  \\
	191 & \ 53 \ (28\%)  \\
	373 & 22 \ (6\%)  
   \end{array}
  & 
  % 557 women accepted, 1835 applied, 30.4% acceptance rate
   \begin{array}{cc}
     \text{Applied} & \text{Accepted} \\
     108 & \ 89 \ (82\%)  \\
	25 & \ 17 \ (68\%) \  \\
	593 & 202 \ (34\%)  \\
	375 & 131 \ (35\%)  \\
	393 & \ 94 \ (24\%)   \\
	341 & \ 24 \ (7\%)  
   \end{array} 
     & 

   \begin{array}{cc}
     \text{Applied} & \text{Accepted} \\
     933 & \ 601 \ (64\%)  \\
	 585 & \ 370 \ (63\%) \  \\
	918 & 322 \ (35\%)  \\
	792 & 269 \ (34\%)  \\
	584 & \ 147 \ (25\%)   \\
	714 & \ 46 \ (6\%)  
  \end{array}
\end{array}
\]

The first argument I give is similar to the final analysis given in \cite{SexBiasBerkeley},\footnote{The paper is written to convey the subtlety of the statistical phenomenon involved (an instance of ``Simpson's Paradox''), and so considers several poor choices of statistical analyses before arriving at the final one.} though it makes weaker assumptions (Assumption \ref{a:intungendsimilar} in particular: their corresponding, implicit assumption is obtained by replacing the parameters .037 and 9 with 0s). The argument resolves the apparent paradox by assuming a sufficiently-precise definition of ``gender discrimination'' and reasoning from there. More specifically, it first fixes a definition of ``gender discrimination\quco  and then defines (in natural language) a hypothetical admissions protocol that prevents gender discrimination by design. Considering then a hypothetical round-of-admissions scenario that has the same set of applications as in the actual round of admissions, if we assume that the ungendered departmental acceptance rates are not much different in the hypothetical scenario, then it can be shown that the overall bias is actually {\it worse} for women in the hypothetical scenario. Since the hypothetical scenario has no gender discrimination by design, and is otherwise as similar as possible to the real scenario, we conclude that the observed bias cannot be blamed on gender discrimination.

The second argument tells us why it is that our vagueness about ``gender discrimination'' resulted in an apparent paradox; namely, we were implicitly admitting definitions of ``gender discrimination'' that allow for the question of the presence/absence of discrimination to depend on whether or not the sexes apply to different departments at different rates. If we forbid such definitions, then to prove that  the gendered departmental acceptance rates {\it do not} constitute gender discrimination, it should suffice to show that there is an overall bias {\it in favour} of women in any hypothetical admissions round in which the gendered departmental acceptance rates are close to what they actually were, and where men and women apply to each department at close to the same rate.

I'll use $g$ to refer to the {\it language interpretation guide} for the language $\la$ of this argument.\\
$\la \sdiff \lama$ consists of:
\begin{PPI}
\item The constant $\AccH$.
\item The propositional variables (i.e. $0$-ary predicate symbols) $\biasonly$,\\ $\dismiss$, $\gendirrel$.
\end{PPI}
$\lama$ consists of:
\begin{PPI}
\item A number of mathematical symbols that have their standard meaning: constants $0, 1, 512, 825, \ldots$, function symbols $|\cdot|, \cap, \cup, +, -, *, /$, predicate symbols $<, =$.
\item The constants $\Apps$, $\Acc$, $\AppM$, $\AppF$, $\App[1], \ldots, \App[6]$. Since the elements of these sets are not in the universe, their semantics are determined by axioms that assert their sizes and the sizes of sets formed by intersecting and unioning them with each other. 
\item The sorts are $\AS$ for application sets and $\QQu$ for the rational numbers with an element for ``undefined''. See below for $g$'s entries for them. 
\end{PPI}
The types of the function/predicate symbols other than the 0-ary predicate symbols (and besides $=$, which is untyped) are as follows. With respect to the definition of {\it interpreted formal proof} from Section \ref{s:VIFP}, they are all {\it assumptions} as opposed to {\it simplifying assumptions}. 
\setcounter{counterone}{\value{footnote}}
\addtocounter{footnote}{1}
\setcounter{countertwo}{\value{footnote}}
\[\begin{array}{rcl}
\begin{array}{l}\Apps, \Acc, \AccH, \AppM, \AppF, \\ \App[1], \ldots, \App[6] \end{array} & : & \AS \\
|\cdot| &: & \AS \to \QQu \\
0, 1, 512, 825, \ldots & : & \QQu \\
\cap, \cup &: & \AS \times \AS \to \AS \\
+,-,* & : & \QQu \times \QQu \to \QQu \\
%/ &: &  \QQu \times \QQu \pto \QQu \footnotemark[\value{counterone}] \\
/ &: &  \QQu \times \QQu \to \QQu  \\
< & : & \QQu \times \QQu \to \BB \footnote{$\BB$ is the type for booleans; technically it is not a sort, so its elements are not in the universe of discourse.}
%< & : & \QQu \times \QQu \to \BB \footnotemark[\value{countertwo}]
\end{array}\]
%\footnotetext[\value{counterone}]{$\pto$ denotes the type of a partial function. The version of many-sorted FOL I use has build-in (AKA ``first-class'') partial functions, based on \cite{farmer-simple}.}
%\footnotetext[\value{countertwo}]{$\BB$ is the type for booleans; technically it is not a sort, so its elements are not in the universe of discourse.}

\section{\Finished{First argument}}

The {\it goal sentence} is the following implication involving propositional variables whose informal meanings, given by the language interpretation guide $g$, will be given next. 
\[ \gendirrel \wedge \biasonly \implies \dismiss \]

$g(\biasonly)$ consists of the above table, and then the assertion: ``The bias shown in the data is the only evidence put forward by the group who accused Berkeley of gender discrimination.''
\ms

$g(\gendirrel)$ we take to be just ``Assumption 1'' from \cite{SexBiasBerkeley}, which I quote here:
\begin{quote}{\it Assumption 1 is that in any given discipline male and female applicants do not differ in respect of their intelligence, skill, qualifications, promise, or other attribute deemed legitimately pertinent to their acceptance as students. It is precisely this assumption that makes the study of "sex bias" meaningful, for if we did not hold it any differences in acceptance of applicants by sex could be attributed to differences in their qualifications, promise as scholars, and so on. Theoretically one could test the assumption, for example, by examining presumably unbiased estimators of academic qualification such as Graduate Record Examination scores, undergraduate grade point averages, and so on. There are, however, enormous practical difficulties in this. We therefore predicate our discussion on the validity of assumption 1.} \cite{SexBiasBerkeley}
\end{quote}
\ms

$g(\dismiss) = $ The judge hearing the suit against Berkeley should dismiss the suit on grounds of lack of evidence.
\ms 

$g(\QQu) = $ The rational numbers plus one extra object for error/undefined.
\ms

$g(\AS) = $ The powerset of $\Apps$. Note that the individual applications are not in the universe of discourse (though each singleton set is), since they are not required for the proof.
\ms

$g$ also says that 
\begin{PPI} 
\item $0,1,512,$ etc are the expected numerals.
\item $|\cdot|$ is the function that gives the size of each set in $\AS$. 
\item $\cap, \cup$ are the expected binary functions on $\AS$.
\item $+, -, *$ are the expected binary functions on the naturals extended so that they equals $\bot$ when either or both of the arguments are $\bot$.
\item $/$ is division on the rationals extended so that it equals $\bot$ iff one or both of the arguments are $\bot$ or the second argument is $0$.
\item $<$ is the usual ordering on the rationals extended by making $\bot$ be neither greater than nor less than any number or itself.
\end{PPI}

Recall that the next 11 symbols are all $0$-ary constant symbols.
\ms \\
$g(\Apps) = $ $\Apps$ is the set of applications. Its size is 4526 (sum of the entries in the two ``Applied'' columns of the table).
\ms \\
$g(\Acc) = $ $\Acc$ is the set of (actual) accepted applications. Its size is 1755 (sum of the entries in the two ``Accepted'' columns of the table).
\ms \\
$g(\AccH) = $ We need a sufficiently-precise, context-specific definition of ``gender discrimination\quco and to get it we imagine a hypothetical scenario. An alternative admissions process is used, which starts with exactly the same set of applications $\Apps$, and then involves an elaborate\footnote{It need not be efficient/economical, since we are only introducing the scenario as a reasoning device.}, manual process of masking the gender on each of them (including any publications and other supporting materials). The application reviewers, while reading the applications and making their decisions, are locked in a room together without access to outside information, except that interviews are done over computer using an instant messaging client (which, of course, is monitored to make sure the gender of the applicant remains ambiguous). Then, $\AccH$ is the set of accepted applications in the hypothetical scenario.
\ms \\
$g(\AppM) = $ $\AppM$ is a subset of $\Apps$ of size 2691 (sum of the first ``Applied'' column in the table), specifically the applications where the applicant is male.
\ms \\
$g(\AppF) = $ $\AppF$ is a subset of $\Apps$ of size 1835 (sum of the second ``Applied'' column in the table), specifically the applications where the applicant is female.
\ms \\
For $d = 1,\ldots,6$:\\
$g(\App[d]) = $ $\App[d]$ is the set of applications for admission into department $d$.\\
\ms 

\begin{defn} \label{d:appsymbdefs}
For $g \in \{m,f\}$ and $d \in \{1,\ldots,6\}$:
\[ \App \ceq \App[m] \dunion \App[f] \]
\[ \App[d][g] \ceq \App[d] \cap \App[g] \]
\[ \Acc[d][g] \ceq \App[d][g] \cap \Acc \]
\[ \AccH[d][g] \ceq \App[d][g] \cap \AccH \]
\end{defn}

\begin{defn} \label{d:plusminusdef}
For $x,y,z \in \QQu$, we write $z \in [x \pm y]$ for $x - y \leq z \leq x+y$.
\end{defn}

\begin{assumption} In the hypothetical scenario, the number of applicants of gender $g$ accepted to department $d$ is as close as possible to what we'd expect assuming that gender is uncorrelated with ability within the set of applicants to department $d$. For $d \in \{1,\ldots,6\}$ and $g \in \{m,f\}$:
\[\hspace{-3in} \gendirrel \implies \]
\[ | \AccH[d][g] | \in \left[ |\AccH[d]| \cdot \frac{|\App[d][g]|}{|\App[d]|} \pm \nicefrac{1}{2} \right]\]
\[ \frac{|\AccH[d][g]|}{|\AccH[d]|} = \frac{|\App[d][g]|}{|\App[d]|} \]
\end{assumption}

\begin{assumption} \label{a:intungendsimilar}
Assuming that gender is uncorrelated with ability within the set of applicants to department $d$,
the number of applicants accepted to department $d$ in the hypothetical scenario is close to the number accepted in the real scenario. That is, the overall, non-gendered departmental acceptance rates do not change much when we switch to gender-blind reviews. We require that a model satisfies at least one of the following two quantifications of that idea. For $d \in \{1,\ldots,6\}$:
\[ 
\hspace{-3in} \gendirrel \implies 
\]
\[
\begin{array}{rl} 
& \biggl( \bigwedge_{1 \leq d \leq 6} |\Acc[d] | \cdot (1-.037) \leq |\AccH[d]| \leq |\Acc[d] | \cdot (1+.037)  \biggr) \\
\vee & \biggl( \bigwedge_{1 \leq d \leq 6} |\AccH[d]| \in \left[ |\Acc[d] | \pm 9 \right] \biggr)
\end{array}
\]
The constants $.037$ and $9$ are roughly the most extreme values that make the proof go through.
To illustrate the first form, the bounds for the departments with the fewest and greatest number of accepted applicants are:
\[ 45 \leq |\AccH[6]| \leq 47 \quad \text{   and  } \quad 579 \leq |\AccH[1]| \leq 623 \]
\end{assumption}

\begin{defn} For $g \in \{m,f\}$:
\[ \accrate{g} \ceq \Acc[g]/\App[g] \quad \text{and} \quad 
 \accrateh{g} \ceq \AccH[g]/\App[g] \]
\end{defn}

\begin{assumption} \label{a:finalaxiom_b} \hspace{0pt} 
If $\biasonly$ and 
\[ 
 \frac{ \accrateh{m}}{ \accrateh{f}} >
   \frac{\accrate{m}}{\accrate{f}} 
\]
then $\dismiss$
\end{assumption}

\begin{simplifyingassumption} \label{sa:biasonly}
$\biasonly$
\end{simplifyingassumption}

\begin{claim} \label{c:berkeleyarg1mainclaim}
\[ \gendirrel \quad  \implies \quad  \frac{\accrateh{m}}{\accrateh{f}} >  \frac{\accrate{m}}{\accrate{f}} \]
\end{claim}
\begin{proof}
It is not hard to formulate this as a linear integer programming problem, where the variables are the sizes of the sets $\AccH[d][g]$. Coming up with inequalities that express the previous axioms and the data axioms from Section \ref{s:dataaxioms} is easy. Reduce the Claim itself to a linear inequality, and then negate it. One can then proof using any decent integer programming solver that the resulting system of equations is unsatisfiable.
\end{proof}

\begin{claim} The goal sentence easily follows from the previous three propositions.
\[ \gendirrel \wedge \biasonly \implies \dismiss \]
\end{claim}

\section{\Finished{Second argument}}\label{ss:berkeley}
\newcommandx{\AppH}[2][1={},2={}]{
\ifthenelse{\isempty{#1} \or \isempty{#2}}
                    {\textsf{App}_{\textsf{hyp}}^{#1#2}}
                    {\textsf{App}_{\textsf{hyp}}^{#1,#2}}  }
\indenton
This second argument better captures the intuition of the usual informal resolution of the apparent paradox; the observed bias is completely explained by the fact that women favored highly-competitive departments (meaning, with higher rejection rates) more so than men. We show that there is an overall bias {\it in favour} of women in any hypothetical admissions round in which the gendered departmental acceptance rates are close to what they actually were, and where men and women apply to each department at close to the same rate.

In this argument, the set of applications in the hypothetical scenario can be different from those in the real scenario, so we introduce the new symbols $\AppH[d] : \AS$ for $1 \leq d \leq 6$.

The hypothetical admissions round is similar to the true admissions round (Axioms \ref{a:numberappssame} and \ref{a:genderedaccratessimilar}) except that men and women apply to each department at close to the same rate  (Assumption \ref{a:genderedappratessimilar}) - meaning the fraction of male applications that go to department $d$ is close to the fraction of female applications that go to department $d$. We need to update the language interpretation guide entries $g(\AppH[d])$ and $g(\AccH)$ to reflect these alternate assumptions.

This proof uses Definitions \ref{d:appsymbdefs} and \ref{d:plusminusdef} from the previous proof.

\begin{assumption} \label{a:numberappssame}
 In the hypothetical round of admissions, the total number of applications to department $d$ is the same as in the actual round of admissions. Likewise for the total number of applications from men and women.\footnote{This axiom could be weakened in principle, by replacing the equations with bounds, but doing so in the obvious way introduces nonlinear constraints, and then I would need to use a different constraint solver.} \\
 For $d \in \{1,\ldots,6\}$ and $g \in \{m,f\}$: 
 \[|\AppH[d]| = |\App[d]|, \quad |\AppH[g]| = |\App[g]|\]
\end{assumption}

\begin{assumption} \label{a:genderedappratessimilar}In the hypothetical scenario, gendered departmental {\it application} rates are close to gender-independent. For $d \in \{1,\ldots,6\}$ and $g \in \{m,f\}$: 
\[ |\AppH[d][g]| \in \left[ |\AppH[g]|\cdot \frac{|\AppH[d]|}{|\AppH|}   \pm 6 \right] \]
\end{assumption}

\begin{assumption} \label{a:genderedaccratessimilar} In the hypothetical scenario, gendered departmental {\it acceptance} rates are close to the same as in the real scenario.\\
For $d \in \{1,\ldots,6\}$ and $g \in \{m,f\}$: 
\[ |\AccH[d][g]| \in \left[ \frac{|\Acc[d][g]|}{|\App[d][g]|} \cdot |\AppH[d][g]| \pm 6 \right] \]
\end{assumption}

\begin{claim} \label{c:genderarg2mainclaim}
$\accrateh{f} > \accrateh{m}$
\end{claim}
\begin{proof}
As in the previous proof, it is easy to reduce this to a linear integer programming problem. Coming up with constraints that express the previous axioms and the data axioms from the next section is easy. Then, add the constraint 
\[ \left( \sum_{1 \leq d \leq 6} |\AccH[d][f]| \right) / |\App[f]| \leq   \left( \sum_{1 \leq d \leq 6} |\AccH[d][m]| \right) / |\App[m]| \]
which expresses the negation of the Claim (recall that $|\App[m]|$ and $|\App[f]|$ are constants). Finally, prove that the resulting system of equations is unsatisfiable.
\end{proof}

\begin{assumption} \label{a:finalaxiom_c} \hspace{0pt} 
If $\biasonly$ and $\accrateh{f} > \accrateh{m}$ \\ then $\dismiss$
\end{assumption}

Simplifying Assumption \ref{sa:biasonly} from the previous proof, which just asserts $\biasonly$, is also used here. From it, Assumption \ref{a:finalaxiom_c}, and Claim \ref{c:genderarg2mainclaim}, the goal sentence \\ $\dismiss$ follows immediately.

\section{\Finished{Data Axioms}} \label{s:dataaxioms}
\begin{assumption} \label{a:tableMainAxiom} 
\[ |\Apps| = 4526, \ \  \bigwedge_{1 \leq d \leq 6} \App[d] \subeq \App, \ \ \Acc \subeq \App, \ \  \AccH \subeq \App \]
\[ |\App[1][m] | = 825, \ \ | \Acc[1][m] | = 512 , \ \ |\App[1][f] | = 108, \ \ | \Acc[1][f] | = 89  \]
\[ |\App[2][m] | = 560, \ \ | \Acc[2][m] | = 353 , \ \ |\App[2][f] | = 25 , \ \ |\Acc[2][f] | = 17  \]
\[ |\App[3][m] | = 325, \ \ | \Acc[3][m] | = 120 , \ \ |\App[3][f] | = 593 , \ \ | \Acc[3][f] | = 202  \]
\[ |\App[4][m] | = 417, \ \ | \Acc[4][m] | = 138 , \ \ |\App[4][f] | = 375 , \ \ | \Acc[4][f] | = 131  \]
\[ |\App[5][m] | = 191, \ \ | \Acc[5][m] | = 53 , \ \ |\App[5][f] | = 393 , \ \ | \Acc[5][f] | = 94  \]
\[ |\App[6][m] | = 373, \ \ | \Acc[6][m] | = 22 , \ \ |\App[6][f] | = 341 , \ \ | \Acc[6][f] | = 24  \]
That $\Apps$ is the disjoint union of $\App[1],\ldots,\App[6]$ follows from the previous sentences (under the standard interpretation of numbers and sets).
\end{assumption}

%% file: LeightonHayExampleFeb2014.tex
\providecommand{\evidenceneutral}{\pair{\text{the newspaper hair evidence is neutral or exculpatory}}}
\newcommand{\From}[2]{\text{From}_{#1}(#2)}
\newcommand{\FromRV}[1]{\text{From}_{#1}}

\newcommand{\Clipped}[1]{\text{Clipped}_{#1}}

\newcommand{\widthRV}{\text{width}}
\newcommand{\binof}[1]{\text{bin}(#1)}
\newcommand{\InNewsRV}{X_{\text{news}}}
\newcommand{\NewsHairs}{\text{NewsHairs}}

\newcommand{\ProbSymb}[1]{\underset{#1}{\text{P}}}
\newcommandx{\Prob}[2][1={}]{\ProbSymb{#1}(#2)}

\newcommand{\NewsEvid}{\text{Clippings}}
\newcommand{\BeardScalpTrim}{\text{Beard\&ScalpTrim}}
\newcommand{\BeardTrim}{\text{BeardTrim}}

\newcommand{\mubeard}{\mu_{\text{b}}}
\newcommand{\muscalp}{\mu_{\text{s}}}
\newcommand{\sigbeard}{\sigma_{\text{b}}}
\newcommand{\sigscalp}{\sigma_{\text{s}}}

\newcommand{\binsbeard}{f_{\text{b}}}
\newcommand{\binsscalp}{f_{\text{s}}}

\newcommand{\FromBeard}[1]{\text{Beard}(#1)}
\newcommand{\FromScalp}[1]{\text{Scalp}(#1)}
\newcommand{\InNews}[1]{\text{News}(#1)}
\newcommandx{\width}[1][1={}]{\ifthenelse{\isempty{#1}}{\text{width}}{\text{width}(#1)}}
\newcommand{\BParamsRV}{\textsf{BParams}}
\newcommand{\MixRV}{\textsf{Mix}}
\newcommand{\GRV}{\textsf{G}}
\newcommand{\NGRV}{\overline{\textsf{G}}}
\newcommand{\WidthsRV}{\textsf{Widths}}
\newcommand{\HRV}{\textsf{H}}
\newcommand{\ClippedRV}{\textsf{Clipped}}
\newcommand{\vb}{\vec{b}}
\newcommand{\vp}{\vec{p}}
\newcommand{\PBeard}{\text{P}_{\text{beard}}}
\newcommand{\PBeardR}[1]{R_{#1}}
\renewcommandx{\Pr}[2][2={}]{\ifthenelse{\isempty{#2}}{\text{P}(#1)}{\underset{#2}{\text{P}}(#1)}}
\newcommand{\amin}{\alpha_{\text{min}}}
\newcommand{\amax}{\alpha_{\text{max}}}
\newcommand{\bin}[1]{\text{bin}_#1}
\newcommand{\bd}{\ensuremath{\text{B}}}
\newcommand{\densp}[1]{d_{\BParamsRV}(#1)}
\newcommand{\densa}[1]{d_{\MixRV}(#1)}
\newcommand{\denspa}[1]{d_{\pair{\BParamsRV,\MixRV}}(#1)}
\newcommand{\interv}[2]{I_{#1,#2}}
\newcommand{\region}[2]{R_{#1,#2}}

\newcommand{\true}{\text{true}}
\newcommand{\false}{\text{false}}

\newcommand{\likeratio}{\text{{\it likelihood-ratio}}}

\newcommand{\Bins}{\text{Bins}}
\newcommand{\RandVar}[1]{\text{RandVar}(#1)}
\newcommand{\ProbTerm}{\text{ProbTerms}}
\newcommand{\DiscrProbTerm}{\text{DiscrProbTerms}}
\newcommand{\HairFragments}{\text{HairFragments}}
\newcommand{\Interval}{\text{Interval}_{[0,1]}}
\newcommand{\CubicRegion}{\text{CubicRegion}_{[0,1]^3}}
\newcommand{\Bool}{\BB}
\newcommand{\Pow}[1]{2^{#1}}

\newcommand{\nquad}{\!\!\!\!}

{\bf Note: There is complete version of the proof in this chapter now, finished post-graduation, which this document has not been updated with. Contact dustin.wehr@gmail.com for details.}

Leighton Hay is one of two men convicted of murdering a man in an Ontario nightclub in 2002. The other man, Gary Eunich, is certainly guilty, but evidence against Hay is weak-- much weaker, in my opinion and in the opinion of the Association in Defense of the Wrongly Accused (AIDWYC)\footnote{Thanks to Joanne McLean and Deryck Ramcharitar for making the case files available to me.}, than should have been necessary to convict. A good, short summary about the case can be found here:\\
\href{http://www.theglobeandmail.com/news/national/defence-prosecution-split-on-need-for-forensic-hair-testing/article1367543/}{\texttt{http://www.theglobeandmail.com/news/national/defence-prosecution-split-on}\\
\texttt{-need-for-forensic-hair-testing/article1367543/}}

The prosecution's case relies strongly on the testimony of one witness, Leisa Maillard, who picked (a 2 year old picture of) Hay out of a photo lineup of 12 black men of similar age, and said she was 80\% sure that he was the shooter. There were a number of other witnesses, none of whom identified Hay as one of the killers. Ms. Maillard's testimony is weak in a number of ways (e.g. she failed to identify him in a lineup a week after the shooting, and at two trials when she picked out Gary Eunich instead), but here we will be concerned with only one of them: she described the unknown killer as having 2-inch ``picky dreads,'' whereas Hay had short-trimmed hair when he was arrested the morning after the murder. Thus, the police introduced the theory that Hay cut his hair during the night,
%\footnote{Indeed, in the dark; police staked out Hay's house all night, and reported that the lights were off the entire time.} 
between the murder and his arrest the following morning. In support of the theory, they offered as evidence a balled-up newspaper containing hair clippings that was found at the top of the garbage in the bathroom used by Hay. Their theory, in more detail, is that the known killer, Gary Eunich, cut Hay's hair and beard during the night between the murder and the arrests, using the newspaper to catch the discarded hair, then emptied most of the discarded hair into the toilet; and crucially, a hundred-or-so short hair clippings remained stuck to the newspaper (Due perhaps to being lighter than the dreads? It was not explained why.). It is the origin of those hair clippings that we are concerned with in this argument; Hay has always said that the clippings were from a recent beard-only trim. If that is so, then the newspaper clippings are not at all inculpatory, and knowing this could very well have changed the jury's verdict, since the clippings --as hard as this is to believe-- were the main corroborating evidence in support of Ms. Maillard's eye witness testimony.

Both sides, defense and prosecution, agree that the newspaper clippings belong to Hay, and that either they originated from his beard and scalp (prosecution's theory), or just his beard (defense's theory). We will try to prove, from reasonable assumptions, that it is more likely that the hair clippings were the product of a beard-only trim than it is that they were the product of a beard and scalp trim.  

On 8 Nov 2013 the Supreme Court of Canada granted Hay a new trial in a unanimous decision, based on the new expert analysis of the hair clippings that we use in this argument. \sout{We do not yet know whether the Crown will choose to prosecute Hay again, or if they do, whether they will attempt to again use the hair clippings as evidence against him.} On 28 Nov 2014, the Crown dropped its murder charges against Hay, declining to prosecute him again, and he was freed. As usual in these cases, there was no pronouncement of innocence, and Hay and his lawyers will have to fight for monetary compensation.

\begin{table} \label{t:news}
    \begin{tabular}{|l|l|l|} 
    \hline
    Name in proof & Max width (micrometers) & Count \\ \hline
    $\bin{1}$        & 0 to 112.5                  & 10 \\ \hline
    $\bin{2}$        & 112.5 to 137.5              & 20 \\ \hline
    $\bin{3}$        & 137.5 to 162.5              & 40 \\ \hline
    $\bin{4}$        & 162.5 to 187.5              & 19 \\ \hline
    \end{tabular}
    \caption{Measurements of 89 hairs found in a balled-up newspaper at the top of Hay's bathroom garbage. Forensic experts on both sides agreed that the hairs in $\bin{3}$ and $\bin{4}$ are very likely beard hairs, and that the hairs in $\bin{1}$ and $\bin{2}$ could be either beard or scalp hairs.}
\end{table}

\begin{table} \label{t:scalp}
   \begin{tabular}{|l|l|}
    \hline
    Max width (micrometers) & Count \\ \hline
    12.5 to 37.5                & 3     \\ \hline
    37.5 to 62.5                & 28    \\ \hline
    62.5 to 87.5                & 41    \\ \hline
    87.5 to 112.5               & 17    \\ \hline
    112.5 to 137.5              & 1     \\ \hline
    \end{tabular}
    \caption{Measurements of Hay's scalp hairs obtained at the request of AIDWYC in 2010. Note that the first 4 bins are contained in $\bin{1}$ from Table 1. Samples of Hay's beard hairs were not taken and measured in 2010 because the forensic hair experts advised that beard hairs get thicker as a man ages.}
\end{table}

\section{\Finished{High-level argument}}
In 2002, the prosecution introduced the theory that Hay was the second gunman and must have had his dreads cut off and hair trimmed short during the night following the murder. It is clear that they did this to maintain the credibility of their main witness. In 2012, after the new forensic tests ordered by AIDWYC proved that at least most of the hairs found in Hay's bathroom were (very likely) beard hairs, the prosecution changed their theory to accommodate, now hypothesizing that the hairs came from the combination of beard and scalp trims with the same electric razor, using the newspaper to catch the clipped hairs for both trims. Intuitively, that progression of theories is highly suspicious.

On the other hand, perhaps the hairs {\it did} come from the combination of a beard and scalp trim, and the prosecution was simply careless in formulating their original theory. We cannot dismiss the newspaper hairs evidence just because we do not respect the reasoning and rhetoric employed by the prosecution. The argument below takes the prosecution's latest theory seriously. At a high level, the argument has the following structure:
\begin{PPE}
\item There are {\it many} distinct theories of how the hypothesized beard and scalp trims could have happened. In the argument below, we introduce a family of such theories indexed by the parameters $\amin$ and $\amax$. 
\item Most of the theories in that family are bad for the prosecution; they result in a model that predicts the data worse than the defense's beard-trim-only theory.
\item The prosecution cannot justify choosing from among just the theories that are good for them, or giving such theories greater weight.
\end{PPE} 
%We formulate the prosecution's theory in terms of the parameters  such that, for at least some settings of the parameters, the assumptions of the theory would have been reasonable before seeing the hair measurements. 
We will deduce how the parameters $\amin$ and $\amax$ must be set in order for the prosecution's theory to have predictive power as good as the defense's theory, and we will find that the parameters would need to be set to values that have no (supplied) reasonable justification (without referring to the measurements, which would be using the data to fit the model that the data is supposed to predict). If the assumptions from which we derive the parametric theory are reasonable (e.g. the fixed prior over distributions for Hay's beard hair widths, and the fixed distribution for Hay's scalp hair widths), then we can conclude that the newspaper hair evidence is not inculpatory.

 %We allow the prosecution to use post-hoc reasoning to come up with their vague beard\&scalp trim theory, but we do not allow them to use post-hoc reasoning to optimize the quantitative parameters of said theory.
\bigskip

Though the argument to follow is unquestionably an example of Bayesian analysis, I prefer to use the language of frequencies and repeatable events rather than degrees of belief. One could just as well use the language of degrees of belief, with only superficial changes to the axioms. 

We posit constraints on a randomized simulation model of the crime and evidence, which is applicable not just to Hay's case, but also to a number of very-similar hypothetical cases (in some of which the suspect is guilty) taken from an implicitly-constrained distribution $D$. The probabilities are just parameters of the model, and in principle we judge models according to how often they make the correct prediction when a case is chosen at random from $D$. In the argument below, we don't use $D$ directly, but rather use a distribution over a small number of random variables that are meaningful in $D$, namely the joint distribution for the random variables:
\[ \GRV, \ClippedRV, \MixRV, \BParamsRV, \HRV, \WidthsRV \]
Some of the most significant assumptions for the argument are as follows:

\begin{LPPE}
%\item We posit, quite conservatively given the principle of ``beyond a reasonable doubt,'' that before the newspaper hair evidence is created in the simulation, the suspect is at least as likely to be innocent as guilty. 
\item The prior chosen for the suspect's beard hair-width distribution is fair and reasonable.\footnote{The reason we use a prior for the suspect's beard hair width distribution is that Leighton Hay's beard hair widths were never sampled; that decision was on the advice of one of the hair forensics experts, who said that a man's beard hairs tend to get thicker as he ages.} This is Simplifying Assumption \ref{sa:beardprior}. It is probably the most disputable of the assumptions. I give some criticisms of it in Section \ref{s:criticismOfLHArg}.
\item The distribution for the suspect's scalp hair widths, based on the samples taken in 2010, is fair and reasonable (Simplifying Assumption \ref{sa:scalpdistr}). This assumption may be disputable in that it does not assume that Hay's scalp hairs thinned by an average amount for a man of his age and race in the 8 years between the crime and when his scalp hair sample was taken. Of course, it may be that his hairs have not thinned at all. Unfortunately it appears that we cannot know this, as samples were not taken in 2002.
\item The simulation model, on runs where the suspect is guilty (and thus the newspaper hair evidence comes from a combined beard and scalp trim), chooses uniformly at random (Simplifying Assumption \ref{sa:mixprior}) from a sufficiently large range the ratio 
\begin{equation}  \label{e:hairlocationratio}
\frac{\text{P(random clipped hair came from beard, given only that it ended up in the newspaper)}}{\text{P(random clipped hair came from the scalp, given only that it ended up in the newspaper)}} 
\end{equation}
Specifically that range is $[\frac{\amin}{1-\amin},\frac{\amax}{1-\amax}]$ (but note: no axiom of the argument requires that symbolic form to be meaningful to the reader). The axioms enforce no constraints about $\amin$ and $\amax$ except for $0 < \amin < \amax < 1$, but the hypotheses of Claims \ref{c:lhMainClaim1} and \ref{c:lhMainClaim2} assert significant constraints; it turns out that in order for the likelihood ratio $\frac{\Pr{\WidthsRV = \vb \mid \GRV}}{\Pr{\WidthsRV = \vb \mid \NGRV}}$ to be $\geq 1$, the prosecution needs to make an extreme assumption about $\amin$ and $\amax$.
Intuitively, assuming the suspect is guilty, both prosecution and defense are still very ignorant (before seeing the newspaper hair measurements) of how exactly the suspect trimmed his beard and scalp, e.g. in what order, how exactly he used the newspaper, and how exactly he emptied most of the clippings into the toilet, all of which would influence the above ratio (\ref{e:hairlocationratio}). The hypotheses of Claims \ref{c:lhMainClaim1} and \ref{c:lhMainClaim2} formalize that intuition in different ways, which are close to equivalent, but nonetheless I think Claim \ref{c:lhMainClaim2} is significantly easier to understand and accept. 
\item The suspect in the simulation model does not have an unusually low ratio of scalp hairs to beard hairs. This is Assumption \ref{a:betabound}. We can improve the current argument, if we wish, by having the simulation model choose that ratio from some prior distribution, and doing so actually results in a version of Claim \ref{c:lhMainClaim2} that is {\it better} for the defense. I don't do this simply because the extra complexity would reduce the pedagogical value of this example.
\end{LPPE}

\section{\Finished{Argument}}
\indentoff

Because this argument is written in \LaTeX, I present it more-informally than is required by the definition {\it interpreted formal proof}. In particular, I do not explicitly name the types of most symbols, and I don't explain how exactly random variables and the $\Pr{\text{proposition} \mid \text{proposition}}$ syntax are formalized. 
%A completely-formal version of this argument, which strictly adheres to the definition of {\it interpreted formal proof}, will be included in my thesis. That includes explicit types for each symbol, with each type labeled as an assumption or simplifying assumption.  The formalization is mostly straight-forward; the only part that requires some thought is the formalization of random variables and the $\Pr{\text{proposition} \mid \text{proposition}}$ syntax.
\medskip

I will often use the following basic facts. In the completely-formal proof they would be axioms in $\Gassum$ that use only symbols in $\lama$, and thus should be accepted by any member in the intended audience of the proof.
\begin{PPI}
\item For $t_1,t_2,t_3$ boolean-valued terms:
\[\Pr{t_1,t_2 \mid t_3} = \Pr{t_1 \mid t_2, t_3} \Pr{t_2 \mid t_3}\] 
\item For $X$ a continuous random variable with conditional density function $d_X$ whose domain $S$ is a polygonal subset of $\RR^n$ for some $n$:
\[\Pr{t_1 \mid t_2} = \int_{x \in S} \Pr{t_1 \mid t_2, X\!=\!x}\, d_X(x \mid t_2) \]
\end{PPI}

{\bf$\bin{1}, \bin{2}, \bin{3}, \bin{4}$} are constants denoting the four micrometer-intervals from Table 1. Formally, they belong to their own sort, which has exactly 4 elements in every model. We do not actually have micrometer intervals in the ontology of the proof, so we could just as well use $\{1,2,3,4\}$, but I think that would be confusing later on. {\bf $\Bins$} is the sort $\{\bin{1}, \bin{2}, \bin{3}, \bin{4}\}$. 

Throughout this writeup, $\vb = b_1,\ldots,b_{89}$ is a fixed ordering of the newspaper hair measurements shown in Table 1. Specifically, each $b_i$ is one of the constants $\bin{1}, \bin{2}, \bin{3}$, or $\bin{4}$;  $\bin{1}$ appears 10 times, $\bin{2}$ 20 times, $\bin{3}$ 40 times, and $\bin{4}$ 19 times.\\

$\vec{\textbf{p}}$ abbreviates $\pair{p_1,p_2,p_3}$.\\
$\textbf{p}_4$ abbreviates $1 - p_1 - p_2 - p_3$ (except in Claim \ref{c:fourintegrals}, as noted there also).\\

{\bf $\GRV$} is the boolean simulation random variable that determines if the suspect in the current run is guilty. I write just $\GRV$ to abbreviate $\GRV \!=\! \true$ and $\NGRV$ to abbreviate $\GRV \!=\! \false$. \\

{\bf $\ClippedRV$} is a simulation random variable whose value is determined by $G$. When $G$ is false, $\ClippedRV$ is the set of beard hair fragments that fall from the suspect's face when he does a full beard trim with an electric trimmer\footnote{The police collected an electric trimmer that was found, unhidden, in Hay's bedside drawer, which Hay has always said he used for trimming his beard.} several days before the murder took place. When $G$ is true, $\ClippedRV$ is the set of beard and scalp hair fragments that fall from the suspect's head when he does a full beard trim and a full scalp trim (the latter after cutting off his two-inch dreds) with the same electric trimmer. This includes any such fragments that were flushed down the sink or toilet, but not including --in the case that the suspect is guilty-- hair fragments that were part of his 2-inch ``picky dreads.'' \\

{\bf $\HRV$} is a simulation random variable whose distribution is the uniform distribution over $\ClippedRV$, i.e. it is a random hair clipping.\\

{\bf $\BParamsRV$} is the simulation random variable that gives the parameters of the suspect's {\bf b}eard hair width distribution.\\

{\bf $\MixRV$} is the simulation random variable that gives the the mixture parameter that determine's the prosecution's newspaper hair width distribution given the beard and scalp hair width distributions.\\

{\bf NOTATION:} $\BParamsRV$ and $\MixRV$ will usually be hidden in order to de-clutter equations and to fit within the page width. Wherever you see $\vp$ or $\pair{p_1,p_2,p_3}$ where a boolean-valued term is expected, that is an abbreviation for $\BParamsRV \!=\! \vp$ or $\BParamsRV \!=\! \pair{p_1,p_2,p_3}$, respectively. Similarly, I write just $\alpha$ as an abbreviation for $\MixRV=\alpha$.\\

{\bf$\bd$} is the set from which our prior for the suspect's beard hair width distribution is defined. It is the set of triples $\pair{p_1,p_2,p_3} \in [0,1]^3$ such that $p_1 \leq p_2,p_3,p_4$ and $\pair{p_1,p_2,p_3,p_4}$ is unimodal when interpreted as a discrete distribution where $p_i$ is the probability that the width of a hair randomly chosen from the suspect's scalp (in 2002) falls in bin $i$. \\

$\Pr{t_1 \mid t_2}$ is the notation we use for the Bayesian/simulation distribution over the random variables $\GRV, \ClippedRV, \MixRV, \BParamsRV, \HRV, \WidthsRV$, where $t_1$ and $t_2$ are terms taking on boolean values; it is the probability over runs of the simulation that $t_1$ evaluates to true given that $t_2$ evaluates to true. \\

{\bf $\WidthsRV$} is the simulation random variable that gives the approximate widths (in terms of the 4 intervals $\bin{j}$) of the 89 hair clippings that end up in the balled-up newspaper. \\

{\bf NOTATION:} When the variables $\vp$ and $\alpha$ appear unbound in an axiom, I mean for them to be implicitly quantified in the outermost position like so: $\forall \vp \in \bd$ and $\forall \alpha \in [\amin,\amax]$.\\

When $X$ is a continuous random variable with a density function, $d_X$ denotes that function.\\

\begin{defn}\label{d:likeratio} We are aiming to show that from reasonable assumptions, the following likelihood ratio is less than 1, meaning that the defense's theory explains the newspaper hairs evidence at least as well as the prosecution's theory. The notation $\likeratio(\amin,\amax)$ is used just to highlight the dependence on the parameters $\amin,\amax$.
\[ \likeratio(\amin,\amax) \ceq \frac{\Pr{\WidthsRV = \vb \mid \GRV}}{\Pr{\WidthsRV = \vb \mid \NGRV}} \]
\end{defn}

\begin{assumption} The values of $\BParamsRV$ and $\MixRV$ are chosen independently of each other and $\GRV$ (whether or not the suspect is guilty). Hence the defense and prosecution have the same prior for the suspect's beard hair width distribution.\\
For $t \in \{ \true, \false \}$:
\[ \denspa{\vp,\alpha \mid \GRV \!=\! t} = \densp{\vp} \cdot \densa{\alpha} \]
\end{assumption}

$\amin$ and $\amax$ are constants in $(0,1)$ such that $\amin < \amax$.

\begin{simplifyingassumption}  \label{sa:mixprior}
The prior distribution for the mixture parameter $\MixRV$  is the uniform distribution over $[\amin,\amax]$. 
\[ \densa{\alpha} = 
\begin{cases} 
1 / (\amax - \amin) & \text{if $\alpha \in [\amin,\amax]$} \\ 
0 & \text{otherwise}
\end{cases}
\]
\end{simplifyingassumption}

\begin{simplifyingassumption} \label{sa:beardprior}
The prior distribution for the parameters of the suspect's beard hair width distribution is the uniform distribution over the set $\bd \subeq [0,1]^3$ defined above.
\[
\densp{\vp} = 
\begin{cases} 
1 / \| \bd \| & \text{if $\vp \in \bd$} \\ 
0 & \text{otherwise}
\end{cases}
\]
\end{simplifyingassumption}

\medskip

{\bf $\InNews{h}$} $\!=\! \true$ iff the hair clipping $h$ ends up in the balled-up newspaper.\\
{\bf $\FromBeard{h}$} $\!=\! \true$ (respectively {\bf $\FromScalp{h}$} $\!=\! \true$)  iff hair clipping $h$ came from the suspect's beard (respectively scalp). 
\begin{assumption} \label{a:scalpIsNegBeard}Both prosecution and defense agreed that all the hairs in the newspaper came from the suspect's beard or scalp, and not both.\footnote{``Not both'' actually ignores the issue of sideburn hairs, whose widths can be intermediate between scalp and beard hair widths. Doing this is favourable for the prosecution.}
\[\FromScalp{h} = \neg \FromBeard{h}\]
\end{assumption}

{\bf $\width$} is the function from $\ClippedRV$ to $\{ \bin{1}, \bin{2}, \bin{3}, \bin{4}\}$ such that $\width[h]$ is the interval in which the maximum-width of hair clipping $h$ falls.

\begin{simplifyingassumption} \label{sa:hairindependence}
In the simulation model, the hairs that ended up in the newspaper are chosen independently at random with replacement from some hair-width distributions. 
\[ \Pr{\WidthsRV \!=\! \vb \mid \GRV, \vp, \alpha} = \prod\limits_{i=1}^{89} \Pr{\width[\HRV] \!=\! b_i \mid \InNews{\HRV}, \GRV, \vp, \alpha} \]
\[ \Pr{\WidthsRV \!=\! \vb \mid \NGRV, \vp} = \prod\limits_{i=1}^{89} \Pr{\width[\HRV] \!=\! b_i \mid \InNews{\HRV}, \NGRV, \vp}\]
\end{simplifyingassumption}

\begin{claim} \label{c:scalpbeardfactor}
We can write the width distribution of newspaper hairs in terms of the width distributions of beard and scalp hairs, together with the probability that a random newspaper hair is a beard hair. 
\[
\begin{array}{rl}
  & \Pr{ \width[\HRV] \!=\! b_i \mid \InNews{\HRV}, \GRV, \vp, \alpha}  \\
= & \Pr{ \width[\HRV] \!=\! b_i \mid \FromBeard{\HRV}, \InNews{\HRV}, \GRV, \vp, \alpha } \ \Pr{ \FromBeard{\HRV} \mid \InNews{\HRV}, \GRV, \vp, \alpha} \\
+ &  \Pr{ \width[\HRV] \!=\! b_i \mid \FromScalp{\HRV},  \InNews{\HRV}, \GRV, \vp, \alpha} \ \Pr{ \FromScalp{\HRV} \mid \InNews{\HRV}, \GRV, \vp, \alpha} 
\end{array} 
\]
\end{claim}
\begin{proof}
Follows from Assumption \ref{a:scalpIsNegBeard}.
\end{proof}

\begin{assumption} In the defense's model (not guilty $\NGRV$), all the newspaper hair came from a beard trim, and so the mixture parameter is irrelevant.
\[
\begin{array}{rl}
  & \Pr{ \width[\HRV] \!=\! b_i \mid \InNews{\HRV}, \NGRV, \vp, \alpha}  \\
= & \Pr{ \width[\HRV] \!=\! b_i \mid \FromBeard{\HRV}, \InNews{\HRV}, \NGRV, \vp } \\ 
\end{array}
\]
\end{assumption}

\begin{assumption} Given that a clipped hair came from the suspect's beard, the hair's width is independent of whether the suspect is guilty in this run of the simulation. Thus the defense and prosecution models use the same distribution of hair widths for the suspect's beard.
\[
\begin{array}{rl}
 & \Pr{ \width[\HRV] \!=\! b_i \mid \FromBeard{\HRV}, \InNews{\HRV}, \NGRV, \alpha, \vp } \\ 
= & \Pr{ \width[\HRV] \!=\! b_i \mid \FromBeard{\HRV}, \InNews{\HRV}, \GRV, \alpha, \vp } \\
= & \Pr{ \width[\HRV] \!=\! b_i \mid \FromBeard{\HRV}, \InNews{\HRV}, \alpha, \vp }
\end{array}
\]
\end{assumption}

\begin{assumption} \label{a:alphameaning}
We finally give the precise meaning of the simulation's mixture parameter random variable $\MixRV$. It is the probability, when the suspect is guilty, that a randomly chosen hair clipping came from the suspects beard {\it given} that it ended up in the newspaper.
\[ \alpha = \Pr{ \FromBeard{\HRV} \mid \InNews{\HRV}, \GRV, \vp, \MixRV \!=\! \alpha} \]
\[ 1 - \alpha = \Pr{ \FromScalp{\HRV} \mid \InNews{\HRV}, \GRV, \vp, \MixRV \!=\! \alpha} \]
\end{assumption}

\begin{assumption} \label{a:pmeaning}
The precise meaning of the simulation random variable $\BParamsRV$. Recall that $p_4$ abbreviates $1-p_1-p_2-p_3$. For $j \in\{1,2,3,4\}$:
\[ p_j = \Pr{ \width[\HRV] \!=\! \bin{j} \mid \FromBeard{\HRV} , \BParamsRV \!=\! \pair{p_1,p_2,p_3}, \InNews{\HRV} }  \]
\end{assumption}

\begin{simplifyingassumption} \label{sa:scalpdistr}We use a completely-fixed distribution for the suspect's scalp hair, namely the one that maximizes the probability of obtaining the hair sample measurements from Table 2 when 90 hairs are chosen independently and uniformly at random from the suspect's scalp.
\[
 \Pr{ \width[\HRV] = b_i \mid \FromScalp{\HRV}, \GRV, \alpha, \vp } = 
      \begin{cases} \nicefrac{89}{90} & \text{if } i=1 \\ 
      						\nicefrac{1}{90} & \text{if } i=2 \\ 
						     0 & \text{if } i=3,4 \\ 
      \end{cases}
\]
\end{simplifyingassumption}

{\bf The next axiom and claim give the main result, and the later Claim \ref{c:lhMainClaim2} is (almost) a corollary of Claim \ref{c:lhMainClaim1}.}

\begin{assumption} \label{a:LHtopaxiom}
If $\frac{\Pr{\WidthsRV = \vb \mid \GRV}}{\Pr{\WidthsRV = \vb \mid \NGRV}} \leq 1$ (i.e. $\likeratio \leq 1$), then \\ 
$\evidenceneutral$.\footnote{The text in brackets is a constant predicate symbol.}
\end{assumption}

\begin{claim} \label{c:lhMainClaim1} 
%\TODO[Verify that this is true. VERIFIED]  
If $\amin \leq .849$ then $\frac{\Pr{\WidthsRV = \vb \mid \GRV}}{\Pr{\WidthsRV = \vb \mid \NGRV}} < 1$
\end{claim}

The proof of Claim \ref{c:lhMainClaim1} is outlined formally below, after Claim \ref{c:lhMainClaim2}.

%\begin{claim} This should follow from the definition of $\MixRV$ above; since $\MixRV$ is the probability that some event $E'$ occurs given that some events $\vec{E}$ occur
%\[ \bothPr{ \width[\HRV] = b_i \mid \FromBeard{\HRV}, \InNews{\HRV}, \GRV, \alpha, \vp } = \bothPr{ \width[\HRV] = b_i \mid \FromBeard{\HRV}, \InNews{\HRV}, \GRV, \vp } \] 
%\end{claim}

\bigskip

With the introduction of a new parameter and a mild assumption about its values (Assumption \ref{a:betabound}, the ratio on the left side being the new parameter), we will obtain a corollary of Claim \ref{c:lhMainClaim1} that is easier to interpret.
\medskip

We do not know what the ratio of beard to scalp hairs on Hay's head was on the date of the murder, and it is not hard to see that a higher value of $\Pr{ \FromBeard{\HRV} \mid \GRV, \vp, \alpha}$ is favourable for the prosecution.\footnote{Raising the value makes both models worse, but it hurts the prosecution's model less since the prosecution's model can accommodate by lowering $\amin$ and $\amax$.} We do, however, know that the unknown shooter's beard was described as ``scraggly'' and ``patchy'' by eye witnesses, and we have no reason to think that LH had a smaller than average number of scalp hairs. Thus it is a conservative approximation (from the perspective of the prosecution) to assume that Hay had a great quantity of beard hairs for a man (40,000), and an average quantity of scalp hairs for a man with black hair (110,000).\footnote{Trustworthy sources for these numbers are hard to find. 40,000 is just the largest figure I found amongst untrustworthy sources, and 110,000 is a figure that appears in a number of untrustworthy sources. If this troubles you, consider the ratio a parameter whose upper bound we can argue about later.} Thus we assume:
\begin{assumption} \label{a:betabound} 
\[ \frac{\Pr{ \FromBeard{\HRV} \mid \GRV, \vp, \alpha}}
{\Pr{ \FromScalp{\HRV} \mid \GRV, \vp, \alpha}} \leq 4/11 \]
\end{assumption}

\begin{claim} \label{c:lhMainClaim2}
The hypothesis of Assumption \ref{a:LHtopaxiom} also follows if we assume Assumption \ref{a:betabound} and that {\it the uniform prior over $\MixRV$ gives positive density to a model where a random clipped beard hair is $\leq 15$ times more likely to end up in the newspaper as a random clipped scalp hair}:

If  there exists $\alpha \in [\amin,\amax]$ and $\vp \in \bd$ such that
\[
\frac{\Pr{ \InNews{\HRV} \mid \FromBeard{\HRV}, \GRV, \vp, \alpha}}
           {\Pr{ \InNews{\HRV} \mid \FromScalp{\HRV}, \GRV, \vp, \alpha}} \leq 15
           \]
then  \[\frac{\Pr{\WidthsRV \!=\! \vb \mid \GRV}}{\Pr{\WidthsRV \!=\! \vb \mid \NGRV}} < 1\]          
\end{claim}
\begin{proof}
Let $\alpha,\vp$ be as in the hypothesis. 

From basic rules about conditional probabilities:
\begin{equation}  \label{e:alpharatios}
 \frac{\alpha}{1-\alpha} 
= \frac{\Pr{ \FromBeard{\HRV} \mid \InNews{\HRV}, \GRV, \vp, \alpha}}{\Pr{ \FromScalp{\HRV} \mid \InNews{\HRV}, \GRV, \vp, \alpha}} 
= \frac{\Pr{ \InNews{\HRV} \mid \FromBeard{\HRV} \GRV, \vp, \alpha} \ \Pr{ \FromBeard{\HRV} \mid \GRV, \vp, \alpha}}
  		  {\Pr{ \InNews{\HRV} \mid \FromScalp{\HRV}, \GRV, \vp, \alpha} \ \Pr{ \FromScalp{\HRV} \mid \GRV, \vp, \alpha}} 
\end{equation}

Using the inequality from the hypothesis and Assumption \ref{a:betabound}, solve for $\alpha$ in (\ref{e:alpharatios}). This gives $\alpha \leq 0.84507$. Since $\amin \leq \alpha$ we have $\amin \leq .84507$, so we can use Claim \ref{c:lhMainClaim1} to conclude that the likelihood ratio is less than 1.
\end{proof}

\begin{simplifyingassumption}[hypothesis of Claim \ref{c:lhMainClaim2}] \label{sa:mainsa}
There exists $\alpha \in [\amin,\amax]$ and $\vp \in \bd$ such that
\[
\frac{\Pr{ \InNews{\HRV} \mid \FromBeard{\HRV}, \GRV, \vp, \alpha}}
           {\Pr{ \InNews{\HRV} \mid \FromScalp{\HRV}, \GRV, \vp, \alpha}} \leq 15
           \]
\end{simplifyingassumption}

\bigskip 

\begin{goalsentence} \label{c:lhGoalSentence} 
$\evidenceneutral$
\end{goalsentence}
\begin{proof} From Simplifying Assumption \ref{sa:mainsa}, Claim \ref{c:lhMainClaim2}, and Assumption \ref{a:LHtopaxiom}.
\end{proof}

\bigskip

{\bf Proof of Claim \ref{c:lhMainClaim1}}\\

{\it Note: there is nothing very interesting about this proof; it is basically just a guide for computing the $\likeratio$ as a function of $\amin, \amax$.}\\

To compute the integrals, I will break up the polygonal region $\bd$ into several pieces which are easier to handle with normal Riemann integration over real intervals.

Let $\bd_{1}$ be the subset of $\bd$ where $p_2 > p_3 \geq p_4$\\ % R2
$\bd_{2}$ the subset of $\bd$ where $p_3 > p_2 > p_4$\\  % R32
$\bd_{3}$ the subset of $\bd$ where $p_3 > p_4 \geq p_2$\\ % R34
$\bd_{4}$ the subset of $\bd$ where $p_4 > p_3 \geq p_2$  % R4

\begin{claim} \label{c:bdpartition}
$\bd$ is the disjoint union of $\bd_{1}, \bd_{2}, \bd_3, \bd_4$.
\end{claim}

\begin{claim} \label{c:fourintegrals} In the scope of this claim, $p_4$ is a normal variable, not an abbreviation for $1-p_1-p_2-p_3$.
\[\int\limits_{\vp = \pair{p_1,p_2,p_3} \in \bd_{1}} \nquad\nquad\nquad t(p_1,p_2,p_3,1-p_1-p_2-p_3) d\vp
= \int\limits_{p_1 = 0}^{\nicefrac{1}{4}} \int\limits_{p_4 = p_1}^{\frac{1-p_1}{3}} \int\limits_{p_3 = p_4}^{\frac{1-p_1-p_4}{2}} \nquad t(p_1,1-p_1-p_3-p_4, p_3, p_4') dp_1 dp_4 dp_3\]
\[\int\limits_{\vp=\pair{p_1,p_2,p_3} \in \bd_{2}} \nquad\nquad\nquad t(p_1,p_2,p_3,1-p_1-p_2-p_3) d\vp
= \int\limits_{p_1 = 0}^{\nicefrac{1}{4}} \int\limits_{p_4 = p_1}^{\frac{1-p_1}{3}} \int\limits_{p_2 = p_4}^{\frac{1-p_1-p_4}{2}} \nquad t(p_1,p_2, 1-p_1-p_2-p_4, p_4) dp_1 dp_4 dp_2\]
\[\int\limits_{\vp=\pair{p_1,p_2,p_3} \in \bd_{3}} \nquad\nquad\nquad t(p_1,p_2,p_3,1-p_1-p_2-p_3) d\vp
= \int\limits_{p_1 = 0}^{\nicefrac{1}{4}} \int\limits_{p_2 = p_1}^{\frac{1-p_1}{3}} \int\limits_{p_4 = p_2}^{\frac{1-p_1-p_2}{2}} \nquad t(p_1,p_2, 1-p_1-p_2-p_4, p_4) dp_1 dp_2 dp_4\]
\[\int\limits_{\vp=\pair{p_1,p_2,p_3} \in \bd_{4}} \nquad\nquad\nquad t(p_1,p_2,p_3,1-p_1-p_2-p_3) d\vp
= \int\limits_{p_1 = 0}^{\nicefrac{1}{4}} \int\limits_{p_2 = p_1}^{\frac{1-p_1}{3}} \int\limits_{p_3 = p_2}^{\frac{1-p_1-p_2}{2}} \nquad t(p_1,p_2, p_3,1-p_1-p_2-p_3) dp_1 dp_2 dp_3\]
\end{claim}

\begin{claim}
$\| \bd \| = 1/36$
\end{claim}
\begin{proof}
The measure of $\bd_j$ can be computed by standard means by substituting 1 in for $t(\ldots)$ in the right side of the $j$-th equation of Claim \ref{c:fourintegrals}. We find that $\|\bd_1\| = \| \bd_2 \| = \| \bd_3 \| = \| \bd_4 \| = 1/144$. Hence $\| \bd \| = 1/36$ follows from Claim \ref{c:bdpartition}.
\end{proof}

%\begin{assumption} \TODO[Do I actually use this?] Given that a clipped hair came from suspect's beard, the hair's width is independent of  the value of the random variable $\MixRV$ (recall $\alpha$ abbreviates $\MixRV = \alpha$ in this context)
%\[
% \Pr{ \width[\HRV] = b_i \mid \FromBeard{\HRV}, \InNews{\HRV}, \alpha, \vp }
% =  \Pr{ \width[\HRV] = b_i \mid \FromBeard{\HRV}, \InNews{\HRV}, \vp } 
%\]
%\end{assumption}

%\begin{simplifyingassumption} \TODO[Do I need this?] Given the origin (beard or scalp) of a clipped hair, its width is independent of whether it ends up in the newspaper or gets flushed down the toilet. The accuracy of this assumption can certainly be challenged, but this would likely not be in the best interest of the prosecution.\footnote{I can think of reasons why the thinner scalp hairs might be more likely to stick to the newspaper than the thicker beard hairs, e.g. their being more influenced by weak electrostatic forces. We will see why that would be bad for the prosecution.}
%%For all $\vp \in \bd$:
%\[\Pr{ \width[\HRV] = b_i \mid \FromScalp{\HRV} , \InNews{\HRV}, \GRV }  =  \Pr{ \width[\HRV] = b_i \mid \FromScalp{\HRV}, \GRV } \]
%\[\Pr{ \width[\HRV] = b_i \mid \FromBeard{\HRV} , \InNews{\HRV}, \vp  }  =  \Pr{ \width[\HRV] = b_i \mid \FromBeard{\HRV}, \vp } \]
%\end{simplifyingassumption}

\begin{claim} Simplified forms amenable to efficient computation:
\[\Pr{\WidthsRV \!=\! \vb \mid \NGRV, \pair{p_1,p_2,p_3}} = p_1^{10} p_2^{20} p_3^{40} p_4^{19}\]
\[\Pr{\WidthsRV \!=\! \vb \mid \GRV, \pair{p_1,p_2,p_3}, \alpha} = (p_1 \alpha + \nicefrac{89}{90}(1-\alpha))^{10} (p_2\alpha + \nicefrac{1}{90}(1-\alpha))^{20} (p_3 \alpha)^{40} (p_4 \alpha)^{19}\]
\end{claim}
\begin{proof}
The first equation follows easily from Simplifying Assumption \ref{sa:hairindependence} and Assumption \ref{a:pmeaning}. The second follows easily from Simplifying Assumption \ref{sa:hairindependence}, Axioms \ref{a:pmeaning} and \ref{a:alphameaning}, and Claim \ref{c:scalpbeardfactor}.
\end{proof}

\medskip

From the next fact and Claim \ref{c:fourintegrals} we can compute the two terms of the likelihood ratio for fixed $\amin$ and $\amax$.  
\begin{claim} \label{c:computingIntegrals}
\[ \Pr{\WidthsRV \!=\! \vb \mid \GRV}  
= \int\limits_{\alpha \in [\amin,\amax]} \int\limits_{\vp\in \bd} \Pr{\WidthsRV \!=\! \vb \mid \GRV, \vp, \alpha}\, \denspa{\vp,\alpha \mid \GRV}  \]
\[ =  \frac{1}{(\amax - \amin) \|\bd\|}\sum\limits_{i \in \{1,2,3,4\}} \int\limits_{\alpha \in [\amin,\amax]} \int\limits_{\vp\in \bd_{i}} \Pr{\WidthsRV \!=\! \vb \mid \GRV, \vp, \alpha} \]
\[ \Pr{\WidthsRV \!=\! \vb \mid \NGRV} 
= \int\limits_{\vp\in \bd} \Pr{\WidthsRV \!=\! \vb \mid \NGRV, \vp}\, \densp{\vp \mid \NGRV}  \]
\[
= \frac{1}{ \|\bd\|} \sum\limits_{i \in \{1,2,3,4\}} \int\limits_{\vp\in \bd_{i}} \Pr{\WidthsRV \!=\! \vb \mid \NGRV, \vp} \]
\end{claim}
\begin{proof}
The first equation follows just from $\vp,\alpha \mapsto \Pr{\WidthsRV \!=\! \vb \mid \GRV, \vp, \alpha}$ being an integrable function and $\denspa{\vp,\alpha \mid \GRV}$ being the conditional density function for $\pair{\MixRV,\BParamsRV}$ given $\GRV = \true$. 

The second equation follows from Claim \ref{c:bdpartition}, Simplifying Assumptions \ref{sa:mixprior} and \ref{sa:beardprior}, and the fact that $\vp, \alpha \mapsto \Pr{\WidthsRV \!=\! \vb \mid \GRV, \vp, \alpha}$ is bounded. The first and fourth of those facts suffice to show that the integral over $\bd$ is equal to the sum of the integrals over the sets $\bd_j$.

Justifications for the third and fourth equations are similar to those for the first and second.
\end{proof}
\bigskip

As of now I've mostly used Mathematica's numeric integration, which doesn't provide error bounds, to evaluate the intervals, but there are also software packages one can use that provide error bounds. 
\indenton

The likelihood ratio (Definition \ref{d:likeratio}) achieves its maximum of $\approx 1.27$ when $\amin$ and $\amax$ are practically equal  (unsurprising, as that allows the prosecution model to choose the best mixture parameter) and around  $.935$; Plot \ref{lhThinAlphaVarying} illustrates this, showing the likelihood ratio as a function of $\amin$ when $\amax-\amin = 10^{-6}$. To prove Claim \ref{c:lhMainClaim1} we need to look at parameterizations of $\amin,\amax$ similar to the one depicted in Plot \ref{lhAlphaMaxVarying}, which shows the likelihood ratio as a function of $\amax$ when $\amin = .849$ (the extreme point in the hypothesis of Claim \ref{c:lhMainClaim1}), in which case the likelihood ratio is maximized at $\approx .996$ when $\amax=1$. In general, for smaller fixed $\amin$, the quantity 
\[ \max\limits_{\amax \in (\amin,1)}(\likeratio(\amin,\amax)) \]
decreases as $\amin$ does. More precisely, Claim \ref{c:lhMainClaim1} follows from the following three propositions in Claim \ref{c:unprovedclaims}. The first has been tested using Mathematica's numerical integration; if it is false, it is unlikely to be false by a wide margin (i.e. taking a value slightly smaller than .849 should suffice). The remaining two have also not been proved, but one can gain good confidence in them by testing plots similar to Figure \ref{lhAlphaMaxVarying} for values of $\amin < .849$. Proving Claim \ref{c:unprovedclaims} or a slightly weaker version of it is just a matter of spending more time on it (or enlisting the help of an expert to do it quickly). But we will see in the next section that the argument is more-vulnerable to attack in other ways.

\begin{claim} \label{c:unprovedclaims} 
Here, the notation $\likeratio(\alpha_1,\alpha_2)$ is short for ``the real number taken on by the defined term $\likeratio$ when $\amin = \alpha_1$ and $\amax = \alpha_2$.''
\begin{PPE}
\item $\likeratio(.849,1) < .997$
\item For $\alpha_1 < .849$ have $\likeratio(\alpha_1,1) < \likeratio(.849,1)$
\item For $\alpha_1 < .849$ and $\alpha_1 < \alpha_2 < 1$ have $\likeratio(\alpha_1,\alpha_2) < \likeratio(\alpha_1,1)$
\end{PPE}
\end{claim}

\IMG{
\begin{figure}[h!] \label{lhThinAlphaVarying}
  \caption{Likelihood ratio as a function of $\amin$ when $\amax-\amin = 10^{-6}$, obtained by numerical integration.}
  \centering
  \includegraphics[scale=.8]{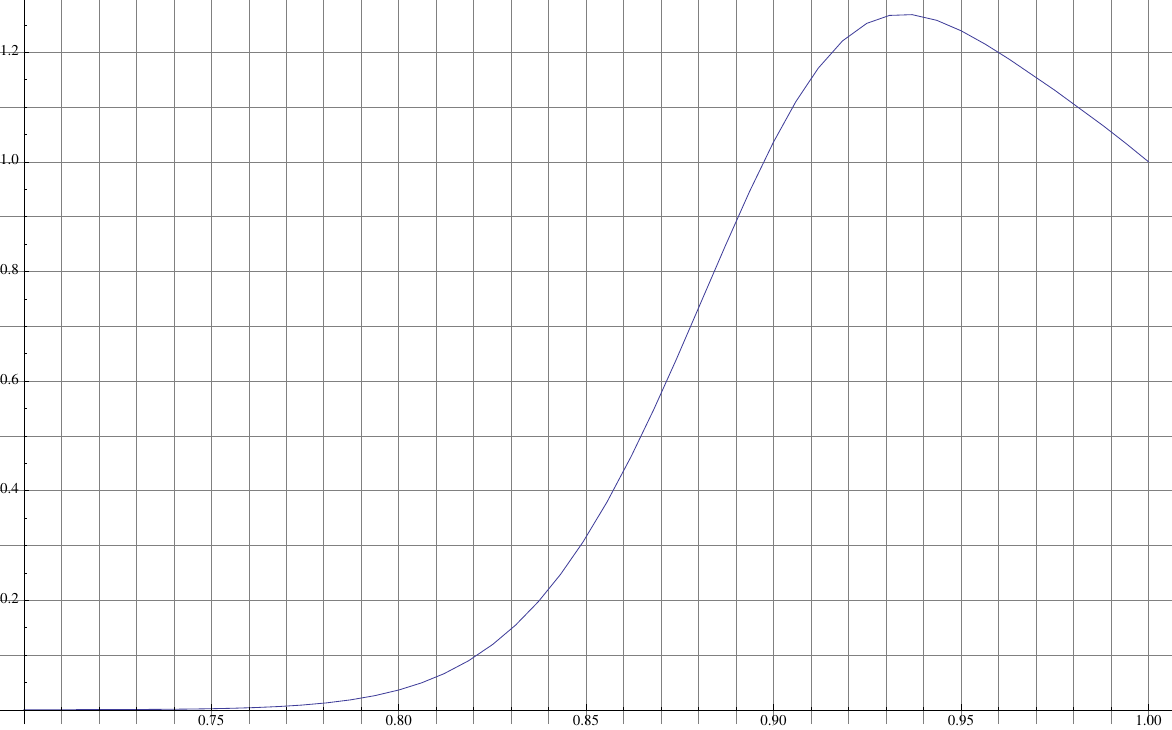}\\
\end{figure}
}

\IMG{
\begin{figure}[h!] \label{lhAlphaMaxVarying}
  \caption{Likelihood ratio as a function of $\amax$ when $\amin = .849$, obtained by numerical integration. The shape of this plot is similar for smaller values of $\amin$, being maximized when $\amax = 1$, which is what parts 2 and 3 of Claim \ref{c:unprovedclaims} express. }
  \centering
\includegraphics[scale=.8]{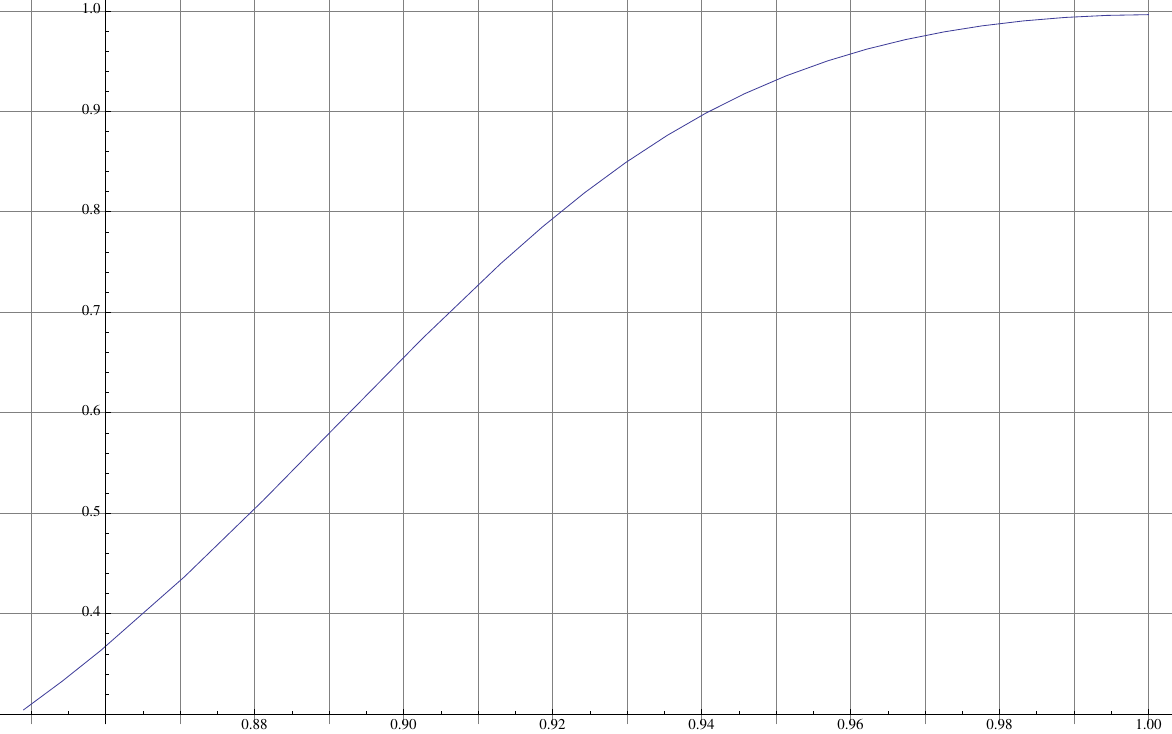}
\end{figure}
}

\section{\Finished{Criticism of argument}} \label{s:criticismOfLHArg}
\indenton
\subsection{\Finished{Criticism 1}}
It is arguable that the prior for the suspect's beard hair width distribution is slightly biased in favor of the defense, in which case the prosecution could {\bf reject Simplifying Assumption \ref{sa:beardprior}}. In particular, the average value of the component of $\BParamsRV$ for $\bin{1}$, the bin corresponding to the thinnest hairs, is $0.0625$.\footnote{Compute by substituting $p_1$ in for $t$ in each of the four equations of Claim \ref{c:fourintegrals}, and sum the results.} It is best for the defense when the value of that component is $11/89$, and best for the prosecution when it is $0$, so the prosecution could reasonably insist that a prior is not fair unless the average is at most the mean of those two extremes, which is $\approx 0.0618$.
\smallskip

We can raise this criticism in a disciplined way, for example by suggesting an axiom that expresses the above; if $x$ is the value of $p_1$ that maximizes the probability of the evidence given $\GRV = \true$, and $y$ is the value of the $p_1$ that maximizes the probability of the evidence given $\GRV = \false$, then $\int_{\vp \in \bd} p_1 \leq (x+y)/2$.

The defense can respond to the criticism, and I will explain how in Section \ref{s:response}. Doing so requires slightly strengthening the hypotheses of Claims \ref{c:lhMainClaim1} and \ref{c:lhMainClaim2}.

\subsection{\Finished{Criticism 2}}
\indent\indent The second criticism says that the prior for $\BParamsRV$ is unreasonable, with respect to measurements of beard hair widths of black men in the literature, in that it never yields a beard hair width distribution that has hairs of width greater than 187.5 micrometers. 
\smallskip
In terms of the argument, the critic should reject the (implicit) axioms that constitute the types of the symbols $\width$ and $\WidthsRV$; according to the semantics of those symbols, their types assert that all the hairs in Leighton Hay's beard and scalp had thickness at most 187.5 micrometers, which is unjustified. Formally, according to Section \ref{s:criticizingVIproofs}, one way to do this would be for the critic to suggest new definitions of $\Bins, \width$, and $\WidthsRV$. The critic can do this by suggesting new axioms (some of which are type constraints). Most importantly, the critic should suggest redefining the sort $\Bins$ as $\{\bin{1},\ldots,\bin{5}\}$, where $\bin{5}$ is a new constant. The results of that approach are discussed in the next section.

%So far, every prior I've used that allows such thick hairs actually makes the likelihood ratio {\it smaller} (better for the defense).
\medskip

\subsection{\Finished{Response to criticisms}} \label{s:response}
\indent\indent We can address both criticisms at once; if we introduce a fifth component of $\BParamsRV$ corresponding to the interval $(187.5, \infty)$, and like the first component (probability width is in $\bin{1}$) of $\BParamsRV$ constrain it to be less than the middle three components (for $\bin{2},\bin{3},\bin{4}$), then the average value of the $\bin{1}$ component of $\BParamsRV$ goes down to $< .057$. We then need to slightly strengthen the hypotheses of the two main claims, changing the parameter $.85$ in Claim \ref{c:lhMainClaim1} to $.835$ and the parameter $15$ in Claim \ref{c:lhMainClaim2} to $13.9$. Then, the proof works as before.

\subsection{\Finished{An open problem}}
\indent\indent Though I do not have such a criticism in mind, the prosecution could potentially argue that the prior for Hay's beard hair distribution is still biased, in the sense that it does not take into account everything we know about the beard hair width distributions of young black men or Hay himself, say by referring to literature such as \cite{tolgyesi} (cited in the documents submitted by expert witnesses from both sides of the trial), or by taking samples of Hay's current beard hair width distribution and somehow adjusting for the increase in width that expert witnesses said is likely, since Hay was only 19 at the time of the murder. Or they could criticize my choice of prior by claiming that it assumes {\it too much}.\footnote{Although I expect that would be a bad idea. For example, I found that if we take the prior to be the completely uniform prior over finite distributions for 5 bins, then the results are significantly worse for the prosecution.}

\indenton
Given that, an ideal proof would have the following form. We would first come up with some relation $R$ over priors for 5-bin distributions, such that $R(f)$ expresses as well as possible (given the constraint of having to complete the proof of the following proposition) that $f$ is ``fair and reasonable\quco. Then, we would find the largest constant $\alpha_0 \in (0,1)$ such that we can prove:
\begin{quote}
For any $f \in R$, if $f$ is used as the prior for the suspect's beard hair width distribution, and $\amin < \alpha_0$, then $\likeratio < 1$
\end{quote}

The same goes for Hay's scalp hair width distribution; it would be better to have a broader set of distributions that an adversary can choose from. At the very least, the argument should accommodate the possibility that Hay's scalp hairs have thinned over time, in which case we would make use of the fact that Hay is {\it not} balding (male pattern balding makes hair follicles, and the hairs they produce, gradually thinner, until the hair follicle is blocked completely).

%% file: SmokingAndCancerExample_feb2014.tex
%\providecommand{\smokingAndCancerExampleOnly}[1]{#1}
%
%\smokingAndCancerExampleOnly{
%\documentclass[12pt]{article} 
%\usepackage{dustin_fav_packages}
%\usepackage{dustin_utilities_etc}
%\usepackage{dustin_fonts}
%\usepackage{dustin_environments}
%\usepackage{dustin_macros}
%\usepackage{fullpage}
%\usepackage{../thesis_proposal/thesis_proposal_macros}
%\newcommand{\NoteToSelf}[1]{\Blue{#1}}
%\begin{document}
%\tableofcontents
%}

%Sorts  
\newcommand{\co}[1]{\overline{#1}}
\newcommand{\aPeeps}{A}
\newcommand{\bPeeps}{B}
\newcommand{\Peeps}{\text{P}}
\newcommand{\FS}[1]{\text{FS}[#1]}
\newcommand{\FnSortOp}[2]{\text{Fn}[#1,#2]}
\newcommand{\RV}[1]{\text{RV}[#1]}
\newcommand{\OutcomeDistn}{\text{StudyOutcomeDistn}}
\newcommand{\CM}{\text{CM}}
\newcommand{\BoolRV}{\text{BoolRV}}
\newcommand{\Strings}{\text{Str}}

% Fn symbols 
\newcommand{\AllCM}{\text{AllCM}}
\newcommand{\outcomes}{\text{StudyOutcomes}}
\newcommand{\fmin}[1]{\text{min}(#1)}
\newcommand{\fmax}[1]{\text{max}(#1)}

\newcommand{\sampA}{A^{\text{samp}}}
\newcommand{\sampB}{B^{\text{samp}}}
\newcommand{\LCA}{\text{LC}_{A}^{\text{samp}}}
\newcommand{\SA}{\text{S}_{A}^{\text{samp}}} 
\newcommand{\LCB}{\text{LC}_{B}^{\text{samp}}}
\newcommand{\SB}{\text{S}_{B}^{\text{samp}}}

\newcommand{\popA}{A^{\text{pop}}}
\newcommand{\popB}{B^{\text{pop}}}
\newcommand{\popLCB}{\text{LC}^{\text{pop}}_B}
\newcommand{\dpopLCB}{\text{LC}^{\text{pop}}_{B,d}}
\newcommand{\popcoLCB}{\co{\text{LC}}^{\text{pop}}_B}
\newcommand{\dpopcoLCB}{\co{\text{LC}}^{\text{pop}}_{B,d}}
\newcommand{\dpopSB}{\text{S}^{\text{pop}}_{B,d}}
\newcommand{\dpopcoSB}{\co{\dpopSB}}
\newcommand{\ipopSB}{\text{S}^{\text{pop}}_{B,i}}
\newcommand{\popcoS}{\co{\text{S}}^{\text{pop}}_B}
\newcommand{\trueoutcome}{\text{trueOutcome}}
\newcommand{\testinterval}{\text{testInterval}}
\newcommand{\maxTestIntervalSize}{\text{maxTestIntervalSize}}
\newcommand{\test}[1]{\text{test}(#1)}
\newcommand{\testp}[2]{\text{test}_{#1}(#2)}

\newcommand{\rvS}{\text{S}}
\newcommand{\rvLC}{\text{LC}}
\renewcommand{\Pr}[2]{\text{Pr}_{#1}(#2)}
\newcommand{\prCaus}[1]{\Pr{\text{caus}}{#1}}
\newcommand{\idealCausDistr}{\text{idealCausalOutcomeDistr}}

\newcommand{\BinomDist}[3]{\text{binDistr}_{#1,#2}(#3)}
\newcommand{\fbinomial}[2]{{#1 \choose #2}}
\newcommand{\probability}{[0,1]}
\newcommand{\openprobability}{(0,1)}
\newcommand{\hypergeo}[4]{\text{hyper}(#1,#2,#3,#4)}
\newcommand{\condhyper}[7]{\text{condHyper}(#1,#2,#3,#4,#5,#6,#7)}
\newcommand{\condhyperalt}[5]{{\text{condHyper}(#1,#2,#3,#4,#5)}}
\newcommand{\condbinom}[3]{{\text{condBinom}(#1,#2,#3)}}

\newcommand{\pdf}[2]{\text{outcomeDistr}(#1,#2)}
\newcommand{\cpdf}[1]{\text{pdf}(#1)}
\newcommand{\irv}[1]{1[#1]}

\newcommand{\subsets}[2]{{#1 \choose #2}}
\newcommand{\size}[1]{\vert #1 \vert}
\newcommand{\fsum}[3]{\Sigma_{#1}^{#2} #3}

\newcommand{\ShouldUse}[1]{\text{ShouldRequire}(#1)}
\newcommand{\productwarning}[1]{\text{productWarning}(#1)}
\newcommand{\predictor}[1]{\text{predictor}(#1)}
\newcommand{\dModel}{\textsf{dependModel}}
\newcommand{\iModel}{\textsf{indepModel}}

\newcommand{\Beats}[2]{\text{Beats}(#1,#2)}
\newcommand{\BeatsAll}[1]{\text{BeatsAll}(#1)}

\newcommand{\RRu}{\RR^{\bot}}
\newcommand{\NNu}{\NN^{\bot}}

%
%Defined fn symbols
%    for each X ∈ {LC_A, S_A} there is a symbol \co{X} defined by
%        \co{X} := A \ X
%    for each X ∈ {LC_B, S_B} there is a symbol \co{X} defined by
%        \co{X} := B \ X

{\bf Note: There is complete version of the proof in this chapter now, finished post-graduation, which this document has not been updated with. Contact dustin.wehr@gmail.com for details.}

%What would a good deductive argument look like for the now-well-accepted fact that cigarette smoking causes lung cancer via its delivery of carcinogenic tar to the lungs? 
In 1950 two landmark papers were published giving some of the first strong statistical evidence \textit{in the English-speaking world}\cite{smokinghistorycriticism}\footnote{In Smith and Egger's short letter to the editors of the Bulletin of the World Health Organization\cite{smokinghistorycriticism}, they give a very interesting account of how the history of this scientific progress is poorly known. In fact there were already \textit{reviews} of the literature on the connection between smoking and lung cancer as early as 1929! Even the theory of second-hand smoking is at least as old as 1928. } that tobacco smoking causes cancer, the first in the United States\cite{americansmokingstudy} and the second in England\cite{britishsmokingstudy}. Yet it was not until 1965 that cigarette packages were required to have health warnings in the United States. Michael J. Thun, in his article {\it When truth is unwelcome: the first reports on smoking and lung cancer}, argued that 15 years was much too long given the strength of those studies: %In my thesis I will compare his quantitative conclusion to mine.

\begin{quote}
In retrospect, the strength of the association in the two largest and most influential of these studies -- by Ernest Wynder \& Evarts Graham in the {\it Journal of the American Medical Association (JAMA)}\ldots and by Richard Doll \& Austin Bradford Hill (both of whom were later knighted for their work) in the {\it British Medical Journal}-- should have been sufficient to evoke a much stronger and more immediate response than the one that actually occurred. Had the methods for calculating and interpreting odds ratios been available at the time, the British study would have reported a relative risk of 14 in cigarette smokers compared with never-smokers, and the American study a relative risk of nearly 7,\footnote{Note that these relative risk calculations treat the two studies separately, whereas both versions of the argument in this chapter use the earlier study to fit a model for the later study.} too high to be dismissed as bias. \cite{thun}
\end{quote}

I will give part of an argument here that the health warnings policy was well-justified already in the early 1950s. The full argument involves introducing two more {\it candidate models} (see below),  the cigarette companies' {\it unknown genotype model} and the statistician R.A. Fisher's {\it soothing herb model}. How to refute those models is discussed in the next section. The part of the argument given here simply compares a weak version $\dModel$ of the standard, causal model, to the naive null-hypothesis model $\iModel$, which posits that smoking and lung cancer are independent. I call $\dModel$ the ``dependent-variables'' model, since it doesn't actually formalize {\it why} it predicts that smoking and cancer are dependent variables.

%For the duration of this section, pretend we are back in 1950, just after publication of the publication of the American study. Suppose, counterfactually, that the British study had not yet been performed.

This argument is an instance of the following setup: An experiment to measure some variable is designed and published, with the possible outcomes of the experiment (values of the variable) defined precisely. Sufficient time is given for all the interested parties to publish competing models for predicting the outcome of the experiment, by giving probability distributions over the set of possible outcomes. The experiment is performed. Suppose that one of the models $M$ is "overwhelmingly better" (defined in the experimental design - below, via the definition of $\Beats{\cdot}{\cdot}$ and Axiom \ref{a:beatsall}) at predicting the true outcome (or an outcome near the true one) than the others. Moreover, suppose that $M$ predicts that the use of a certain product may pose a health risk to its users; below, this is $\productwarning{M}$. Then the result of this competition must be communicated to potential users of the product. The warning can be revoked if $M$ loses in a later equally rigorous experiment competition.

The purpose of this example is, in part, to demonstrate that the requirement of deductive reasoning is not a limitation for problems in the domain I specified (Section \ref{s:problemdomain})\footnotemark, provided at least that one is firmly committed to certain ideals of persuasion.
% I said more about this example in Section \ref{sss:symbolsNotRequiring}.
\footnotetext{This example does not today meet the second criteria (contentiousness) that I listed there, but it did in the 1950s.}

Two versions of the argument are given, both dependent on mathematical claims that are unproved, but easily testable, and very likely easily resolvable by an appropriate expert (see footnote \ref{f:unprovedclaims} on page \pageref{f:unprovedclaims} about unproved purely-mathematical claims). The version in Section \ref{s:smokingcancer2} is simpler and more complete than the version in Section \ref{s:smokingcancer1}, but also weaker in that it uses a more idealized, less accurate model of the experiment.

\section{\Finished{Extensions and refinements of the argument}} \label{s:refinements}

In the argument below, the causal scientific model, which motivates the assumptions made by $\dModel$, is not made explicit. With the addition of Fisher's {\it soothing herb model} and the tobacco companies' {\it unknown genotype model} (i.e. adding those models to the set $\AllCM$), it would be necessary to make candidate models derive their experimental outcome distributions from more-qualitative assumptions. The reason is that those models are contrived to fit the data; they have outcome distributions similar to $\dModel$'s, seemingly (but not provably!) just to prevent $\dModel$ from winning on purely quantitative grounds, as it does against $\iModel$. Hence it is necessary to have a test that at least requires that a model's outcome distribution is derived from some more-readily-understandable axioms. In Fisher's model, the readily-understandable axioms essentially say that lung cancer causes smoking. In the unknown genotype model, they say that there is a common genetic cause of both lung cancer and a person's propensity to smoke tobacco. The easiest way to refute those models is to incorporate the data on female smoking and cancer, which neither model is able to explain without making them more elaborate.\footnote{Smoking became popular among men years before it became popular among women, and the lung cancer rates reflect this. The unknown genotype model could explain the earlier, smaller rates of lung cancer and smoking among women by suggesting a sex-linked genotype; however, they would not be able to explain why the rates increased so quickly. As for Fisher's {\it soothing herb model} (lung cancer causes smoking, because of the soothing effect of smoking), it would require an additional hypothesis, unrelated to the purported soothing effect, to explain why there was a delay in the increase of female lung cancer rates. } In fact, the argument could be strengthened in either or both of two ways: make models derive their experimental outcome distribution from qualitative assumptions, or make their experimental outcome distributions explain more data. I would advocate both. 

Another, more technical and subtle way of improving the argument is to elaborate the definition of $\Beats{M_1}{M_2}$ in such a way that, in effect, instantiations of the parameters of the two models $M_1$ and $M_2$ are only compared if they agree on the number of smokers in the British population. This would prevent one model from having an advantage over the other simply by having a better estimate of the total number of smokers (which, intuitively, we don't care about). Unfortunately, it would also very likely make the suitably-modified versions of Conjectures \ref{c:smokingoptimization} and \ref{c:smokingoptimization2} harder to prove.

%\TODO[{\bf Refinements:}]\\
%%simpler version of Axiom \ref{a:dModelConstraint1}. 

\section{\Finished{Proof with hypergeometric distributions contingent on an unproved mathematical claim}} \label{s:smokingcancer1}

%the American study and before publication, but after completion, of the British study. In expectation of further studies on the relationship between tobacco smoking and cancer, the authors of the British study suggest a test:

\medskip
\indentoff

{\bf Vaguely-defined sorts (in $\lav$)}
\begin{LPI}
\item $\CM$  : candidate models for the possible outcomes of the British study\cite{britishsmokingstudy}. In the current version of this argument, a candidate model $M$ is determined by $\pdf{M}{\cdot}$ and $\productwarning{M}$.
\item $\aPeeps$ : set of adult men living in the US at the time when the American study\cite{americansmokingstudy} was done.
\item $\bPeeps$ : set of adult men living in England at the time when the British study was done.
%\item $\OutcomeDistn$ : probability distributions over $\outcomes$, the possible outcomes of the british study.% determined by $\pdf{f}{\cdot}$ \\
%$\ProbDist{\outcomes}$ : probability distn over (the interpretation of) $\outcomes$
%$\StudyOutcomeDistn \ceq \ProbDist{\outcomes}$   probability distn over the possible outcomes of the britist study, determined by pdf(f,·)
\end{LPI}

{\bf Sharply-definable sorts (in $\lama$)} 
\begin{LPI}
\item $\RR$ and $\NN$ - reals and natural numbers.
\item $\RRu$ and $\NNu$ reals and naturals, but each with an extra element for ``undefined'', to serve as the range of division and subtraction. Intuitively structures should make these be supersets of $\RR$ and $\NN$, but technically all the sorts are disjoint. For readability I will not display the unary function symbols that are sometimes necessary to convert between the two sorts.
\item $\FS{\alpha}$  - finite subsets of (the interpretation of) the given sort $\alpha$. This is a sort operator, i.e. a function from sorts to sorts.
\item $\FnSortOp{\alpha}{\beta}$  - the functions from $\alpha$ to $\beta$, another sort operator.
%\item $\BoolRV[\alpha]$ : boolean random variables whose domain is the (interpretation of the) sort $\alpha$.
\item $\Strings$ - strings over the ASCII alphabet
\item $\outcomes$ (informally a ``subsort'' or ``subtype'' of $\FS{\NN}$) - the set $\{ 620,\ldots,649\}$. Before the study is done, we don't know how many of the people with lung cancer are smokers, i.e. $\size{\LCB \cap \SB}$ is unknown. The size of that set is smallest when every person without lung cancer is a smoker, and largest when every person with lung cancer is a smoker, so the set of outcomes of the study (the possible sizes of $\LCB \cap \SB$) is $\{\size{\SB} - \size{\co{\LCB}}, \ldots, \size{\LCB} \} = \{ 620, \ldots, 649\}$.
\end{LPI}

{\bf Function symbols in $\lav$} \\
In the following, a person being a ``smoker'' means that they smoked at least one cigarette per day during the most-recent period when they smoked.
\begin{LPI}
\item $\popB : \FS{\bPeeps}$ is a {\it hypothetical set}; the population that we imagine the British study samples were drawn from.
\item $\popLCB : \FS{\bPeeps} $ is the set of people in $\popB$ with lung cancer. %from which the sample of 649 was drawn.  
\item $\popcoLCB : \FS{\bPeeps}$ is the set of people in $\popB$ without lung cancer. % from which the sample of 649 was drawn.  
\item $\ipopSB : \FS{\bPeeps}$ is $\iModel$'s guess at the set of smokers in $\popB$. 
\item $\dpopSB : \FS{\bPeeps}$ is $\dModel$'s guess at the set of smokers in $\popB$.
\item $\sampA, \sampB : \FS{\aPeeps}$ is the sample of patients used in the American (resp. British) study. 
%$\sampB : \FS{\bPeeps}$ is the sample of patients used in the British study. \\
\item $\LCA, \LCB : \FS{\aPeeps}$ is the set of people in $\sampA$ (resp $\sampB$) who have lung cancer. 
%$\LCB : \FS{\aPeeps}$ is the people in $\sampB$ who have lung cancer. \\
\item $\SA, \SB : \FS{\aPeeps}$ is the set of smokers in $\sampA$ (resp $\sampB$).
%\item $\prCaus{\cdot} : \BoolRV[\bPeeps] \to \RR$  is the probability that the given random variable evaluates to true when a person is chosen uniformly at random from $\bPeeps$. 
%\item $\predictor{\cdot} : \CM \to \OutcomeDistn$ is the given candidate model's distribution over $\outcomes$.
%\item $\cpdf{\cdot} : \CM \to \outcomes \to \RR$ is the given candidate model's distribution over $\outcomes$.
\item $\pdf{\cdot}{\cdot} : \CM \times \outcomes \to \RR$ is the given candidate model's distribution over $\outcomes$.
%\item $\trueoutcome : \NN$ is the true outcome of the British study, i.e. the size of $\LCB \cap \SB$ observed when the study was actually performed. 
\item $\AllCM : \FS{\CM}$ - the set of all candidate models. It should contain a candidate model from every interested party.
\end{LPI}

{\bf Defined function symbols (in $\lad$)} 
\begin{LPI}
%\item $\outcomes : \FS{\NN} \ceq \{ \size{\SB} - \size{\co{\LCB}}, \ldots, \size{\LCB}\}$.
\item $\outcomes : \FS{\NN} \ceq \{ 620, \ldots, 649\}$.
A copy of the sort $\outcomes$ (see above for definition) that resides in the universe. So $\outcomes$ denotes both (1) a sort, and (2) an element of the universe defined to be the set that is the intended interpretation of (1). 
\item Constants for the complements of some sets:
\begin{PPI}
\item For each symbol $X \in \{\LCA, \SA\}$: \quad  $\co{X} \ceq \sampA \sdiff X$ 
\item For each symbol $X \in \{\LCB, \SB\}$: \quad  $\co{X} \ceq \sampB \sdiff X$ 
\item For each symbol $X \in \{\popLCB, \dpopSB, \ipopSB \}$: \quad  $\co{X} \ceq \popB \sdiff X$ 
\end{PPI}
\item $\Pr{x \in U}{x \in V_1 \,|\, x \in V_2} : \FS{\alpha} \times \FS{\alpha} \times \FS{\alpha} \pto \RR \ceq \size{V_1 \cap V_2 \cap U} /\size{V_2 \cap U}$ 
%\item $\testinterval \ceq \{ \size{\SB \cap \LCB} - 2, \ldots, \size{\SB \cap \LCB} + 2 \}$. This is about 17\% of the 29 possible $\outcomes$.
%\item $\maxTestIntervalSize \ceq 5$. This is about 17\% of the 29 possible $\outcomes$.
%\item $\test{M} : \CM \times \NN \pto \RR \ \ceq \ \sum_{x \in \testinterval}  \pdf{M}{x}$
%\item $\testinterval(k) : \{0,1,2\} \to \FS{\outcomes} \ceq \{ \size{\SB \cap \LCB} - k, \ldots, \size{\SB \cap \LCB} + k \}$%, and undefined if the interval of natural numbers would not be a subset of $\outcomes$.
\item For each $k\in \{0,1,2\}$: \\ $\testinterval_k : \FS{\outcomes} \ceq 
%\{ 647 - k, \ldots, 647 + k \}$
\{ \size{\SB \cap \LCB} - k, \ldots, \size{\SB \cap \LCB} + k \}$
\item For each $k\in \{0,1,2\}$: \\ $\testp{k}{M} : \CM  \to \RR \ \ceq \ \sum\limits_{x = \fmin{\testinterval_k}}^{x = \fmax{\testinterval_k}}  \pdf{M}{x}$
%\item For each $k\in \{0,1,2\}$: \\ $\testp{k}{M} : \CM \times \{0,1,2\} \to \RR \ \ceq \ \sum_{x \in \testinterval_k}  \pdf{M}{x}$
%\item $\prCaus{x \ | \ y} : \BoolRV[\bPeeps] \times \BoolRV[\bPeeps] \pto \RR \ceq \prCaus{x \cap y} / \prCaus{y}$
%\item $\bPeeps \ceq \popLCB \cup \popcoLCB$
\end{LPI}
\medskip

{\bf Predicate symbols in $\lad$} 
%\\ \NoteToSelf{Note-to-self maybe $\Beats{M_1}{M_2}$ should be vague predicate symbol, and then I would have an axiom that gives its definition?}
\begin{LPI}
%\item $\Beats{M_1{:}\CM}{M_2{:}\CM} \iff \pdf{M_1}{\size{\SB \cap \LCB}} > 10,000 \cdot \pdf{M_2}{\size{\SB \cap \LCB}}$
\item $\Beats{M_1{:}\CM}{M_2{:}\CM} \iff \bigwedge_{k \in \{0,1,2\}} \testp{k}{M_1} > 1000 \cdot \testp{k}{M_2}$. Model $M_1$ beats model $M_2$ if it assigns much higher probability to the true outcome $\size{\SB \cap \LCB}$, as well as to the intervals of size 3 and 5 around the true outcome. The interval of size 5 is about 17\% of $\outcomes$, and any larger interval would be biased since the interval of size 5 already contains the maximum of $\outcomes$.
\item $\BeatsAll{M_1{:}\CM} \iff \forall M_2{:}\CM. (M_2 \in \AllCM \wedge M_1 \neq M_2) \implies \Beats{M_1}{M_2}$ simply says that $M_1$ beats all the other models in $\AllCM$.
\end{LPI}
\medskip

{\bf Function symbols in $\lama$}%\footnote{I've left out numerals} 
\begin{LPI}
\item $\{ x, \ldots, y\} : \NN \times \NN \to \FS{\NN}$ is the set of naturals from $x$ to $y$ inclusive, or the empty set if $x > y$.
\item $+, \cdot : \NN \times \NN \to \NN $ are addition and multiplication for $\NN$.
\item $+, \cdot : \RR \times \RR \to \RR $ addition and multiplication for $\RR$. These symbols are distinct from the ones for $\NN$, but display in the same way.
\item $ - : \NN \times \NN \to \NNu$ is subtraction, but undefined if the result is negative.
\item $ / : \RR \times \RR \to \RRu$ is division, undefined when the second argument is 0.
\item $\textsf{apply} : \FnSortOp{\NN}{\RR} \times \NN \to \RR$ is the application of a function object to an argument. There are versions of this symbol for a few other types as well. 
\item $\fsum{x=t_1}{t_2}{t_3(x)} : \NN \times \NN \times \FnSortOp{\NN}{\RR} \to \RR$ is the usual summation binder symbol. The formal syntax is $\Sigma(t_1,t_2,\lambda x{:}\NN. t_3)$, where ``$\lambda x{:}\NN. t_3$'' is actually just the name of a constant symbol of sort $\FnSortOp{\NN}{\RR}$ that is implicitly defined in terms of the open term $t_3$ and the function symbol $\textsf{apply}$, but I've hidden those definitions from this writeup for readability.
\item $\cap : \FS{\alpha} \times \FS{\alpha} \to \FS{\alpha}$ is set intersection. As with the other function symbols in this list whose type is presented with a sort variable $\alpha$, there are multiple distinct function symbols, for various instantiations of $\alpha$, that each display as $\cap$. In the HTML presentation of interpreted formal proofs, one can disambiguate the symbol by hovering over it to see its type. %, \ \BoolRV[\alpha] \times \BoolRV[\alpha] \to \BoolRV[\alpha]$ 
\item $\sdiff : \FS{\alpha} \times \FS{\alpha} \to \FS{\alpha}$ is set difference. %, \ \BoolRV[\alpha] \times \BoolRV[\alpha] \to \BoolRV[\alpha]$ 
\item $\size{\cdot} : \FS{\alpha} \to \NN$ is the size of the given finite subset of (the interpretation of) $\alpha$.
\item $\subsets{X}{k} : \FS{\alpha} \times \NN \to \FS{\FS{\alpha}}$ is the set of subsets of $X$ of size $k$. 
\item $\fmin{\cdot}, \fmax{\cdot} : \FS{\NN} \to \NNu$ are the minimum and maximum elements of a finite set of naturals. Undefined if the set is empty.
\item $\hypergeo{k}{s}{N}{s'} : \NN \times \NN \times \NN \times \NN \to \RRu$ is the hypergeometric distribution (in the last argument; the other three arguments are parameters), defined when $s' \leq s \leq N, s \leq k \leq N$; if a population of size $N$ has $s$ smokers and $N-s$ nonsmokers, and $k$ people are chosen uniformly at random {\it without} replacement from the population, then $\hypergeo{k}{s}{N}{s'}$ is the probability that the resulting set contains exactly $s'$ smokers.
%\item $\condhyper{s''}{k}{s_1}{s_2}{X_1}{X_2}{s'} : \NN \times \NN \times \NN \times \NN \times \FS{\bPeeps} \times \FS{\bPeeps} \times \NN \pto \RR$ is a probability distribution, defined when $s' \leq s_1 \leq s'' \leq N$,\, $s_1 \leq \size{X_1}$,\, $s_2 \leq \size{X_2}$.  Suppose we have disjoint sets of people $X_1$ and $X_2$ , with $X_1$ having $s_1$ smokers and $X_2$ having $s_2$ smokers. Uniformly at random we choose size-$k$ subsets $Y_1$ and $Y_2$ of $X_1$ and $X_2$. Then $\condhyper{s''}{k}{s_1}{s_2}{X_1}{X_2}{s'}$ is the conditional probability that $Y_1$ contains exactly $s'$ smokers, given that there are $s''$ smokers in $Y_1 \cup Y_2$.
\item $\condhyperalt{s_1}{s_2}{X_1}{X_2}{s_1'} : \NN \times \NN \times \FS{\bPeeps} \times \FS{\bPeeps} \times \NN \to \RRu$ is a probability distribution (in the last argument; the other four arguments are parameters), defined when $s_1' \leq s_1 \leq \size{\SB} \leq N$,\, $s_1 \leq \size{X_1}$,\, $s_2 \leq \size{X_2}$.  Suppose we have disjoint sets of people $X_1$ and $X_2$ , with $X_1$ having $s_1$ smokers and $X_2$ having $s_2$ smokers. Uniformly at random we choose size-$\size{\LCB}$ subsets $X_1'$ of $X_1$ and $X_2'$ of $X_2$. Then $\condhyperalt{s_1}{s_2}{X_1}{X_2}{s_1'}$ is the conditional probability that $X_1'$ contains exactly $s_1'$ smokers, given that there are $\size{\SB}$ smokers in $X_1' \cup X_2'$.
\end{LPI}

\bigskip

\begin{simplifyingassumption}
We would change this to a normal Assumption if we included formalizations of Fisher's and the tobacco companies' models also (see section \ref{s:refinements} above).
\[ \AllCM = \{ \iModel, \dModel \}\]
\end{simplifyingassumption}

\begin{viqdefn} Sizes of sets from the American study.
\[ \begin{array}{lll}
&\size{\co{\LCA}}  &= 780 \quad \text{patients in sample with conditions other than cancer}\\
&\size{\LCA}  &= 605 \quad \text{patients in sample with lung cancer} \\
&\size{\co{\SA} \cap \co{\LCA}} &= 114 \quad \text{nonsmokers with conditions other than cancer} \\
&\size{\co{\SA} \cap \LCA} &= 8 \quad \text{nonsmokers with lung cancer}  
\end{array}
\]
\end{viqdefn}

\begin{viqdefn} \label{a:productwarnings} \daiskip 
$\productwarning{\dModel} =$ {\it ``Scientific studies have found a correlation between tobacco smoking and lung cancer that is currently best-explained by the hypothesis that smoking causes an increase in the probability that any person will get lung cancer.''}  \\

$\productwarning{\iModel} =$ ``''   (the empty string)
\end{viqdefn}

\begin{viqdefn} This gives the sizes of the sample sets, and certain subsets of those sets, from the British study. %Candidate models are expected to use $\size{\LCB}, \size{\co{\LCB}}$, and $\size{\SB}$ to define their distribution over $\outcomes$. 
We evaluate the different models on how well they predict the size of $\LCB \cap \SB$, given the sizes of $\LCB, \co{\LCB},$ and $\SB$. A model predicts the size well if its distribution over $\outcomes$ assigns high probability to $\size{\LCB \cap \SB}$ or some close number; this is formalized in the definition of $\Beats{\cdot}{\cdot}$.
\[\begin{array}{lll}
&\size{\LCB} = \size{\co{\LCB}} &= 649  \\
&\size{\SB} &= 1269  \\ %\footnote{That's fine ---> This wouldn't be available until after the British study was completed, so I will have to make some decision about the whether or not to phrase the context of the proof as happening before or after} %\TODO[\text{This wouldn't be available until after the British study was completed}]
& \size{\LCB \cap \SB} &= 647 \\
& \size{\co{\LCB} \cap \SB} &= 622 
\end{array} 
\]
\end{viqdefn}

\begin{simplifyingassumption}[$\dModel$ posits a hypergeometric distribution] \label{a:dhypergeo}
Note that the values of the four parameters in \\ $\condhyperalt{\size{\popLCB \cap \dpopSB}}{\size{\popcoLCB \cap \dpopSB}}{\popLCB}{\popcoLCB}{\cdot}$ are only bounded by the other axioms, especially Assumptions (\ref{a:bothModelsConstraint1}), (\ref{a:bothModelsConstraint2}), (\ref{a:dModelConstraint1}), and (\ref{a:dModelConstraint2}), with the latter two distinguishing $\dModel$'s distribution from $\iModel$'s. Still, this and \simpassump\, \ref{a:ihypergeo} are the weakest of the axioms with respect to the standards of accuracy that I strive for. Unlike the others, we cannot seriously claim that this axiom is literally true with respect to the informal intended semantics given by the language interpretation guide, simply because the authors of the British study did not methodically randomize the way that they chose their sample sets of men with and without lung cancer. I would be satisfied to have an axiom that says $\pdf{\dModel}{\cdot}$ is ``close enough'' to a hypergeometric distribution, but I have not yet investigated suitable ways of formalizing ``close enough,'' and it is not clear that there would be a benefit in pedagogy or cogency that warrants the added complexity.
\[
\begin{array}{rl}
& \forall s{:}\outcomes.\ \pdf{\dModel}{s}  \\
%= & \condhyper{\size{\SB}}{\size{\LCB}}{\size{\popLCB \cap \dpopSB}}{\size{\popcoLCB \cap \dpopSB}}{\popLCB}{\popcoLCB}{s'}
= & \condhyperalt{\size{\popLCB \cap \dpopSB}}{\size{\popcoLCB \cap \dpopSB}}{\popLCB}{\popcoLCB}{s}
\end{array}
\]
\end{simplifyingassumption}

\begin{simplifyingassumption}[$\iModel$ posits a hypergeometric distribution]  \label{a:ihypergeo}
Note that the values of the four parameters in \\ 
$\condhyperalt{\size{\popLCB \cap \ipopSB}}{\size{\popcoLCB \cap \ipopSB}}{\popLCB}{\popcoLCB}{\cdot}$ are only constrainted by the other axioms, especially Assumptions (\ref{a:bothModelsConstraint1}), (\ref{a:bothModelsConstraint2}), and (\ref{a:iModelConstraint}), with Assumption (\ref{a:iModelConstraint}) distinguishing $\iModel$'s distribution from $\dModel$'s. 
\[
\begin{array}{rl}
& \forall s{:}\outcomes.\ \pdf{\iModel}{s}  \\
%= & \condhyper{\size{\SB}}{\size{\LCB}}{\size{\popLCB \cap \ipopSB}}{\size{\popcoLCB \cap \ipopSB}}{\popLCB}{\popcoLCB}{s'}
= & \condhyperalt{\size{\popLCB \cap \ipopSB}}{\size{\co{\popLCB} \cap \ipopSB}}{\popLCB}{\co{\popLCB}}{s}
\end{array}
\]
\end{simplifyingassumption}

\begin{assumption} \label{a:bothModelsConstraint1} This is a conservative axiom for $\dModel$; a figure from the British study says that the rate of lung cancer in men was $10.6$ per $100,000$ in 1936-1939, and population data for England in 1951 puts the population at about 38.7 million, hence even if the population from which the British sample was drawn is taken to be the entire nation, if we assume about half the population was male, and that the rate at most trippled from 1939 to 1950, then we should expect at most 6100 men with lung cancer.
\[\size{\popLCB} \leq 7000\]
\end{assumption}

\begin{assumption} \label{a:bothModelsConstraint2}
This is a conservative axiom for $\dModel$; it says that of the hospital patients from which the British scientists drew their sample, at most 1 in 6 had lung cancer (in reality it would have been significantly lower).
\[\size{\popcoLCB} \geq 5*\size{\popLCB}\]
\end{assumption} 

\begin{assumption} \label{a:dModelConstraint1}
Consider the ratio of probabilities of being a British nonsmoker given that you have lung cancer vs. given that you have a hospitalizable illness other than lung cancer. This axiom says that it is not much smaller than the corresponding ratio seen in the American study sample, specifically not more than 3 times smaller.
If we were to define a best-guess version of the dependent-variables model $\dModel$, we would set the (unknown) left side of the below inequality equal to the (known) right side (and similarly for Assumption (\ref{a:dModelConstraint2})), in effect positing that the correlation between smoking and lung cancer in the British population is identical to the correlation in the American sample.  However, the evidence is so strongly in favor of $\dModel$ that this much weaker assumption suffices:
\[ 
\frac{\Pr{x \in \popB}{x \in \co{\dpopSB} \ | \ x \in \popLCB}}
{\Pr{x \in \popB}{x \in \co{\dpopSB} \ | \ x \in \co{\popLCB}}} 
\leq 3 \cdot
\frac{\Pr{x \in \sampA}{x \in \co{\SA} \ | \ x \in \LCA}}
{\Pr{x \in \sampA}{x \in \co{\SA} \ | \ x \in \co{\LCA}}} \quad \footnote{$= 3 (.0132231 / .146154) \approx .27142$ }
\]
% Pr_{B^pop}(¬S_B^{d,pop} | LC_B^pop)               Pr_{A^samp}(¬S_A | LC_A)
%----------------------------------  ≥  (1/5) ----------------------------
% Pr_{B^pop}(¬S_B^{d,pop} | ¬LC_B^pop)              Pr_{A^samp}(¬S_A | ¬LC_A)
\end{assumption}

\begin{assumption} \label{a:dModelConstraint2}
Same comment as in Assumption \ref{a:dModelConstraint1} applies here.
\[ \Pr{x \in \popB}{x \in \co{\dpopSB} \ | \ x \in \co{\popLCB}} 
\geq 1/3 \cdot
 \Pr{x \in \sampA}{x \in \co{\SA} \ | \ x \in \co{\LCA}} \quad \footnote{ $= (1/3) .146154 = .048718$ }\]
\end{assumption}

\begin{assumption}  \label{a:iModelConstraint}
The independent-variables model simply posits that, in the population from which the British sample was drawn, the fraction of smokers among people with lung cancer is the same as the fraction of smokers among people with illnesses other than lung cancer.
%\[ \left| \frac{ \size{  \ipopSB \cap \co{\popLCB}} }{ \size{\co{\popLCB}} } \right| \]
\[ \Pr{x \in \popB}{x \in \ipopSB \ | \ x \in \co{\popLCB}} 
=
 \Pr{x \in \popB}{x \in \ipopSB \ | \ x \in \popLCB} \]
\end{assumption}

%\NoteToSelf{Note to self: maybe $\AllCM$ and $\ShouldUse{\cdot}$ should have a date parameter.}
The next assumption states the intended consequence of one model beating all the others.
\begin{assumption} \label{a:beatsall}
$\forall M{:}\CM.\ \BeatsAll{M} \implies \ShouldUse{\productwarning{M}}$
\end{assumption}

\begin{claim} \label{as:equivalenceOfDistributions}
\[ 
%\condhyper{s''}{k}{s_1}{s_2}{X_1}{X_2}{s'} 
\condhyperalt{s_1}{s_2}{X_1}{X_2}{s_1'} 
\]
\text{equals}
\[
 \frac{ \hypergeo{\size{\LCB}}{s_1}{\size{X_1}}{s_1'} \cdot \hypergeo{\size{\co{\LCB}}}{s_1}{\size{X_2}}{\size{\SB} - s_1'}}
{\sum\limits_{x = \fmin{\outcomes}}^{\fmax{\outcomes}} \hypergeo{\size{\LCB}}{s_1}{\size{X_1}}{x} \cdot \hypergeo{\size{\co{\LCB}}}{s_1}{\size{X_2}}{\size{\SB} - x} 
} 
 \]
\end{claim}
\begin{proof}
This is a standard definition of the conditional hypergeometric distribution. An informal proof is easy from the informal semantics given for $\condhyperalt{\cdot}{\cdot}{\cdot}{\cdot}{\cdot}$. Note that we could alternatively have made $\condhyperalt{\cdot}{\cdot}{\cdot}{\cdot}{\cdot}$ a defined function symbol.
\end{proof}

The above axioms, together with some basic mathematical axioms, prove that for any setting of the free parameters $\size{\popLCB}, \size{\co{\popLCB}}, \size{\ipopSB \cap \popLCB}, \size{\dpopSB \cap \popLCB}$, etc that obeys the constraints given by Axioms (\ref{a:bothModelsConstraint1})-(\ref{a:iModelConstraint}), the dependent-variables model decisively beats the independent-variables model:
\begin{conjecture} \label{c:smokingoptimization}
%\[ \frac{\pdf{\dModel}{\size{\SB \cap \LCB}}}{\pdf{\iModel}{\size{\SB \cap \LCB}}} > 5,000 \]
\[ \bigwedge_{k \in \{0,1,2\}} \testp{k}{\dModel} > 5000 \cdot \testp{k}{\iModel} \]
\end{conjecture}

%\begin{proof}
%It is easy to see (ACTUALLY NO it's not, since those two constraints are coupled) that the quantity on the left hand side is minimized when:
%\begin{LPPI}
%\item the constraint in Axiom (\ref{a:dModelConstraint1}) is an equality. Informally, we maximize the ratio 
%\[ \frac{\text{probability British person is nonsmoker given they have lung cancer}}{\text{probability British person is nonsmoker given they have a condition other than lung cancer}} \]
%subject to the constraint that it is no more than 3 times the analagous ratio that was observed in the American study.
%\end{LPPI}
%We also need the less-obvious facts that the left hand side is minimized when:
%\begin{LPPI}
%\item the constraint in Axiom (\ref{a:dModelConstraint2}) is an equality. Informally, we minimize (probability of a British person being a nonsmoker given that they have a condition other than lung cancer) subject to the constraint that it is not more than a factor of 3 smaller than the analagous frequency observed in the American study. 
%\item the size of $\popLCB$ is maximized subject to Axiom (\ref{a:bothModelsConstraint1}), hence is 7000.
%\item the size of $\popcoLCB$ is minimized subject to Axioms (\ref{a:bothModelsConstraint2}) and (\ref{a:bothModelsConstraint1}), hence is 35000.
%\end{LPPI}
%\end{proof}

From Conjecture \ref{c:smokingoptimization}, the {\bf goal sentence} follows:
\[ \ShouldUse{\productwarning{\dModel}} \]

\section{\Finished{Simpler, more-easily completable proof}} \label{s:smokingcancer2}

\newcommand{\prSdgivenLC}{p_{\rvS_d | \rvLC}}
\newcommand{\prSdgivenNotLC}{p_{\rvS_d | \co{\rvLC}}}
\newcommand{\prSi}{p_{\rvS_i | *}}

We add a symbol for the binomial distribution. 
\[ \BinomDist{\cdot}{\cdot}{\cdot} : \probability \times \NN \times \NN \to \NNu \] 
We may either give $\BinomDist{\cdot}{\cdot}{\cdot}$ a prose definition, and then state the next sentence as a Claim, or we could make $\BinomDist{\cdot}{\cdot}{\cdot}$ a defined function symbol defined by the next sentence. Either way is consistent with the definition of {\it interpreted formal proof}. In the proof from the previous section, the former option was taken. In this section I will leave it ambiguous.
 \[ \forall p{:}\openprobability. \forall n,t{:}\NN.\ (0 \leq t \leq n)  \implies  \BinomDist{p}{n}{t} = {n \choose t} p^t (1-p)^{n-t}\] 
We introduce a family of probability distributions that takes the place of $\condhyperalt{\cdot}{\cdot}{\cdot}{\cdot}{\cdot}$. We give it the following prose definition, and the later two axioms, Claims (\ref{as:equivalenceOfDistributions2}) and (\ref{as:equivalenceOfDistributions3}), are made only for the purpose of calculation.
\begin{LPPI}
\item $\condbinom{p_1}{p_2}{s_1'} : \probability \times \probability \times \NN \to \RR$ is a probability distribution (in the last argument; the other two arguments are parameters). Suppose we sample (with replacement) $\size{\LCB}$ times from each of two binomial distribution, the first having success probability $p_1$ and the second having success probability $p_2$. Then $\condbinom{p_1}{p_2}{s_1'}$ is the conditional probability that we get $s_1'$ successes from the first distribution {\it given} that the sum of successes is $\size{\SB}$.\footnote{Note that this family of distributions is usually given with $\size{\LCB}$ and $\size{\SB}$ as parameters}
\end{LPPI}
We also introduce three new constants $\prSdgivenLC, \prSdgivenNotLC, $ and $\prSi$ of type $\probability$. $\prSdgivenLC$ and $\prSdgivenLC$ are $\dModel$'s estimates of the fraction of smokers in the lung cancer population and in the population of people with conditions other than lung cancer. $\prSi$ is $\iModel$'s estimate of the fraction of smokers in both populations.

\medskip

We drop \simpassump{s}\, (\ref{a:dhypergeo}) and (\ref{a:ihypergeo}), replacing them with the following two:

\begin{simplifyingassumption}[$\dModel$ posits a binomial distribution] \label{a:dbinom}
Note that the values of the two parameters (first two arguments) of $\condbinom{\cdot}{\cdot}{\cdot}$ are only bounded by the axioms from the proof in the previous section and Assumptions (\ref{a:dBinomModelConstraint1}), and (\ref{a:dBinomModelConstraint2}), with the latter two distinguishing $\dModel$'s distribution from $\iModel$'s. 
The remainder of this paragraph (i.e. language interpretation guide entry) is essentially the same as in the description of Simplifying Assumption (\ref{a:dhypergeo}).
This and \simpassump\, (\ref{a:ibinom}) are the weakest of the axioms with respect to the standards of accuracy that I strive for. Unlike the others, we cannot seriously claim that this axiom is literally true with respect to the informal intended semantics, simply because the authors of the British study did not methodically randomize the way that they chose their sample sets of men with and without lung cancer. I would be satisfied to have an axiom that says $\pdf{\dModel}{\cdot}$ is ``close enough'' to a binomial distribution, but I have not yet investigated suitable ways of formalizing ``close enough,'' and it is not clear that there would be a benefit in pedagogy or cogency that warrants the added complexity.
\[
\begin{array}{rl}
%& \exists! p_1,p_2{:}\probability.\ \forall s{:}\outcomes.\  \\
%%= & \condhyper{\size{\SB}}{\size{\LCB}}{\size{\popLCB \cap \dpopSB}}{\size{\popcoLCB \cap \dpopSB}}{\popLCB}{\popcoLCB}{s'}
%& \qquad \pdf{\dModel}{s}  = \condbinom{p_1}{p_2}{s}
& \forall s{:}\outcomes.\  \\
%= & \condhyper{\size{\SB}}{\size{\LCB}}{\size{\popLCB \cap \dpopSB}}{\size{\popcoLCB \cap \dpopSB}}{\popLCB}{\popcoLCB}{s'}
& \qquad \pdf{\dModel}{s}  = \condbinom{\prSdgivenLC}{\prSdgivenNotLC}{s}
\end{array}
\]
\end{simplifyingassumption}

\begin{simplifyingassumption}[$\iModel$ posits a binomial distribution] \label{a:ibinom}
\[
\begin{array}{rl}
%& \exists! p{:}\probability.\ \forall s{:}\outcomes.\  \\
%%= & \condhyper{\size{\SB}}{\size{\LCB}}{\size{\popLCB \cap \dpopSB}}{\size{\popcoLCB \cap \dpopSB}}{\popLCB}{\popcoLCB}{s'}
%& \qquad \pdf{\iModel}{s}  = \condbinom{\prSi}{\prSi}{s}
&\forall s{:}\outcomes.\  \\
%= & \condhyper{\size{\SB}}{\size{\LCB}}{\size{\popLCB \cap \dpopSB}}{\size{\popcoLCB \cap \dpopSB}}{\popLCB}{\popcoLCB}{s'}
& \qquad \pdf{\iModel}{s}  = \condbinom{\prSi}{\prSi}{s}
\end{array}
\]
\end{simplifyingassumption}

The next two axioms bound the frequencies mentioned in the previous two axioms. The description of Axiom (\ref{a:dModelConstraint1}) in the previous section has some motivation that applies here as well. 
\begin{assumption} \label{a:dBinomModelConstraint1}
\[
\frac{1}{2} \cdot  \Pr{x \in \sampA}{x \in \co{\SA} \ | \ x \in \co{\LCA}} \leq 1 - \prSdgivenNotLC \leq  2 \cdot \Pr{x \in \sampA}{x \in \co{\SA} \ | \ x \in \co{\LCA}}
\]
\end{assumption}

\begin{assumption} \label{a:dBinomModelConstraint2}
\[
\frac{1}{2} \cdot  \Pr{x \in \sampA}{x \in \co{\SA} \ | \ x \in \LCA} \leq 1 - \prSdgivenLC \leq  2 \cdot \Pr{x \in \sampA}{x \in \co{\SA} \ | \ x \in \LCA}
\]
\end{assumption}

\medskip

The next two axioms tell us how to compute the distributions
\begin{claim} \label{as:equivalenceOfDistributions2}
For $n = \size{\LCB}$ (and recall $\size{\LCB} = \size{\co{\LCB}}$) have
\[ 
%\condhyper{s''}{k}{s_1}{s_2}{X_1}{X_2}{s'} 
\condbinom{p_1}{p_2}{a} 
\]
\text{equals}
\[
 \frac{ \BinomDist{p_1}{n}{a} \cdot \BinomDist{p_2}{n}{\size{\SB} - a}}
{\sum\limits_{x = \fmin{\outcomes}}^{\fmax{\outcomes}} \BinomDist{p_1}{n}{x} \cdot \BinomDist{p_2}{n}{\size{\SB} - x}
} 
 \]
\end{claim}
\begin{proof}
This is a standard symbolic expression for the conditional binomial distribution. Note that we could alternatively have made $\condbinom{\cdot}{\cdot}{\cdot}$ a defined function symbol.
\end{proof}

\begin{claim} \label{as:equivalenceOfDistributions3}
For $n = \size{\LCB}$ (and recall $\size{\LCB} = \size{\co{\LCB}}$) have
\[ 
\condbinom{p}{p}{a}  = \frac{\fbinomial{n}{a} \cdot \fbinomial{n}{ \size{\SB} - a}}{ \fbinomial{2n}{\size{\SB}} } 
\]
\end{claim}
\begin{proof}
This is a well known fact, but I will provide the proof since it is short.

Let $n = \size{\LCB}$ and $j = \size{SB}$. Referring to its prose definition, $\condbinom{p}{p}{a}$ equals the probability of getting $a$ successes when sampling $n$ times from a distribution with success probability $p$, given that you've also sampled $n$ times from another distribution with success probability $p$ and the total of the two success counts is $j$. Equivalently, $\condbinom{p}{p}{a}$ is
\[
\frac
{\Pr{}{\text{get $a$ out of $n$ successes from first distribution and $j-a$ out of $n$ from second}}}
{\Pr{}{\text{get total of $j$ successes out of $2n$}}}
\]
Because the two distributions are sampled from independently, that equals
\[
\frac
{\Pr{}{\text{get $a$ out of $n$ successes from first}} \ \Pr{}{\text{get $j-a$ out of $n$ from second}}}
{\Pr{}{\text{get total of $j$ successes out of $2n$}}}
\]
i.e. 
\[ 
\frac{\BinomDist{p}{n}{a} \ \BinomDist{p}{n}{j-a}}{\BinomDist{p}{2n}{j}}
\]
Expanding the definition of \text{binDistr} that is:
\[ 
\frac{ {n \choose a} p^a (1-p)^{n-a} \  {n \choose j-a} p^{j-a} (1-p)^{n-j+a}}
{ {2n \choose j} p^j (1-p)^{2n-j}} = 
\frac{ {n \choose a} \  {n \choose j-a}}
{ {2n \choose j}}
\]
%, that is are expanded, one finds that the three terms of the form  p^t (1-p)^{n-t}in the numerator cancel with the terms in the denominator except for the binomial coefficients
\end{proof}

The following Conjecture should be easier to prove than Conjecture \ref{c:smokingoptimization}, but it is still a tough (for a non-specialist) constrained nonlinear optimization problem.
\begin{conjecture} \label{c:smokingoptimization2}
%\[ \frac{\pdf{\dModel}{\size{\SB \cap \LCB}}}{\pdf{\iModel}{\size{\SB \cap \LCB}}} > 5,000 \]
\[ \text{For $k$ in $\{0,1,2\}$ have } \testp{k}{\dModel} > 2000 \cdot \testp{k}{\iModel} \]
\end{conjecture}
\textit{Informal argument in support:}\\
The independent variables model has no parameters, so $\testp{k}{\iModel}$ is a constant for each $k \in \{0,1,2\}$. For each $k \in \{0,1,2\}$, viewing the 3D plot of $\testp{k}{\dModel}$ as a function of the parameters $\prSdgivenLC$ and $\prSdgivenNotLC$ (pictured below), if we make the very plausible assumption that there are no sharp extrema missed by the plot, then it is clear that within the range allowed by Axioms (\ref{a:dBinomModelConstraint1}) and (\ref{a:dBinomModelConstraint2}), the function is minimized at one of the corner points. In fact each of the three functions is minimized when $\prSdgivenLC$ is minimal and $\prSdgivenNotLC$ is maximal. This is because the maximum likelihood model for the American data slightly overestimates the correlation between smoking and lung cancer in the British data. The minimums tell us that for any setting of the parameters of the $\dModel$ that obeys the axioms, for $\testp{0}{\cdot}, \testp{1}{\cdot}$, and $\testp{2}{\cdot}$ it gives probability more than $12,000, 5,000$, and $2,000$ times higher, respectively, than $\iModel$.  

%For $\testp{0}{\dModel}$ and $\testp{1}{\dModel}$ it is minimized when $\prSdgivenLC$ is maximal and $\prSdgivenNotLC$ is minimal; for $\testp{2}{\dModel}$ the minimum is at the opposite point, when $\prSdgivenLC$ is minimal and $\prSdgivenNotLC$ is maximal.\footnote{The reason for the switch of locations of the minimum is technical: The maximum likelihood model for the American data slightly overestimates the correlation between smoking and lung cancer in the British data, which can be used to explain the minimum for $\testp{0}{\dModel}$. However, when we broaden the test interval enough, so that we accept all numbers between 645 and 649 as equally-good predictions of the number of smokers among the lung cancer patients, then all the possible overestimating models (i.e. with means greater than 647) are very close to being as good as the model that maximizes the likelihood of the British data, which results in the worst model being the one that maximally {\it underestimates} the correlation between smoking and lung cancer.}
%
\IMG{
\begin{figure}[h!] \label{smokingpic1}
  \caption{Optimization problems for $k=1,2,3$}
  \centering
\includegraphics[scale=1.2]{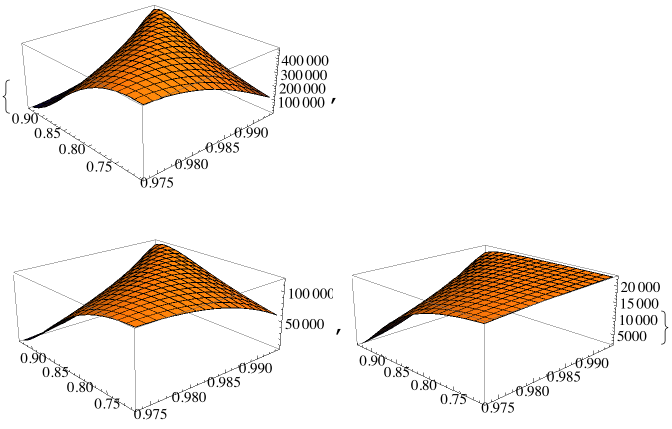}
\end{figure}
}

%% file: thesis.bbl
\newcommand{\etalchar}[1]{$^{#1}$}
\begin{thebibliography}{BCD{\etalchar{+}}11}

\bibitem[AFKP13]{KidneyExchange}
Itai Ashlagi, Felix Fischer, Ian~A Kash, and Ariel~D Procaccia.
\newblock Mix and match: A strategyproof mechanism for multi-hospital kidney
  exchange.
\newblock {\em Games and Economic Behavior}, 2013.

\bibitem[BCD{\etalchar{+}}11]{cvc4}
Clark Barrett, Christopher~L Conway, Morgan Deters, Liana Hadarean, Dejan
  Jovanovi{\'c}, Tim King, Andrew Reynolds, and Cesare Tinelli.
\newblock {CVC}4.
\newblock In {\em Computer aided verification}, pages 171--177. Springer, 2011.

\bibitem[BdRV02]{blackburn}
P.~Blackburn, M.~de~Rijke, and Y.~Venema.
\newblock {\em Modal Logic}.
\newblock Cambridge Tracts in Theoretical Computer Science. Cambridge
  University Press, 2002.

\bibitem[BHO75]{SexBiasBerkeley}
P.~J. Bickel, E.~A. Hammel, and J.~W. O'Connell.
\newblock Sex bias in graduate admissions: Data from {B}erkeley.
\newblock {\em {Science}}, 187(4175):398--404, 1975.

\bibitem[BS13]{knowabilitysep}
Berit Brogaard and Joe Salerno.
\newblock {F}itch's paradox of knowability.
\newblock In Edward~N. Zalta, editor, {\em The Stanford Encyclopedia of
  Philosophy}. Winter 2013 edition, 2013.

\bibitem[CC14]{knowability}
Massimiliano Carrara and Daniele Chiffi.
\newblock The knowability paradox in the light of a logic for pragmatics.
\newblock In {\em Recent Trends in Philosophical Logic}, pages 31--46.
  Springer, 2014.

\bibitem[DF99]{DonnellyDNA}
Peter Donnelly and Richard~D Friedman.
\newblock {DNA} database searches and the legal consumption of scientific
  evidence.
\newblock {\em Michigan Law Review}, pages 931--984, 1999.

\bibitem[DH50]{britishsmokingstudy}
Richard Doll and AB~Hill.
\newblock Smoking and carcinoma of the lung (reprint).
\newblock {\em Bulletin of the World Health Organization}, 77(1):84--93, 1999
  (original 1950).

\bibitem[Don05]{DonnellyRvAdams}
Peter Donnelly.
\newblock Appealing statistics.
\newblock {\em Significance}, 2(1):46--48, 2005.

\bibitem[Efr05]{BayesiansFrequentistsScientists}
Bradley Efron.
\newblock Bayesians, frequentists, and scientists.
\newblock {\em Journal of the American Statistical Association}, 100(469):1--5,
  2005.

\bibitem[F{\etalchar{+}}11]{BayesianPublicPolicy}
Stephen~E Fienberg et~al.
\newblock Bayesian models and methods in public policy and government settings.
\newblock {\em Statistical Science}, 26(2):212--226, 2011.

\bibitem[Far93]{farmer-simple}
William~M. Farmer.
\newblock A simple type theory with partial functions and subtypes.
\newblock {\em Annals of Pure and Applied Logic}, 64(3):211--240, November
  1993.

\bibitem[Gor88]{ImportanceOfNonmonotonicity}
Thomas~F. Gordon.
\newblock The importance of nonmonotonicity for legal reasoning.
\newblock In F.~Haft H.~Fiedler and R.~Traunm\"{u}ller, editors, {\em Expert
  systems in law: impacts on legal theory and computer law}, Neue Methoden im
  Recht. Attempto Verlag, 1988.

\bibitem[Gor13]{CarneadesWebapp}
Thomas~F Gordon.
\newblock Introducing the {C}arneades web application.
\newblock In {\em Proceedings of the Fourteenth International Conference on
  Artificial Intelligence and Law}, pages 243--244. ACM, 2013.

\bibitem[Kad08]{StatisticsInTheLaw}
Joseph~B. Kadane.
\newblock {\em Statistics in the Law : A Practitioner's Guide, Cases, and
  Materials}.
\newblock Oxford University Press, USA, 2008.

\bibitem[KV13]{vampire}
Laura Kov{\'a}cs and Andrei Voronkov.
\newblock First-order theorem proving and {V}ampire.
\newblock In {\em Computer Aided Verification}, pages 1--35. Springer, 2013.

\bibitem[Kva95]{kvanvig1995}
Jonathan Kvanvig.
\newblock The knowability paradox and the prospects for anti-realism.
\newblock {\em No{\^u}s}, pages 481--500, 1995.

\bibitem[LCMJ10]{truthmachine}
M.~Lynch, S.A. Cole, R.~McNally, and K.~Jordan.
\newblock {\em Truth Machine: The Contentious History of {DNA} Fingerprinting}.
\newblock University of Chicago Press, 2010.

\bibitem[LCS11]{leibniz}
G.W. Leibniz, S.~Charlotte, and L.~Strickland.
\newblock {\em Leibniz and the Two {S}ophies: The Philosophical
  Correspondence}.
\newblock Other voice in early modern Europe: Toronto series. Iter
  Incorporated, 2011.

\bibitem[LL76]{leibniz1976}
G.W. Leibniz and L.E. Loemker.
\newblock {\em Philosophical Papers and Letters}.
\newblock Number v. 1 in Synthese Historical Library. D. Reidel Publishing
  Company, 1976.

\bibitem[Man96]{Manzano}
Mar\'i{}a Manzano.
\newblock {\em Extensions of First Order Logic}.
\newblock Cambridge University Press, 1996.

\bibitem[Pea09]{Causality}
Judea Pearl.
\newblock {\em Causality: Models, Reasoning and Inference}.
\newblock Cambridge University Press, New York, NY, USA, 2nd edition, 2009.

\bibitem[Pra10]{Prakken2010}
Henry Prakken.
\newblock An abstract framework for argumentation with structured arguments.
\newblock {\em Argument \& Computation}, 1(2):93--124, 2010.

\bibitem[Res85]{HeritageOfLogicalPositivism}
N.~Rescher.
\newblock {\em The Heritage of Logical Positivism}.
\newblock CPS publications in philosophy of science. University Press of
  America, 1985.

\bibitem[SE05]{smokinghistorycriticism}
George~Davey Smith and Matthias Egger.
\newblock The first reports on smoking and lung cancer: why are they
  consistently ignored?
\newblock {\em Bulletin of the World Health Organization}, 83(10):799--800,
  2005.

\bibitem[Ses07]{sesardic}
Neven Sesardic.
\newblock Sudden infant death or murder? a royal confusion about probabilities.
\newblock {\em {The British Journal for the Philosophy of Science}},
  58(2):299--329, 2007.

\bibitem[Sor13]{sep-vagueness}
Roy Sorensen.
\newblock Vagueness.
\newblock In Edward~N. Zalta, editor, {\em The Stanford Encyclopedia of
  Philosophy}. Winter 2013 edition, 2013.

\bibitem[SPMS09]{ColdHit}
Yun~S Song, Anand Patil, Erin~E Murphy, and Montgomery Slatkin.
\newblock Average probability that a ``cold hit'' in a {DNA} database search
  results in an erroneous attribution.
\newblock {\em Journal of forensic sciences}, 54(1):22--27, 2009.

\bibitem[Sut09]{tptp}
G.~Sutcliffe.
\newblock {The {TPTP} Problem Library and Associated Infrastructure: The {FOF}
  and {CNF} Parts, v3.5.0}.
\newblock {\em Journal of Automated Reasoning}, 43(4):337--362, 2009.

\bibitem[SWC00]{snark}
Mark~E Stickel, Richard~J Waldinger, and Vinay~K Chaudhri.
\newblock A guide to {SNARK}.
\newblock Technical report, SRI International, 2000.

\bibitem[Tal13]{sep-epistemology-bayesian}
William Talbott.
\newblock Bayesian epistemology.
\newblock In Edward~N. Zalta, editor, {\em The Stanford Encyclopedia of
  Philosophy}. Fall 2013 edition, 2013.

\bibitem[TCFK83]{tolgyesi}
Eva Tolgyesi, DW~Coble, FS~Fang, and EO~Kairinen.
\newblock A comparative study of beard and scalp hair.
\newblock {\em J Soc Cosmet Chem}, 34:361--382, 1983.

\bibitem[Thu05]{thun}
Michael~J Thun.
\newblock When truth is unwelcome: the first reports on smoking and lung
  cancer.
\newblock {\em Bulletin of the World Health Organization}, 83(2):144--145,
  2005.

\bibitem[Ven98]{ManzanoReview}
Yde Venema.
\newblock Review of ``{E}xtensions of first order logic''.
\newblock {\em The Journal of Symbolic Logic}, 63(3):pp. 1194--1196, 1998.

\bibitem[Wal08]{InformalLogic}
D.N. Walton.
\newblock {\em Informal Logic: A Pragmatic Approach}.
\newblock Cambridge University Press, 2008.

\bibitem[Wal11]{WaltonFindingTheLogic}
D.N. Walton.
\newblock Finding the logic in argumentation.
\newblock {\em International Colloquium, Inside Arguments}, 2011.

\bibitem[WG85]{americansmokingstudy}
Ernest~L Wynder and Evarts~A Graham.
\newblock Tobacco smoking as a possible etiologic factor in bronchiogenic
  carcinoma: A study of six hundred and eighty-four proved cases.
\newblock {\em Jama}, 253(20):2986--2994, 1985.

\bibitem[WK95]{Commitment}
D.~Walton and E.~Krabbe.
\newblock {\em Commitment in Dialogue: Basic concepts of interpersonal
  reasoning}.
\newblock State University of New York Press, Albany NY, 1995.

\end{thebibliography}
